\documentclass[a4paper,11pt]{article}
\pdfoutput=1
\usepackage{jheppub}

\usepackage{amsmath}
\usepackage{amsfonts}
\usepackage{booktabs}
\usepackage[T1]{fontenc}
\usepackage[utf8]{inputenc}
\usepackage{lmodern}
\usepackage{pifont}
\usepackage[group-digits=false]{siunitx}
\usepackage{graphicx}
\usepackage{color}
\usepackage[dvipsnames]{xcolor}
\usepackage{colortbl}
\usepackage[subrefformat=parens]{subcaption}
\usepackage{multirow}
\usepackage{csquotes}
\usepackage{enumerate}
\usepackage{centernot}
\usepackage{cleveref}
\usepackage{xspace} 		

\allowdisplaybreaks

\usepackage{dcolumn}
\newcolumntype{d}[1]{D{.}{.}{#1}}

\usepackage{hyperref}

\definecolor{EmeraldGreen}{HTML}{1ea78d}
\definecolor{EnglishRed}{HTML}{b02427}
\hypersetup{colorlinks=true,urlcolor=EmeraldGreen,citecolor=EmeraldGreen,linkcolor=EnglishRed}

\makeatletter
\newcommand\myparagraph{\@startsection{paragraph}{4}{\z@}%
  {-15\p@ \@plus 6\p@ \@minus 3\p@}%
  {3\p@}%
  {\normalfont\itshape}%
}
\makeatother

\def\draftdate{\relax}
\def\mda{\relax}
\def\mua{\relax}
\def\mla{\relax}
\def\Mda{\relax}
\def\Mua{\relax}
\def\Mla{\relax}
\def\draft{
\def\thtystars{******************************}
\def\sixtystars{\thtystars\thtystars}
\typeout{}
\typeout{\sixtystars**}
\typeout{* Draft mode!
         For final version remove \protect\draft\space in source file *}
\typeout{\sixtystars**}
\typeout{}
\def\draftdate{\today}
\def\mua{\marginpar[\boldmath\hfil$\uparrow$]%
                   {\boldmath$\uparrow$\hfil}%
                    \typeout{marginpar: $\uparrow$}\ignorespaces}
\def\mda{\marginpar[\boldmath\hfil$\downarrow$]%
                   {\boldmath$\downarrow$\hfil}%
                    \typeout{marginpar: $\downarrow$}\ignorespaces}
\def\mla{\marginpar[\boldmath\hfil$\rightarrow$]%
                   {\boldmath$\leftarrow $\hfil}%
                    \typeout{marginpar: $\leftrightarrow$}\ignorespaces}
\def\Mua{\marginpar[\boldmath\hfil$\Uparrow$]%
                   {\boldmath$\Uparrow$\hfil}%
                    \typeout{marginpar: $\uparrow$}\ignorespaces}
\def\Mda{\marginpar[\boldmath\hfil$\Downarrow$]%
                   {\boldmath$\Downarrow$\hfil}%
                    \typeout{marginpar: $\downarrow$}\ignorespaces}
\def\Mla{\marginpar[\boldmath\hfil$\Rightarrow$]%
                   {\boldmath$\Leftarrow $\hfil}%
                    \typeout{marginpar: $\leftrightarrow$}\ignorespaces}
\overfullrule 5pt
\oddsidemargin 15mm
\marginparwidth 29mm
}

\colorlet{mygood}{green!50!black}
\colorlet{mybad}{red!80!black}

\newcommand{\bonsay}{\textsc{Bonsay}\xspace}
\newcommand{\recola}{\textsc{Recola}\xspace}
\newcommand{\openloops}{\textsc{OpenLoops2}\xspace}
\newcommand{\collier}{\textsc{Collier}\xspace}
\newcommand{\mocanlo}{\textsc{MoCaNLO}\xspace}

\newcommand{\lhapdf}{\textsc{LHAPDF6}\xspace}
\newcommand{\powheg}{\textsc{Powheg}\xspace}
\newcommand{\sherpa}{\textsc{Sherpa}\xspace}
\newcommand{\mathematica}{\textsc{Mathematica}\xspace}
\newcommand{\feynarts}{\textsc{FeynArts1}\xspace}

\newcommand{\alphas}{\alpha_\mathrm{s}}
\newcommand{\ci}{\mathrm{i}}
\newcommand{\eg}{\text{e.g.}\xspace}
\newcommand{\ie}{\text{i.e.}\xspace}
\newcommand{\Pgne}{\nu_\mathrm{e}}
\newcommand{\Pgmp}{\mu^+}
\newcommand{\Pgngm}{\nu_\mu}
\newcommand{\Pqu}{\mathrm{u}}
\newcommand{\Pqc}{\mathrm{c}}
\newcommand{\Pqd}{\mathrm{d}}
\newcommand{\Pqs}{\mathrm{s}}
\newcommand{\Paqu}{\bar{\mathrm{u}}}
\newcommand{\Paqc}{\bar{\mathrm{c}}}
\newcommand{\Paqd}{\bar{\mathrm{d}}}
\newcommand{\Paqs}{\bar{\mathrm{s}}}

\def\mathswitchr#1{\relax\ifmmode{\mathrm{#1}}\else$\mathrm{#1}$\fi}

\newcommand{\PW}{\mathswitchr W}
\newcommand{\Pw}{\mathswitchr w}
\newcommand{\PZ}{\mathswitchr Z}

\newcommand{\Pg}{\mathswitchr g}
\newcommand{\PH}{\mathswitchr H}

\newcommand{\Pe}{\mathswitchr e}

\newcommand{\Pp}{\mathswitchr p}
\newcommand{\Pd}{\mathswitchr d}
\newcommand{\Pu}{\mathswitchr u}

\newcommand{\Pb}{\mathswitchr b}

\newcommand{\Pep}{\mathswitchr {e^+}}
\newcommand{\Pem}{\mathswitchr {e^-}}
\newcommand{\PWp}{\mathswitchr {W^+}}
\newcommand{\PWm}{\mathswitchr {W^-}}
\newcommand{\PWpm}{\mathswitchr {W^\pm}}

\newcommand{\jets}{\mathrm{jets}}

\def\mathswitch#1{\relax\ifmmode#1\else$#1$\fi}

\newcommand{\MW}{\mathswitch {M_\PW}}

\newcommand{\MZ}{\mathswitch {M_\PZ}}
\newcommand{\MH}{\mathswitch {M_\PH}}
\newcommand{\Me}{\mathswitch {m_\Pe}}

\newcommand{\lsim}
{\;\raisebox{-.3em}{$\stackrel{\displaystyle <}{\sim}$}\;}
\newcommand{\gsim}
{\;\raisebox{-.3em}{$\stackrel{\displaystyle >}{\sim}$}\;}

\def\refeq#1{\mbox{(\ref{#1})}}

\def\reffi#1{\mbox{Fig.~\ref{#1}}}
\def\reffis#1{\mbox{Figs.~\ref{#1}}}
\def\refta#1{\mbox{Table~\ref{#1}}}

\def\refse#1{\mbox{Section~\ref{#1}}}
\def\refses#1{\mbox{Sections~\ref{#1}}}
\def\refapp#1{\mbox{App.~\ref{#1}}}
\def\citere#1{\mbox{Ref.~\cite{#1}}}
\def\citeres#1{\mbox{Refs.~\cite{#1}}}

\newcommand{\TeV}{\unskip\,\mathrm{TeV}}
\newcommand{\GeV}{\unskip\,\mathrm{GeV}}

\DeclareMathOperator{\real}{Re} 

\title{Like-Sign W-Boson Scattering at the LHC --
	Approximations and Full Next-to-Leading-Order Predictions}

\preprint{{\small FR-PHENO-2023-09, IRMP-CP3-23-42}}

\author[a]{Stefan Dittmaier,}
\author[a]{Philipp Maierh\"ofer,}
\author[b]{Christopher Schwan,}
\author[c]{Ramon Winterhalder}

\affiliation[a]{Physikalisches Institut, Albert-Ludwigs-Universit\"at Freiburg, \\
	Hermann-Herder-Straße 3, 79104 Freiburg, Germany}
\affiliation[b]{Institut f\"ur Theoretische Physik und Astrophysik, Universit\"at W\"urzburg, \\
	Emil-Hilb-Weg 22, 97074 W\"urzburg, Germany}
\affiliation[c]{Centre for Cosmology, Particle Physics and Phenomenology (CP3),\\  Universit\'e catholique de Louvain, 
	Chemin du Cyclotron 2, B-1348 Louvain-la-Neuve, Belgium}

\emailAdd{stefan.dittmaier@physik.uni-freiburg.de}
\emailAdd{christopher.schwan@physik.uni-wuerzburg.de}
\emailAdd{ramon.winterhalder@uclouvain.be}

\abstract{We present a new calculation of next-to-leading-order corrections of the strong and electroweak interactions to like-sign W-boson scattering at the Large Hadron Collider, implemented in the Monte Carlo integrator \bonsay. 
The calculation includes leptonic decays of the \PW~bosons. 
It comprises the whole tower of next-to-leading-order contributions to the cross section, which scale like $\alphas^3\alpha^4$, $\alphas^2\alpha^5$, $\alphas\alpha^6$, and $\alpha^7$ in the strong and electroweak couplings $\alphas$ and $\alpha$. 
We present a detailed survey of numerical results confirming the occurrence of large pure electroweak corrections of the order of $\sim-12\%$ for integrated cross sections and even larger corrections in high-energy tails of distributions. 
The electroweak corrections account for the major part of the complete next-to-leading-order correction, 
which amounts to $15{-}20\%$ in size, depending on the details of the event selection chosen for analysing vector-boson-scattering. 
Moreover, we compare the full next-to-leading-order corrections to approximate results based on the neglect of contributions that are not enhanced by the vector-boson scattering kinematics 
\textit{(VBS approximation)} and on resonance expansions for the \PW-boson decays 
\textit{(double-pole approximation)}; 
the quality of this approximation is good within $\lsim1.5\%$ for integrated cross sections and the dominating parts of the differential distributions. 
Finally, for the leading-order predictions, we construct different versions of \textit{effective vector-boson approximations}, which are based on cross-section contributions that are enhanced by collinear emission of \PW~bosons off the initial-state (anti)quarks; in line with previous findings in the literature, it turns out that the approximative quality is rather limited for applications at the LHC.
}

\begin{document}

\maketitle
\flushbottom
\clearpage

\section{Introduction}
\label{sec:introduction}

The experimental investigation of the electroweak (EW) symmetry-breaking mechanism is one of the major physics goals for which the Large Hadron Collider (LHC) was built. The discovery of a neutral ``Higgs-like'' boson in 2012 is undoubtedly a milestone in reaching this goal. The analysis of existing LHC data shows that the new particle behaves very much like the Higgs boson of the Standard Model (SM). However, the empirical consistency of observations with the SM does not rule out the possibility of new physics being at work in the sector of EW
symmetry breaking (EWSB). This possibility is phenomenologically investigated in two complementary ways: (i) direct (on-shell) Higgs-boson production, including the determination of coupling strengths and structures of the observed Higgs boson and the search for additional Higgs bosons, and (ii) the investigation of indirect (off-shell) effects of the Higgs boson in processes like EW vector-boson scattering (VBS) or EW multi-boson production.

In this work, we focus on like-sign \PW-boson scattering with the signature of two equally charged leptons and two jets, which is the most prominent VBS channel at the LHC due to its favourable signal to (irreducible) background ratio. 
Apart from its role as a window to EWSB, VBS is also phenomenologically interesting because it is dominated by self-interactions of the EW gauge bosons at an energy level in the TeV range where the LHC explores uncharted territory. 
However, given the consistency of the discovered Higgs-like particle with the SM Higgs boson and the absence of spectacular signals of new physics, potential deviations from the SM are small and subtle. 
In fact, cross-section measurements of like-sign \PW-pair production~\cite{Aad:2014zda,Khachatryan:2014sta,Aaboud:2016ffv,Sirunyan:2017ret,Aaboud:2019nmv,Aad:2019xxo,Sirunyan:2019der,Sirunyan:2020gyx} still have significant statistical errors and are compatible with SM predictions. 
In view of the higher precision expected from measurements in the upcoming high-luminosity phase of the LHC, very precise predictions for VBS reactions are required, \ie higher-order corrections of the strong and EW interactions have to be known with sufficient precision.
The current state of the art in precision calculations for VBS at the LHC is, \eg, summarized in \citere{Covarelli:2021gyz}, and prospects for VBS analyses at the LHC and beyond are discussed in \citere{BuarqueFranzosi:2021wrv}.

Already in leading-order (LO) predictions, there are two categories of diagrams contributing to the amplitude of $\Pp\Pp\to\ell^\pm\ell^{\prime\pm}\nu_\ell\nu_{\ell'}+2\jets+X$: 
diagrams with EW couplings only and diagrams involving gluon exchange, leading to LO contributions to the cross section proportional to the factors $\alphas^2\alpha^4$, $\alphas\alpha^5$, and
$\alpha^6$ in the EW and strong fine-structure constants
$\alpha$ and $\alphas$, respectively.
At next-to-leading order (NLO), there are cross-section contributions with four different combinations of strong and weak couplings:
$\alphas^3\alpha^4$, $\alphas^2\alpha^5$, $\alphas\alpha^6$, and $\alpha^7$.
A summary and a tuned comparison of results of precision calculations for $\Pp\Pp\to\ell^\pm\ell^{\prime\pm}\nu_\ell\nu_{\ell'}+2\jets+X$ can be found in \citere{Ballestrero:2018anz}. 
Already a long time ago, two categories of corrections with
QCD couplings were considered in the literature: the QCD corrections at $\alphas\alpha^6$ to VBS in {\it VBS approximation}~\cite{Jager:2009xx,Denner:2012dz}, including its matching to a QCD parton shower within \powheg~\cite{Frixione:2007vw,Alioli:2010xd,Jager:2011ms}, and the $\alphas^3\alpha^4$ contribution, which can be considered as
$\PW^\pm\PW^\pm+2\jets$ production via the strong 
interaction~\cite{Melia:2010bm,Greiner:2012im,Campanario:2013gea}.
However, for the remaining two categories, which involve EW or combined QCD--EW corrections, there still only exist results from a single group in the literature~\cite{Biedermann:2016yds,Biedermann:2017bss,Chiesa:2019ulk}.
The primary motivation of the present paper is to provide a survey of independent results on the whole tower of NLO corrections, together with a comparison to the results given in \citeres{Biedermann:2016yds,Biedermann:2017bss,Chiesa:2019ulk}.
Since NLO calculations to multi-particle processes of the type $2\to6$ particles are very challenging and by no means standard, this cross-validation is an essential step in establishing the full NLO prediction for like-sign W-pair production at the LHC.

The presented NLO results are produced with the Monte Carlo program \bonsay, which has already been used in some NLO calculations for many-particle processes.
QCD corrections of orders $\alphas\alpha^6$ and $\alphas^3\alpha^4$ to $\PW^\pm\PW^\pm+2\jets$ production evaluated with \bonsay were already included in the tuned comparison of \citere{Ballestrero:2018anz},
and EW and combined QCD--EW corrections to WZ~scattering calculated with \bonsay were discussed and successfully compared to another independent calculation in \citere{Denner:2019tmn}.
Moreover, the calculation of the full NLO QCD and EW corrections to 
$\Pp\Pp\to\PW\PW\PW\to3\ell3\nu_\ell+X$ presented in \citere{Dittmaier:2019twg} was performed with \bonsay.
Technically, \bonsay is a standalone multi-channel Monte Carlo integrator that can be linked to amplitude generators like \openloops~\cite{Cascioli:2011va,Buccioni:2019sur} and \recola~\cite{Actis:2016mpe,Denner:2017wsf} and other libraries like \lhapdf~\cite{Buckley:2014ana}.

Apart from presenting results based on a full NLO calculation, we delve into approximations for the considered cross section that hold phenomenological or practical significance.
Firstly, we explore the combined application of the \textit{double-pole approximation} (DPA) and the \textit{VBS approximation}. This aims to investigate if, or to what extent, the full NLO prediction can be reproduced within a theoretically simpler and computationally more efficient framework.
Such approximations are desirable for adding further corrections or 
effects from physics beyond the SM (BSM) where a full $2\to6$ particle
NLO calculation might be computationally too expensive and most likely not necessary.
The VBS approximation reduces the number of partonic channels contributing to $\Pp\Pp\to\ell^\pm\ell^{\prime\pm}\nu_\ell\nu_{\ell'}+2\jets+X$ by selecting only those enhanced by the VBS selection cuts, whilst neglecting suppressed interference terms, as done in the calculations of \citeres{Jager:2009xx,Denner:2012dz}.
The DPA further slims down the number of diagrams within the selected channels by considering only the leading term from expanding the amplitudes around the resonances of the produced \PW~bosons.
The DPA was established in cross-section predictions for
\PW-pair production LEP2 in different variants by various groups~~\cite{Beenakker:1998gr,%
Jadach:1996hi,Jadach:1998tz,%
Denner:1999kn,Denner:2000bj,Denner:2002cg,%
Kurihara:2001um}
(see also \citere{Denner:2019vbn} for the general concept).
Our approach aligns closely with the version proposed in \citeres{Denner:1999kn,Denner:2000bj,Denner:2002cg}, which 
applies the resonance expansion only to the infrared-finite part of the virtual corrections and which
was extended to more general final states in \citere{Dittmaier:2015bfe}.

As a second type of approximation, we formulate an \textit{effective vector-boson approximation} (EVA)~\cite{KANE1984367,Dawson:1984gx,CHANOWITZ198485,Lindfors:1985yp} for LO predictions. 
The EVA is based on the enhancement of vector-boson emission off the incoming (anti)quarks in the collinear limit, which leads to the picture of vector bosons being partons of the protons with extra jets pointing into the forward/backward regions.
It is well known that the approximation quality of the EVA is rather limited~\cite{GUNION198657, Kuss:1995yv,Kuss:1996ww,%
ACCOMANDO200674, Accomando:2006hq, Borel2012, Brehmer201490,Bernreuther16, Ruiz:2021tdt}, because the requirement that the partonic scattering energy $\sqrt{\hat s}$ relative to the \PW-boson mass $\MW$ obeys $\ln(\hat s/\MW^2)\gg1$ is not perfectly fulfilled at the LHC.
Nevertheless, the quality of the EVA gives an impression of the extent to which the considered cross section is dominated by the VBS mechanism that is the target of the experimental analysis of the VBS signature. 
Moreover, the EVA might be a promising ansatz for quick qualitative studies of corrections or BSM effects in VBS channels.

The article is organized as follows: 
In \refse{sec:process-definition}, we illustrate all ingredients of the NLO calculation in terms of sample diagrams and describe the structure of the calculation and the employed mathematical techniques.
Subsequently, in \refse{sec:approximations}, we explain the underlying concepts of the VBS approximation and the DPA, which are combined to an approximative NLO prediction as an alternative to the full NLO calculation. 
Since the EVA for LO predictions is less important, it is only briefly sketched in that section, and the full details are given in \refapp{sec:evba-const}.
In \refse{sec:validation}, we describe the internal checks on our calculation and perform a brief comparison to existing NLO results in the literature.
\refse{sec:results} presents a detailed survey of numerical
results tailored to analyses performed in the ATLAS experiment and comparisons of the full LO and NLO predictions to the constructed approximations.
Our conclusions are given in \refse{sec:conclusion}.

\section{Like-sign W-boson scattering at the LHC}
\label{sec:process-definition}


We consider the production of two equally charged leptons of different flavour, together with at least two jets:
\begin{equation}
\Pp \Pp \to \Pep \Pgne \, \Pgmp \Pgngm \, \mathrm{j} \mathrm{j} + X \text{.}
\label{eq:VBSprocess}
\end{equation}
The cross sections for the production of $\Pep \Pgne \, \Pep \Pgne \, \mathrm{j} \mathrm{j}$ and  $\Pgmp \Pgngm \, \Pgmp \Pgngm \, \mathrm{j} \mathrm{j}$ are approximately half of the cross section for $\Pep \Pgne \, \Pgmp \Pgngm \, \mathrm{j} \mathrm{j}$ production. 
The small deviations from the factor $1/2$ are due to the interference effects between diagram types
related by the exchange of identical leptons in the final state, which are suppressed.
Note that in the SM the production of any pair $\ell^+\ell^{\prime +}$ of equally charged leptons comes with a corresponding neutrino pair $\nu_\ell \nu_{\ell'}$ and at least two QCD partons which are quarks or antiquarks. The additional (anti)quarks in the final state are necessary, since the partonic initial states cannot provide the overall charge of $+2$ of the produced leptons.
This means that it would be phenomenologically sound to define a total
cross section for $\Pe^\pm\mu^\pm+X$ production, where $X$ accounts for any
hadronic activity in the detector without explicit jet detection;
demanding two tagging jets, however, rules out events with only soft hadronic 
activity as, \eg, induced by double-parton scattering, which would
contaminate the analysis of EW VBS.

\subsection{LO contributions}
\label{sec:lo_part}

\begin{figure}
\centering
\begin{subfigure}{0.33\textwidth}
\centering
\includegraphics{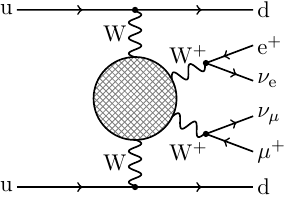}
\caption{VBS, doubly-resonant}
\label{fig:born_qq_vbs}
\end{subfigure}
\begin{subfigure}{0.33\textwidth}
\centering
\includegraphics{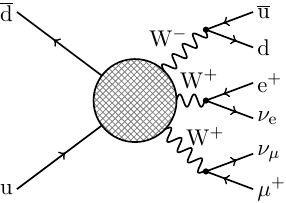}
\caption{triple-boson production}
\label{fig:born_qq_www}
\end{subfigure}%
\begin{subfigure}{0.33\textwidth}
\centering
\includegraphics{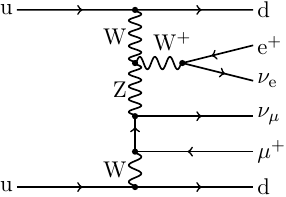}
\caption{singly-resonant}
\label{fig:born_qq_W}
\end{subfigure}%
\par\bigskip
\begin{subfigure}{0.33\textwidth}
\centering
\includegraphics{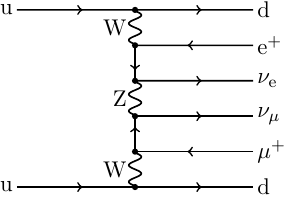}
\caption{non-resonant}
\label{fig:born_qq_nonres}
\end{subfigure}%
\begin{subfigure}{0.33\textwidth}
\centering
\includegraphics{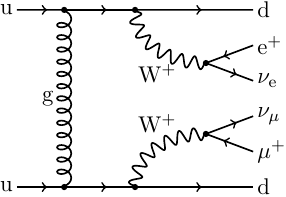}
\caption{QCD contribution}
\label{fig:born_qq_gluon}
\end{subfigure}%
\caption{Examples of LO Feynman diagrams illustrating the various $\PWp\PWp$ production mechanisms.
The shaded blobs represent tree-level subdiagrams for 
$\PWp \PWp\to \PWp \PWp$ and $\Pu \Paqd \to \PWp \PWp \PWm$, respectively.}
\label{fig:born-diagrams}
\end{figure}

\begin{figure}
\centering
\begin{subfigure}{0.33\textwidth}
\centering
\includegraphics{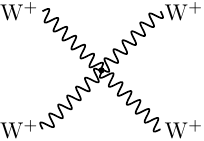}
\caption{4-point interaction}
\label{fig:vbs_4}
\end{subfigure}%
\begin{subfigure}{0.33\textwidth}
\centering
\includegraphics{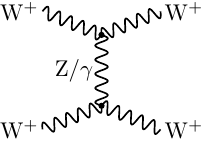}
\caption{$t$ channel}
\label{fig:vbs_t}
\end{subfigure}%
\begin{subfigure}{0.33\textwidth}
\centering
\includegraphics{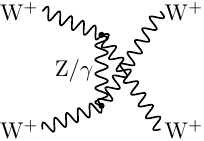}
\caption{$u$ channel}
\label{fig:vbs_u}
\end{subfigure}%
\par\bigskip
\begin{subfigure}{0.33\textwidth}
\centering
\includegraphics{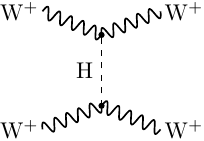}
\caption{$t$-channel Higgs}
\label{fig:vbs_t_higgs}
\end{subfigure}%
\begin{subfigure}{0.33\textwidth}
\centering
\includegraphics{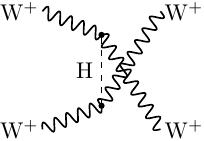}
\caption{$u$-channel Higgs}
\label{fig:vbs_u_higgs}
\end{subfigure}%
\caption{VBS subdiagrams for $\PWp \PWp\to \PWp \PWp$ as contained in the shaded blob of diagram Fig.~\ref{fig:born_qq_vbs}.}
\label{fig:vbs-diagrams}
\end{figure}

\begin{figure}
        \centering
        \begin{subfigure}{0.49\textwidth}
                \centering
                \includegraphics[width=0.95\textwidth]{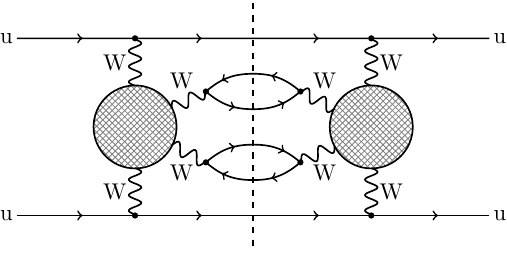}
                \caption{purely EW contribution of order $\mathcal{O}(\alpha^6)$}
                \label{fig:int_born_ew_ew}
        \end{subfigure}
        \begin{subfigure}{0.49\textwidth}
                \centering
                \includegraphics[width=0.95\textwidth]{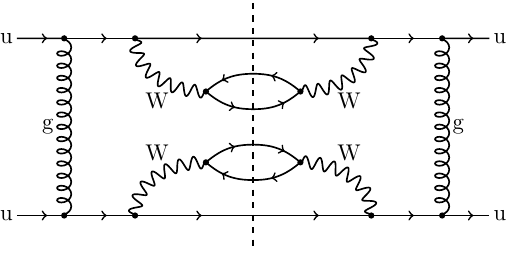}
                \caption{QCD contribution of order $\mathcal{O}(\alpha_s^2\alpha^4)$}
                \label{fig:int_born_qcd_qcd}
        \end{subfigure}

        \par\bigskip
        \begin{subfigure}{0.49\textwidth}
                \centering
                \includegraphics[width=0.95\textwidth]{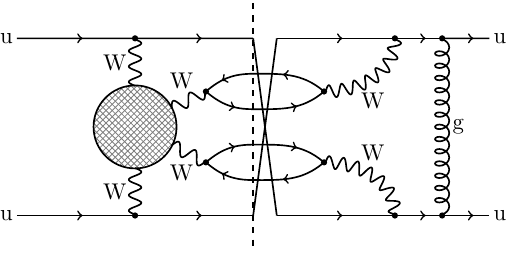}
                \caption{interference contribution of order $\mathcal{O}(\alpha_s\alpha^5)$}
                \label{fig:int_born_ew_qcd}
        \end{subfigure}
        \caption{Examples of squared LO Feynman diagrams illustrating the various $\PWp \PWp$ production mechanisms.
                The shaded blobs represent tree-level subdiagrams for $\PWp \PWp\to \PWp \PWp$.}
        \label{fig:int_born-diagrams}
\end{figure}

At LO, the process \refeq{eq:VBSprocess}
is special in the sense that it has no gluons in the initial states, because the two final-state (anti)quarks cannot fully compensate the double-charge of the leptons in the final state.
This also restricts the number of partonic processes which we group according to total charge $Q_{\mathrm{in}}$ of the initial state, suppressing trivial transpositions of the initial state.
For $Q_{\mathrm{in}}=4/3$, the partonic contributions are
\begin{subequations}
\begin{alignat}{2}
\Pqu \Pqu   &\to \Pep \Pgne \, \Pgmp \Pgngm \, \Pqd \Pqd   \text{,} &\qquad
\Pqc \Pqc   &\to \Pep \Pgne \, \Pgmp \Pgngm \, \Pqs \Pqs   \text{,} \label{eq:q43-lo-me-u-and-t} \\
\Pqc \Pqu   &\to \Pep \Pgne \, \Pgmp \Pgngm \, \Pqd \Pqs   \text{,}
\end{alignat}
\label{eq:q43-lo-mes}%
\end{subequations}
the following processes have charge $Q_{\mathrm{in}}=1$,
\begin{subequations}
\begin{alignat}{2}
\Paqd \Pqu  &\to \Pep \Pgne \, \Pgmp \Pgngm \, \Pqd \Paqu  \text{,} &\qquad
\Paqs \Pqc  &\to \Pep \Pgne \, \Pgmp \Pgngm \, \Pqs \Paqc  \text{,} \label{eq:q33-lo-me-u-and-t} \\
\Paqd \Pqu  &\to \Pep \Pgne \, \Pgmp \Pgngm \, \Pqs \Paqc  \text{,} &\qquad
\Paqs \Pqc  &\to \Pep \Pgne \, \Pgmp \Pgngm \, \Pqd \Paqu  \text{,} \\
\Paqd \Pqc  &\to \Pep \Pgne \, \Pgmp \Pgngm \, \Pqs \Paqu  \text{,} &\qquad
\Paqs \Pqu  &\to \Pep \Pgne \, \Pgmp \Pgngm \, \Pqd \Paqc  \text{,}
\end{alignat}
\label{eq:q33-lo-mes}%
\end{subequations}
and finally there are processes with $Q_{\mathrm{in}}=2/3$,
\begin{subequations}
\begin{alignat}{2}
\Paqd \Paqd &\to \Pep \Pgne \, \Pgmp \Pgngm \, \Paqu \Paqu \text{,} &\qquad
\Paqs \Paqs &\to \Pep \Pgne \, \Pgmp \Pgngm \, \Paqc \Paqc \text{,} \label{eq:q23-lo-me-u-and-t} \\
\Paqs \Paqd &\to \Pep \Pgne \, \Pgmp \Pgngm \, \Paqu \Paqc \text{.}
\end{alignat}
\label{eq:q23-lo-mes}%
\end{subequations}
Here we have ignored the possibility of mixing between the quark generations, 
\ie the CKM matrix is taken as the unit matrix, so that in particular
\Pb-quarks neither appear in the initial nor in the final state, because top-quarks in the initial or final state are not relevant. Taking a trivial CKM matrix is an excellent approximation for $\PWpm\PWpm$ scattering
for the following reasons.
Ignoring the negligible quark mixing with the third generation and 
taking the quarks of the first two generations as massless, all mixing effects drop out in the dominating VBS amplitudes, so that a non-trivial CKM matrix would not have any effect.
A non-trivial CKM matrix would only influence 
the quark--antiquark annihilation channels which deliver only a small contribution to the cross section.
Denoting the EW and the strong coupling constants as $e$ and $g_\mathrm{s}$, respectively, the Feynman diagrams for the partonic 
LO contributions are either of $\mathcal{O}(e^6)$, as illustrated in Figs.~\ref{fig:born_qq_vbs}--\ref{fig:born_qq_nonres}, or of $\mathcal{O}(g_\mathrm{s}^2 e^4)$ with an internal gluon, as illustrated in Fig.~\ref{fig:born_qq_gluon}.
Only the contribution of $\mathcal{O} (e^6)$ contains the actual scattering of W~bosons which is induced
by the diagram classes shown in Fig.~\ref{fig:vbs-diagrams}.
Squaring the LO amplitudes leads to cross-section contributions of
\begin{equation}
\mathcal{O} (\alpha_\mathrm{s}^2 \alpha^4) \text{,} \quad
\mathcal{O} (\alpha_\mathrm{s} \alpha^5)   \text{,} \quad
\mathcal{O} (\alpha^6)                     \text{,}
\end{equation}
which are often called strong production, interference, and EW production, respectively.
The three contributions are illustrated in \reffi{fig:int_born-diagrams}.
For the interference contributions to the squared matrix elements, 
the colour structure of this process (see also the discussion in Sec.~\ref{sec:approximations}) leads to vanishing matrix elements except for those where all quarks are from the same 
generation, \ie the partonic contributions \eqref{eq:q43-lo-me-u-and-t}, \eqref{eq:q33-lo-me-u-and-t}, and \eqref{eq:q23-lo-me-u-and-t}.

\subsection{NLO contributions}
\label{sec:nlo_part}

\begin{figure}
\centering
\begin{subfigure}{0.33\textwidth}
\centering
\includegraphics{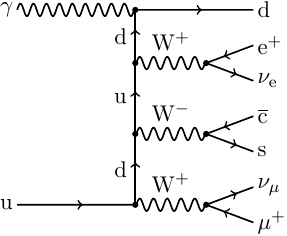}
\caption{initial photon, triple-W prod.}
\label{fig:initial-photon-www}
\end{subfigure}
\begin{subfigure}{0.33\textwidth}
\centering
\includegraphics{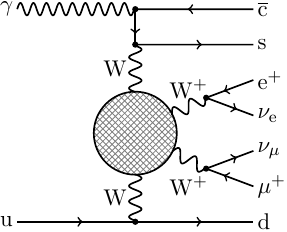}
\caption{initial photon, VBS}
\label{fig:initial-photon-vbs}
\end{subfigure}
\caption{Sample diagrams for real corrections.}
\label{fig:initial-photon}
\end{figure}

\begin{figure}
\centering
\begin{subfigure}{0.33\textwidth}
\centering
\includegraphics{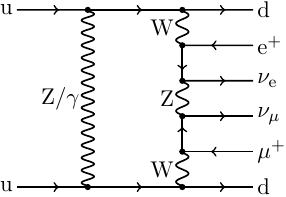}
\caption{8-point function}
\label{fig:virt_8pt}
\end{subfigure}%
\begin{subfigure}{0.33\textwidth}
\centering
\includegraphics{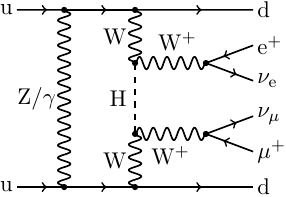}
\caption{6-point function with Higgs}
\label{fig:virt_Higgs}
\end{subfigure}%
\begin{subfigure}{0.33\textwidth}
\centering
\includegraphics{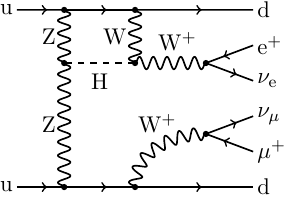}
\caption{Higgs exchange}
\label{fig:virt_Higgsb}
\end{subfigure}\par\bigskip
\begin{subfigure}{0.33\textwidth}
\centering
\includegraphics{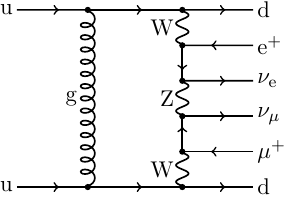}
\caption{gluon between quark lines}
\label{fig:virt_gl}
\end{subfigure}%
\begin{subfigure}{0.33\textwidth}
\centering
\includegraphics{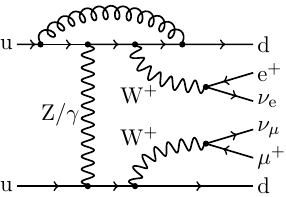}
\caption{gluon at single quark line}
\label{fig:virt_gl_sameline}
\end{subfigure}%
\begin{subfigure}{0.33\textwidth}
	\centering
	\includegraphics{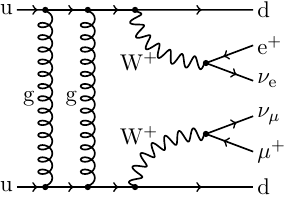}
	\caption{double gluon exchange}
	\label{fig:virt_gl_double_qcd}
\end{subfigure}%
\caption{Examples of one-loop diagrams for EW (a-c) and QCD (d-f) corrections to the partonic subprocess $\Pu \Pu \to \Pd \Pd \, \Pep \Pgne \, \Pgmp \Pgngm$.}
\end{figure}

\begin{figure}
        \centering
        \begin{subfigure}{0.49\textwidth}
                \centering
                \includegraphics[width=0.95\textwidth]{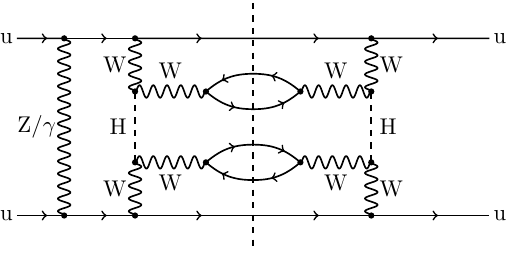}
                \caption{EW correction of order $\mathcal{O}(\alpha^7)$}
                \label{fig:int_virtual_ew_ew}
        \end{subfigure}
        \begin{subfigure}{0.49\textwidth}
                \centering
                \includegraphics[width=0.95\textwidth]{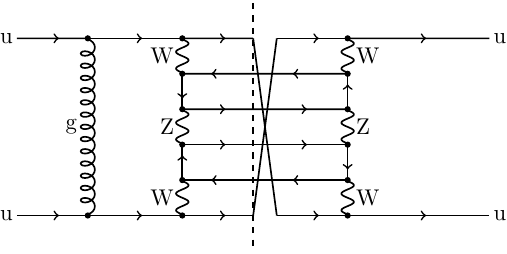}
                \caption{Mixed EW-QCD correction of order $\mathcal{O}(\alpha_s\alpha^6)$}
                \label{fig:int_virtual_qcd_qcd}
        \end{subfigure}

        \par\bigskip
        \begin{subfigure}{0.49\textwidth}
                \centering
                \includegraphics[width=0.95\textwidth]{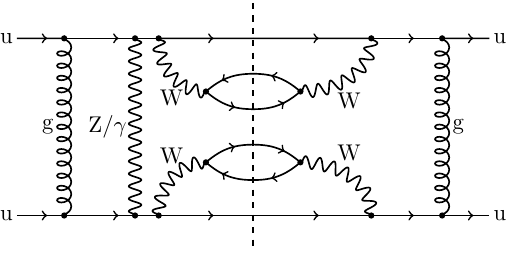}
                \caption{Mixed EW-QCD correction of order $\mathcal{O}(\alpha_s^2\alpha^5)$}
                \label{fig:int_virtual_ew_qcd}
        \end{subfigure}
\begin{subfigure}{0.49\textwidth}
        \centering
        \includegraphics[width=0.95\textwidth]{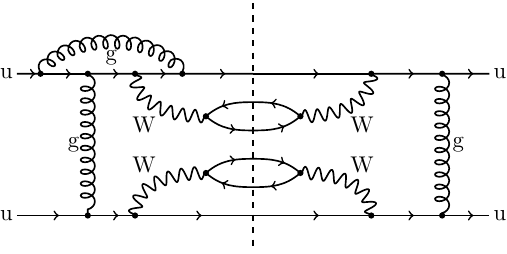}
        \caption{QCD correction of order $\mathcal{O}(\alpha_s^3\alpha^4)$}
        \label{fig:int_virtual_qcd_ew}
\end{subfigure}
        \caption{Examples of squared one-loop Feynman diagrams illustrating the various contributions to the NLO cross section.}
        \label{fig:int_virtual-diagrams}
\end{figure}

At NLO, both virtual and real corrections must be considered.
The partonic processes of the real corrections are simply the ones listed in Eqs.~\eqref{eq:q43-lo-mes}, \eqref{eq:q33-lo-mes}, and \eqref{eq:q23-lo-mes} with either an additional external photon or an additional external gluon.
Thus, there are bremsstrahlung contributions corresponding to each of the partonic channels in  Eqs.~\eqref{eq:q43-lo-mes}, \eqref{eq:q33-lo-mes}, and \eqref{eq:q23-lo-mes}. Moreover, there are quark--gluon-initiated processes with $Q_{\mathrm{in}}=2/3$,
\begin{subequations}
\begin{alignat}{2}
\Pqu \Pg &\to \Pep \Pgne \, \Pgmp \Pgngm \, \Pqd \Pqd \Paqu \text{,} &\quad
\Pqc \Pg &\to \Pep \Pgne \, \Pgmp \Pgngm \, \Pqs \Pqs \Paqc \text{,} \\
\Pqu \Pg &\to \Pep \Pgne \, \Pgmp \Pgngm \, \Pqs \Pqd \Paqc \text{,} &\quad
\Pqc \Pg &\to \Pep \Pgne \, \Pgmp \Pgngm \, \Pqd \Pqs \Paqu \text{,}
\end{alignat}
\end{subequations}
and $Q_{\mathrm{in}}=1/3$,
\begin{subequations}
\begin{alignat}{2}
\Paqd \Pg &\to \Pep \Pgne \, \Pgmp \Pgngm \, \Pqd \Paqu \Paqu \text{,} &\quad
\Paqs \Pg &\to \Pep \Pgne \, \Pgmp \Pgngm \, \Pqs \Paqc \Paqc \text{,} \\
\Paqd \Pg &\to \Pep \Pgne \, \Pgmp \Pgngm \, \Pqs \Paqu \Paqc \text{,} &\quad
\Paqs \Pg &\to \Pep \Pgne \, \Pgmp \Pgngm \, \Pqd \Paqc \Paqu \text{.}
\end{alignat}
\end{subequations}
Similar processes exist with the gluon replaced by the photon,
as illustrated in \reffi{fig:initial-photon}.
Some virtual one-loop Feynman diagrams 
leading to contributions of $\mathcal{O} (e^8)$
are shown in Figs.~\ref{fig:virt_8pt}, \ref{fig:virt_Higgs}, and \ref{fig:virt_Higgsb}. 
Diagrams of $\mathcal{O} (g_s^2 e^6)$ are shown in Figs.~\ref{fig:virt_gl} and \ref{fig:virt_gl_sameline}, and $\mathcal{O} (g_s^4 e^4)$ in Fig.~\ref{fig:virt_gl_double_qcd}.
Interfering these loop diagrams with the 
two possible types of LO graphs, 
as shown in \reffi{fig:int_virtual-diagrams},
we obtain four types of NLO contributions to cross sections:
\begin{equation}
\mathcal{O} (\alpha_\mathrm{s}^3 \alpha^4) \text{,} \quad
\mathcal{O} (\alpha_\mathrm{s}^2 \alpha^5) \text{,} \quad
\mathcal{O} (\alpha_\mathrm{s}   \alpha^6) \text{,} \quad
\mathcal{O} (\alpha^7)                     \text{.}
\end{equation}
Only the first and the last order are corrections of pure QCD and EW origin, respectively,
the other two are corrections of mixed type.

Technically, all calculations are performed within the Monte Carlo program \bonsay, which is based on multi-channel Monte Carlo integration using
adaptive weight optimization~\cite{BERENDS1994308, BERENDS1995437, KLEISS1994141}. Besides calculating cross sections, it further supports the parallel computation of uncertainties induced by different
scale choices and errors in parton distribution functions (PDFs).
The phase-space integration is particularly optimized for multi-particle processes with matrix elements containing multiple peaking or resonance structures. For this, \bonsay extends and implements the algorithm presented in Section~3 of Ref.~\cite{Dittmaier:2002ap}.
Both the tree-level and one-loop matrix elements are provided by
\openloops~\cite{Cascioli:2011va,Buccioni:2019sur} by default,
but have been cross-validated against respective results obtained with
\recola~\cite{Actis:2016mpe,Denner:2017wsf}.
The one-loop integrals are numerically evaluated using
\collier~\cite{Denner:2016kdg}, which is based on the methods and results
described in \citeres{Denner:2002ii,Denner:2005nn,Denner:2010tr}.
In detail, \collier\ provides a sophisticated system of expansion techniques
to stabilize the results for tensor integrals
in the vicinity of exceptional phase-space configurations
and offers various possibilities for further cross-checks, such as two independent 
internal implementations.
Particle resonances are described in the complex-mass 
scheme~\cite{Denner:1999gp,Denner:2005fg,Denner:2019vbn}
to guarantee both gauge independence of amplitudes and NLO accuracy in both resonant and non-resonant phase-space regions.
The extraction and cancellation of (soft and collinear) infrared singularities is accomplished within the dipole subtraction formalism both for
QCD~\cite{Catani:1996vz,Catani:2002hc} and EW~\cite{Dittmaier:1999mb,Dittmaier:2008md} corrections.


%

\section{Approximations for VBS}
\label{sec:approximations}

As an alternative to the full NLO prediction, we construct a simpler and approximate prediction
based on the combination of 1) a VBS approximation and 2) a double-pole approximation.

Independently, we investigate the feasibility of an effective vector-boson approximation (EVA)
for the LO prediction.

\subsection{VBS approximation}
\label{sec:vbsa}

The kinematic mechanism that leads to an enhancement of EW VBS at energies far above the
EW scale is the nearly collinear emission of EW gauge bosons off initial-state (anti)quarks,
where the corresponding gauge-boson propagators receive only a very small virtuality.
In this region, the differential cross section is dominated by the squared sum of
Feynman diagrams with two collinearly enhanced propagators. This set of diagrams
comprises all VBS graphs, but also others, depending in detail on the VBS channel.  
The VBS-enhanced EW contribution to the full cross section, however, competes with
other production mechanisms, among which QCD-induced contributions typically by far dominate.
To sufficiently increase the signal-to-background ratio for an analysis of the EW VBS 
contribution, it is necessary to set up dedicated VBS cuts.

The VBS cuts require a pair of two strongly forward--backward pointing jets, 
called tagging jets $\mathrm{j}_1$ and $\mathrm{j}_2$ in the following.
On the one hand, low transverse momenta $p_{\mathrm{T}}$ 
for $\mathrm{j}_1$ and $\mathrm{j}_2$ are vital for the
VBS enhancement; on the other hand,
the transverse momenta still have to be large enough for tagging.
In practice, only lower limits on $p_{\mathrm{T},\mathrm{j}_1}$
and $p_{\mathrm{T},\mathrm{j}_2}$ of the order of $30{-}50\GeV$ are set.
However, further cuts on the tagging jets are required to successfully 
suppress cross-section contributions that do not feature the $2 \to 2$ VBS subprocess.
The two observables that are used to achieve this discrimination
are the invariant mass of the two leading jets, $M_{\mathrm{j}_1\mathrm{j}_2}$, and the rapidity separation between
them, $\Delta y_{\mathrm{j}_1\mathrm{j}_2}=|y_{\mathrm{j}_1} - y_{\mathrm{j}_2}|$.
Requiring both observables
to be large (see Eq.~\eqref{eq:vbs-cuts} for specific values) achieves this and identifies the VBS region.
In this VBS region we can define a gauge-invariant subset of Feynman diagrams that are not suppressed. 
This simplifies the calculation, and we call this approximation the \textit{VBS approximation}.
This type of approximation was already used in the early calculations~\cite{Jager:2009xx,Denner:2012dz} of QCD corrections to the EW channel of $\PWpm\PWpm$ scattering, which are part of the full $\mathcal{O}(\alphas\alpha^6)$ contribution to the VBS cross section.

\subsubsection{VBS approximation at LO}
\label{sec:vbsa-at-lo}

Although later in Sec.~\ref{sec:dpa_vbs_results} we will not use the VBS approximation for LO matrix elements, we define it here for the purpose of defining the approximation at NLO in Sec.~\ref{sec:vbsa-at-nlo}.
We colour-decompose the matrix elements in terms of sums 
over colour factors multiplied with colour-stripped matrix elements. Considering the partonic processes
\begin{equation}
q_i(p_1) q_k(p_2 ) \to q_j(p_3) q_l(p_4 ) + \Pep \Pgne \, \Pgmp \Pgngm \text{,}
\label{eq:qq2qqenmn}
\end{equation}
where the indices $i$ and $k$ denote the colours of the initial-state 
(anti)quarks and $j$ and $l$ the colours of the final-state (anti)quarks.
Three colour/fermion flows are possible for the $\mathcal{O}(e^6)$ (EW) and $\mathcal{O} (g_\mathrm{s}^2 e^4)$ (QCD) matrix elements,
\begin{align}
\mathcal{M}_\text{EW}  &= \delta_{ik} \delta_{jl} \mathcal{M}_\text{EW}^s  + \delta_{ij} \delta_{kl} \mathcal{M}_\text{EW}^t  + \delta_{il} \delta_{kj} \mathcal{M}_\text{EW}^u \text{,}  \label{eq:me-ew} \\
\mathcal{M}_\text{QCD} &= T_{ik}^a T_{jl}^a       \mathcal{M}_\text{QCD}^s + T_{ij}^a T_{kl}^a       \mathcal{M}_\text{QCD}^t + T_{il}^a T_{kj}^a       \mathcal{M}_\text{QCD}^u \text{,} \label{eq:me-qcd}
\end{align}
where $\mathcal{M}_\text{EW/QCD}^{r}$ with $r= s,t,u$ are separately gauge-invariant sets of Feynman diagrams.
The superscripts $s$, $t$, and $u$ refer to the fermion flow in an equivalent $2 \to 2$ process containing only the (anti)quarks.
They do not denote the propagator structure of the matrix elements, which is more complicated due to the additional colourless particles.
The VBS approximation exploits the fact that the squares of the $s$, $t$, $u$ contributions
dominate over their interferences, which will be neglected, so that only
$|\mathcal{M}^r|^2$ with $r=s,t,u$ will be needed in the
VBS approximation.
Note also that not all of the $s$, $t$, $u$ parts contribute simultaneously in a single channel;
the three parts are introduced to keep the treatment generic here.

At LO, the VBS approximation generates all relevant amplitudes out of one master
amplitude via crossing and flavour symmetries, which are valid due to the masslessness of the fermions.
Starting from the matrix element $\mathcal{M}^t$
for $\Pqc \Pqu \to \Pqd \Pqs \, \Pep \Pgne \, \Pgmp \Pgngm$ with the momentum assignment as in Eq.~\refeq{eq:qq2qqenmn}, 
crossing symmetries lead to the following identities:
\begin{align}
\bigl| \mathcal{M}^s (p_1, p_2, p_3, p_4) \bigr|^2 &= \bigl| \mathcal{M}^t (p_4, p_2, p_3, p_1) \bigr|^2, \\
\bigl| \mathcal{M}^u (p_1, p_2, p_3, p_4) \bigr|^2 &= \bigl| \mathcal{M}^t (p_1, p_2, p_4, p_3) \bigr|^2. 
\label{eq:t-u-crossing}
\end{align}
\refta{tab:zero-matrix-elements} lists how the required contributions
$|\mathcal{M}^r|^2$ for each partonic channel are constructed from the $|\mathcal{M}^t|^2$
of the $\Pqc \Pqu \to \Pqd \Pqs \, \Pep \Pgne \, \Pgmp \Pgngm$ channel and its crossed variants.
\begin{table}
\centering
\begin{tabular}{@{}llcccl@{}}
\toprule
Charge & Process & $\mathcal{M}^s$ & $\mathcal{M}^t$ & $\mathcal{M}^u$ & VBS approximation \\
\midrule
\multirow{2}{*}{$Q_{\mathrm{in}}=4/3$}
&$\Pqu \Pqu \to \Pep \Pgne \, \Pgmp \Pgngm \, \Pqd \Pqd$     & 0 & * & *
&$\Pqc \Pqu \to \Pep \Pgne \, \Pgmp \Pgngm \, \Pqd \Pqs$ \\
&$\Pqc \Pqu \to \Pep \Pgne \, \Pgmp \Pgngm \, \Pqd \Pqs$     & 0 & * & 0
&$\Pqc \Pqu \to \Pep \Pgne \, \Pgmp \Pgngm \, \Pqd \Pqs$ \\
\midrule
\multirow{3}{*}{$Q_{\mathrm{in}}=1$}
&$\Paqd \Pqu \to \Pep \Pgne \, \Pgmp \Pgngm \, \Pqd \Paqu$   & * & * & 0
&$\Paqd \Pqc \to \Pep \Pgne \, \Pgmp \Pgngm \, \Pqs \Paqu$ \\
&$\Paqd \Pqu \to \Pep \Pgne \, \Pgmp \Pgngm \, \Pqs \Paqc$   & * & 0 & 0
&0 \\
&$\Paqd \Pqc \to \Pep \Pgne \, \Pgmp \Pgngm \, \Pqs \Paqu$   & 0 & * & 0
&$\Paqd \Pqc \to \Pep \Pgne \, \Pgmp \Pgngm \, \Pqs \Paqu$ \\
\midrule
\multirow{2}{*}{$Q_{\mathrm{in}}=2/3$}
&$\Paqd \Paqd \to \Pep \Pgne \, \Pgmp \Pgngm \, \Paqu \Paqu$ & 0 & * & *
&$\Paqs \Paqd \to \Pep \Pgne \, \Pgmp \Pgngm \, \Paqu \Paqc$ \\
&$\Paqs \Paqd \to \Pep \Pgne \, \Pgmp \Pgngm \, \Paqu \Paqc$ & 0 & * & 0
&$\Paqs \Paqd \to \Pep \Pgne \, \Pgmp \Pgngm \, \Paqu \Paqc$ \\
\bottomrule
\end{tabular}
\caption{Fermion flow structure for each matrix element for each process listed, see Eqs.~\eqref{eq:me-ew} and \eqref{eq:me-qcd} for definition.
Stars * indicate non-vanishing contributions to the matrix element, and the
last column denotes which process is used in the VBS approximation.}
\label{tab:zero-matrix-elements}
\end{table}
The $s$-fermion-flow matrix elements $\mathcal{M}^s$, which do not contain the VBS subprocess but instead \enquote{semi-leptonic triple-W production} (see also Fig.~\ref{fig:born_qq_www}), are zero whenever $Q_{\mathrm{in}} \neq 1$ 
and, at the same time, the initial-state quarks are 
not from the same generation, because we assume a diagonal CKM matrix.
Only one partonic process has a vanishing contribution for the $t$-fermion-flow, because the quark generations in the initial and final states are different and thus do not mix.
If the final state contains two identical (anti)quarks, Fermi symmetry implies that for each
Feynman graph there is a twin graph with the two final-state (anti)quarks interchanged;
in such a pair of graphs, one has $t$-channel and the other $u$-channel fermion flow, \ie
there are non-vanishing contributions to $\mathcal{M}^t$ and $\mathcal{M}^u$.

We are now ready to define the VBS approximation for the squared matrix elements at $\mathcal{O} (\alpha_\text{s}^2 \alpha^4)$ and $\mathcal{O} (\alpha^6)$:
\begin{enumerate}
\item 
We neglect all $s$-channel fermion flows, since they
are not of VBS type and do not contain the W-boson
propagators with the enhancement for collinear emission.
At LO, those channels are suppressed by the lower cut $M_{\mathrm{j}_1\mathrm{j}_2,\mathrm{cut}}$
on $M_{\mathrm{j}_1\mathrm{j}_2}$, which is the invariant mass of an outgoing
\PWm~boson, so that a cut value $M_{\mathrm{j}_1\mathrm{j}_2,\mathrm{cut}}\gg\MW$ forces this
\PWm~boson into its far off-shell region.
Note that owing to $M_{\mathrm{j}_1\mathrm{j}_2,\mathrm{cut}}>\MH$
this cut also excludes the Higgs-boson resonance at LO, which occurs
in the $s$-channel $\PWp\PWp\PWm$ subprocess for low virtualities $M_{\mathrm{j}_1\mathrm{j}_2}$.

\item
If both $t$- and $u$-channels are present, which is the case for two identical
(anti)quarks in the final state, the squared amplitude involves the two
diagonal parts $|\mathcal{M}^t|^2$ and $|\mathcal{M}^u|^2$ as well as an interference term
between $\mathcal{M}^t$ and $\mathcal{M}^u$.
Making use of the crossing identity Eq.~\eqref{eq:t-u-crossing}
and the fact that the phase-space integral extends over the whole phase space,
the integrals over $|\mathcal{M}^t|^2$ and $|\mathcal{M}^u|^2$ are equal,
so that we simply can drop the $|\mathcal{M}^u|^2$ part and keep the $|\mathcal{M}^t|^2$ part
multiplied by a factor~2 (which compensates the identical-particle factor $1/2$).
%

\item This leaves the interference between the $t$- and $u$-colour flow.
These contributions,
\begin{align}
2 \real \left\{ \delta_{ij} \delta_{kl} \mathcal{M}_\text{EW}^t \delta_{il} \delta_{kj} \left( \mathcal{M}_\text{EW}^u \right)^* \right\} &=
2 N_\text{c} \real \left\{ \mathcal{M}_\text{EW}^t \left( \mathcal{M}_\text{EW}^u \right)^* \right\} \text{,} \\
2 \real \left\{ T_{ij}^a T_{kl}^a \mathcal{M}_\text{QCD}^t \left( T_{il}^b T_{kj}^b \mathcal{M}_\text{QCD}^u \right)^* \right\} &=
- 2 T_\text{F} C_\text{F} \real \left\{ \mathcal{M}_\text{QCD}^t \left( \mathcal{M}_\text{QCD}^u \right)^* \right\} \text{,}
\end{align}
are colour-suppressed by $1/N_\text{c}$ with respect to the squared matrix elements, which have either the colour factor $\delta_{il} \delta_{il} \delta_{kj} \delta_{kj} = N_\text{c}^2$ (EW) 
or $T_{ij}^a T_{kl}^a (T_{ij}^b T_{kl}^b)^* = T_\text{F} N_\text{c} C_\text{F}$ (QCD).
They are also kinematically suppressed, because $\mathcal{M}^u$ and $\mathcal{M}^t$ 
show their collinear enhancements in different parts of phase space, which therefore do not accumulate.
Interference contributions between $t$- and $u$-channels are, thus, neglected in VBS approximation.
\end{enumerate}
The interference contribution at $\mathcal{O} (\alpha_\text{s} \alpha^5)$ does not contain \enquote{squared} fermion flows of the form
\begin{equation}
2 \real\left\{ \mathcal{M}_\text{EW}^t ( \mathcal{M}_\text{QCD}^t )^* \right\} \quad \text{or} \quad
2 \real\left\{ \mathcal{M}_\text{EW}^u ( \mathcal{M}_\text{QCD}^u )^* \right\} \text{,}
\end{equation}
because $\operatorname{Tr} (T^a) = 0$.
The only non-vanishing terms of $\mathcal{O} (\alpha_\text{s} \alpha^5)$ are interference terms 
of the form $\mathcal{M}_\text{EW}^r ( \mathcal{M}_\text{QCD}^{r'} )^*$ with $r\ne r'$
which we set to zero in the VBS approximation, according to steps one and three above.

Effectively, the VBS approximation replaces each matrix element by a simpler one only having the $t$-fermion flow, \eg $\Pqc \Pqu \to \Pep \Pgne \, \Pgmp \Pgngm \, \Pqd \Pqs$, or, if the original matrix element does not have a $t$-fermion flow, by zero.
Table~\ref{tab:zero-matrix-elements} summarises the replacements in its last column.
This definition of the VBS approximation 
agrees with the one used before in Ref.~\cite{Denner:2012dz}.

\subsubsection{VBS approximation at NLO}
\label{sec:vbsa-at-nlo}

At NLO we proceed in the same spirit, applying the same rules described in Sec.~\ref{sec:vbsa-at-lo} to matrix elements with $n$-particle kinematics, \ie the one-loop 
matrix elements and Born-type matrix elements like colour-correlated matrix elements used in the construction of subtraction terms.
For the real matrix elements we distinguish between two cases:
\begin{enumerate}
\item Matrix elements with final-state 
gluons/photons are treated similarly to their Born-type matrix elements without the extra gluon/photon.
This is required by the fact that all subtraction contributions add up to 
zero, \ie that what is subtracted from real matrix elements is exactly cancelled by the integrated 
subtraction terms on the virtual side.
\item Matrix elements with initial-state 
gluons/photons are \emph{not} approximated. They (and all corresponding 
subtraction terms) are instead calculated exactly as in the full off-shell calculation,
because a VBS approximation cannot be defined properly for these contributions.
Note also that
the contributions with initial-state gluons/photons are separately infrared (soft and collinear) safe.
\end{enumerate}
The last point is a choice, and at least for $O(\alpha^6 \alpha_\mathrm{s})$ it is possible to also approximate contributions with an initial-state gluon.
For a discussion of this specific approximation and its quality see Ref.~\cite{Ballestrero:2018anz}.

\subsection{Double-pole approximation}
\label{sec:dpa}

The double-pole approximation (DPA) further reduces the set of relevant diagrams by taking into account only
the leading cross-section contributions in an expansion about the two produced
W-boson resonances, which decay into leptonic final states.
Different variants for carrying out such an expansion have been proposed 
in the literature~\cite{Beenakker:1998gr,%
Jadach:1996hi,Jadach:1998tz,%
Denner:1999kn,Denner:2000bj,Denner:2002cg,%
Kurihara:2001um} (see also \citere{Denner:2019vbn}),
but mostly applied only to pure pair production processes with
no other particles in the final state.
In our calculation, we employ the version described in
\citeres{Denner:1999kn,Denner:2000bj,Denner:2002cg}
for $\Pep\Pem\to\PW\PW\to4\,$fermions and in \citere{Billoni:2013aba}
for $\Pp\Pp\to\PW\PW\to4\,$leptons, 
which was generalized to
more general final states in \citere{Dittmaier:2015bfe} and applied to a
triple-resonance process in \citere{Dittmaier:2019twg}.

In detail, the full off-shell cross section is kept at LO,
because off-shell effects in the LO contribution are generically of 
$\mathcal{O}(\Gamma_\PW/\MW)\sim\mathcal{O}(\alpha)$, \ie of the
order of EW NLO corrections. Thus, we apply the DPA only to the
genuine NLO corrections.

\begin{figure}
	\centering
	\includegraphics[width=0.33\textwidth]{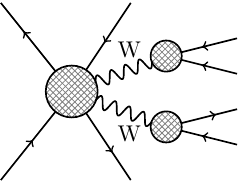}\hspace{1cm}
	\includegraphics[width=0.33\textwidth]{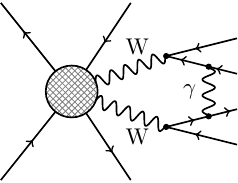}
	\caption{Structure of diagrams contributing to the {\it factorizable} (left) and {\it non-factorizable} (right) NLO virtual corrections in the DPA. The circles indicate either a tree-like substructure or a loop subdiagram.}
	\label{fig:dpa}
\end{figure}

The application of the DPA to virtual NLO corrections splits this
contribution into {\it factorizable} and {\it non-factorizable} parts,
as illustrated in \reffi{fig:dpa}.
The factorizable contribution is based on the selection of all one-loop diagrams that are
enhanced by the explicit appearance of at least one resonant W-boson propagator for each
resonance; all other one-loop diagrams are dropped in this part.
The factorizable contribution, thus, comprises those one-loop corrections that can be uniquely
attributed to the production or decay subprocesses.
To render the resulting factorizable contribution
gauge independent, the kinematics entering the residue of the double
resonance pole is ``on-shell projected'' in such a way that the two nearly resonant
W~bosons become on shell. In this on-shell projection the final-state momenta
are deformed by contributions of the order 
$\mathcal{O}((k_i^2-\MW^2)/\MW^2))$, where $k_i$ are the off-shell momenta
of the nearly resonant W~bosons, so that the deformation of the momenta formally changes
the whole cross-section contribution only by subleading terms.
Specifically, we employ an on-shell projection as described in App.~A of \citere{Denner:2000bj},
which modifies only the momenta of the produced W~bosons and the final-state leptons.
By default, we keep the direction of the first W~boson and the directions of the charged final-state
leptons fixed, but we have checked that chosing different lepton pairs with 
fixed directions does not lead to significant changes in cross-section predictions.
Finally, we mention that our calculation of all double-pole-approximated loop amplitudes was performed in a completely independent way from \openloops and \recola
using in-house \mathematica routines, which reduce amplitudes generated with \feynarts~\cite{Kublbeck:1990xc} to a standard form suitable for numerical evaluation.

The non-factorizable corrections by definition comprise all doubly-resonant
corrections that are not contained in the factorizable corrections.
As can be shown by power-counting of resonance factors (see, \eg, 
\citeres{Denner:1997ia,Dittmaier:2015bfe}), such contributions can only result from
soft-photon exchange between the different production/decay subprocesses, between
the subprocesses and one of the resonances, or between the resonances.
Note that one-loop diagrams with an internal soft photon coupling to at least one of the W~resonances
contribute to both factorizable and non-factorizable corrections, because on-shell
projection of the W-boson momenta and the soft-photon limit in the loop integral
do not commute.
More details on the structure of non-factorizable corrections, on their calculation,
and explicit results can be found in 
\citeres{Beenakker:1997ir,Denner:1997ia,Denner:2000bj,%
Denner:2002cg,Accomando:2004de,Dittmaier:2015bfe}.
Our calculation is based on the results given in \citere{Dittmaier:2015bfe}.

Real NLO corrections can be evaluated in the DPA as well, leading again to
factorizable and non-factorizable parts
(see, \eg, \citeres{Beenakker:1997ir,Denner:1997ia}).
Following the strategy of previous applications, we, however, base the
real corrections on full matrix elements without DPA; more precisely,
we use the VBS approximation as described in the previous section.
The reason for this approach is twofold:
Firstly, we hereby avoid a possible enhancement of uncertanities that might
result from the splitting into real factorizable and real non-factorizable parts in the
region of low-energy photon emission.%
\footnote{See comments in \citere{Denner:2000bj} on this point.}
Secondly, the evaluation of real corrections involves tree-level amplitudes only
and is, thus, less CPU costly than one-loop amplitudes, where the DPA provides great
calculational simplifications.

In order not to spoil the cancellation of infrared singularities between the virtual 
and real corrections, we have to slightly modify the subcontributions of the dipole
subtraction approach~\cite{Catani:1996vz,Catani:2002hc,Dittmaier:1999mb,Dittmaier:2008md}.
While the contribution of the dipole-subtracted real emission matrix elements as well as the
``convolution part'' of the re-added subtraction part are based on full 
(VBS-approximated) matrix elements, the subtraction part with LO kinematics,
called $I$-operator in \citeres{Catani:1996vz,Catani:2002hc} and endpoint part in
\citeres{Dittmaier:1999mb,Dittmaier:2008md}, is added back in the DPA to cancel
the infrared singularities of the DPA-approximated virtual corrections. In this sense, we apply the 
DPA to the infrared-finite part of the virtual corrections only.

\subsection{Effective vector-boson approximation}
\label{sec:evba}

The original idea of the 
effective vector-boson approximation (EVA) extends the idea of partons inside
hadrons to the case of weak vector 
bosons~\cite{KANE1984367,Dawson:1984gx,CHANOWITZ198485,Lindfors:1985yp,Han:2020uid}.
The weak vector bosons are considered as partons in (anti)quarks,
just like (anti)quarks and gluons are partons in hadrons.
This concept is achieved by approximating the vector-boson emission $q\to qV$ by its asymptotic behaviour in the collinear limit, where it is logarithmically enhanced, and considering the splitting process as part of the hadron that breaks up.
This is similar to the Weizs\"acker--Williams approximation \cite{vonWeizsacker:1934nji,Williams:1934} of QED, which can be used to describe photon--photon processes, \eg at $\Pe^+\Pe^-$ colliders, and is motivated by logarithmic enhancements $\log(\Me^2/s)$ originating from collinear photon emission off $\Pe^\pm$ in QED.
We want to apply the EVA to quantify the contributions arising from the actual VBS process.
It is, however, known in the literature~\cite{GUNION198657,Kuss:1995yv,Kuss:1996ww,%
ACCOMANDO200674, Accomando:2006hq, Borel2012, Brehmer201490, Bernreuther16, Ruiz:2021tdt}
that the approximation quality of the EVA is rather limited.
We, therefore, do not advocate it for precision predictions and only consider it for a 
qualitative discussion at LO.
A comprehensive description of our construction of the LO EVA matrix elements is 
given in \refapp{sec:evba-const}. Note that we do not merely take over
existing proposals from the literature, but compare various formulations that differ 
in the details of handling intermediate (off-shell) polarization vectors and
external currents describing the W~radiation off the (anti)quarks and the W~decays
into leptons, in order to account for spin correlations and off-shell effects
as much as possible.

\section{Validation of the calculation}
\label{sec:validation}

\subsection{Internal consistency checks}

\begin{itemize}
\item
The virtual matrix elements for the largest partonic channel, $\Pqu \Pqu \to \Pep \Pgne \, \Pgmp \Pgngm \, \Pqd \Pqd$, have been validated by comparing our default results from \openloops~\cite{Cascioli:2011va,Buccioni:2019sur} for all four orders with results from \recola~\cite{Actis:2016mpe,Denner:2017wsf}.
See also Sec.~\ref{sec:comparison-against-biedermann-et-al}, where we compare the integrated cross sections obtained with \openloops and \recola.
\item
The $\mathcal{O} (\alpha^5 \alphas^2)$ results, specifically integrated dipoles and one-loop matrix elements, have been calculated once with the BLHA convention (see Ref.~\cite{Alioli:2013nda}, Eqs.~(1) and (2)) and another time with the COLI convention (see Ref.~\cite{Denner:2016kdg}, Eq.~(9)), which shifts finite contributions between \openloops and our implementation of the soft- and collinear subtraction.
This constitutes another check of the handling of infrared singularities, between the two mentioned program parts.
\item
The DPA for the infrared-finite part of the virtual corrections, which uses matrix elements that are
based on in-house \mathematica routines, has
been reproduced with  \recola~\cite{Actis:2016mpe,Denner:2017wsf}.
\item
Some of the EVA variants have been derived and implemented 
in two completely independent ways and checked against each other.
\end{itemize}

\subsection{Comparison against Biedermann et al.}
\label{sec:comparison-against-biedermann-et-al}

In this section we compare some cross sections calculated in 
\citeres{Biedermann:2016yds,Biedermann:2017bss,Chiesa:2019ulk},
which are based on matrix elements from the amplitude generator 
\recola~\cite{Actis:2016mpe,Denner:2017wsf} and the Monte Carlo integrator
\mocanlo (see also \citere{Proceedings:2018jsb})
with results from our Monte Carlo program \bonsay,
which uses amplitudes from \openloops~\cite{Buccioni:2019sur}.
We consistently adopt the input parameters and setup described in \citere{Biedermann:2017bss}
in this comparison, \ie this calculation uses the PDF set \texttt{NNPDF\_30\_nlo\_as\_0118\_qed}~\cite{Ball:2013hta, NNPDF:2014otw}, which is not available in \lhapdf and therefore has to be downloaded manually from  \url{http://nnpdf.mi.infn.it/nnpdf3-0qed/}.
Table~\ref{tab:comparison-against-biedermann-et-al} lists the integrated cross sections, separately for each contributing order.
The numbers correspond to Tables~2, 3, and 4 of \citere{Biedermann:2017bss}, 
\ie they include the photon-initiated contribution, which only significantly
contributes in $\mathcal{O} (\alpha^7)$.
\begin{table}
\centering
\begin{tabular}{rl|S[table-format=+1.8]S[table-format=+1.8]}
\toprule
& Order & \multicolumn{1}{c}{Ref.~\cite{Biedermann:2017bss} [\si{\femto\barn}]} & \multicolumn{1}{c}{our calculation [\si{\femto\barn}]} \\
\midrule
LO & $\mathcal{O} (\alpha^6)$                     &  1.4178(2)  & 1.41773(5) \\
   & $\mathcal{O} (\alpha^5 \alpha_\mathrm{s})$   &  0.04815(2) & 0.048138(3) \\
   & $\mathcal{O} (\alpha^4 \alpha_\mathrm{s}^2)$ &  0.17229(5) & 0.17233(2) \\
\midrule
NLO & $\mathcal{O} (\alpha^7)$                     & -0.1732(3)   & -0.1728(6) \\
    & $\mathcal{O} (\alpha^6 \alpha_\mathrm{s})$   & -0.0568(5)   & -0.0560(8) \\
    & $\mathcal{O} (\alpha^5 \alpha_\mathrm{s}^2)$ & -0.00032(13) &  0.0047(2) \\
    & $\mathcal{O} (\alpha^4 \alpha_\mathrm{s}^3)$ & -0.0063(4)   & -0.0073(2) \\
\bottomrule
\end{tabular}
\caption{Comparison of our cross-section predictions
against the integrated cross sections from Ref.~\cite{Biedermann:2017bss}, extracted from Tables~2, 3, and 4 given there.}
\label{tab:comparison-against-biedermann-et-al}
\end{table}

Overall, we observe agreement between the two sets of results, typically within \numrange{0}{2} multiples of the combined Monte Carlo integration uncertainty $\sigma = \sqrt{\sigma_1^2 + \sigma_2^2}$.
We also checked the agreement in differential distributions, for which a similar statement holds, with larger 
deviations for some bins in extreme phase-space regions as expected.

An exception to the previous statements are the numbers 
for the contribution of $\mathcal{O}(\alpha^5\alphas^2)$, for which we find a significant disagreement of
$21\sigma$.
This disagreement originates from a bug in \recola 1.4.3, which is triggered whenever the two contributions at $\mathcal{O}(\alphas^2 \alpha^5)$,
\begin{equation}
2 \operatorname{Re} \left\{ ( \mathcal{M}^{(0)}_{e^6} )^* \mathcal{M}^{(1)}_{g_\mathrm{s}^4 e^4} \right\} + 2 \operatorname{Re} \left\{ ( \mathcal{M}^{(0)}_{g_\mathrm{s}^2 e^4} )^* \mathcal{M}^{(1)}_{g_\mathrm{s}^2 e^6} \right\} \text{,}
\label{eq:recola-bug}
\end{equation}
are calculated separately, as done in \citere{Biedermann:2017bss}.
Here $\mathcal{M}^{(l)}_{g_\mathrm{s}^n e^m}$ are the matrix elements of $\mathcal{O} (g_\mathrm{s}^n e^m)$ with $l$ loops.
This bug will be fixed in an upcoming release of \recola.\footnote{A.\ Denner, S.\ Uccirati, private communication.}
We were able to circumvent the bug by calculating the two contributions in Eq.~\eqref{eq:recola-bug} in a single phase-space integration, thereby reproducing our result obtained with \openloops.

\section{Numerical results}
\label{sec:results}

\subsection{Setup}
\label{sec:setup}

For all calculations we employ the complex-mass scheme~\cite{Denner:1999gp,Denner:2005fg,Denner:2019vbn},
for which the real on-shell masses and decay widths are chosen as follows:
\begin{equation}
\begin{aligned}
M_{\mathrm{W}}^\mathrm{OS} &= \SI{80.379}{\giga\electronvolt}  & \Gamma_\mathrm{W}^\mathrm{OS} &= \SI{2.0845}{\giga\electronvolt} \text{,} \\
M_{\mathrm{Z}}^\mathrm{OS} &= \SI{91.1876}{\giga\electronvolt} & \Gamma_\mathrm{Z}^\mathrm{OS} &= \SI{2.4952}{\giga\electronvolt} \text{,} \\
M_{\mathrm{H}}^\mathrm{OS} &= \SI{125.0}{\giga\electronvolt}   & \Gamma_\mathrm{H}^\mathrm{OS} &= \SI{4.07e-03}{\giga\electronvolt} \text{,} \\
m_{\mathrm{t}}^\mathrm{OS} &= \SI{173.0}{\giga\electronvolt}   & \Gamma_\mathrm{t}^\mathrm{OS} &= \SI{0.0}{\giga\electronvolt} \text{.}
\end{aligned}
\end{equation}
These values are converted to their pole values using the relations (see, \eg, \citere{Denner:2019vbn})
\begin{equation}
M_\mathrm{V}      = M_\mathrm{V}^\text{OS}      / \sqrt{1 + (\Gamma_\mathrm{V}^\text{OS} / M_V^\text{OS})^2} \text{,} \quad
\Gamma_\mathrm{V} = \Gamma_\mathrm{V}^\text{OS} / \sqrt{1 + (\Gamma_\mathrm{V}^\text{OS} / M_V^\text{OS})^2} \text{.}
\end{equation}
Quarks and leptons not mentioned above are treated as massless.
The remaining free EW parameter is the Fermi coupling constant,
\begin{equation}
G_\mu = \SI{1.1663787e-5}{\per\giga\electronvolt\squared} \text{,}
\end{equation}
from which the electromagnetic coupling constant is determined
according to the $G_\mu$ input-pa\-ram\-e\-ter scheme (see, \eg, \citere{Denner:2019vbn}),
\begin{equation}
\alpha = \frac{\sqrt{2}}{\pi} G_\mu \MW^2 \left( 1 - \frac{\MW^2}{\MZ^2} \right) \text{,}
\end{equation}
using the real pole masses.

The setup used for all results presented in this paper are for the LHC operating at a centre-of-mass energy of $\sqrt{s} = \SI{13}{\tera\electronvolt}$.
We use the NNPDF 3.1 LUXQED PDF set \cite{Bertone:2017bme} with a strong coupling constant $\alpha_\mathrm{s} (M_\mathrm{Z}) = 0.118$.
This PDF set employs the LUXQED \cite{Manohar:2016nzj,Manohar:2017eqh} formalism to determine the photon distribution function which is needed for the real photon-initiated processes.

Finally, we set the renormalization and factorization scale to the geometric mean of the transverse momenta of the tagging jets $\mathrm{j}_1$ and $\mathrm{j}_2$,
\begin{equation}
\mu_\text{R}^2 = \mu_\text{F}^2 = p_{\mathrm{T}, \mathrm{j}_1} \cdot p_{\mathrm{T}, \mathrm{j}_2} \text{,}
\end{equation}
and estimate higher-order uncertainties (from QCD) using a seven-point scale variation by taking the envelope, \ie the minimum and maximum of the cross sections evaluated at the following scales:
\begin{equation}
(\mu_\text{R}, \mu_\text{F}) \text{,} \;
(2 \mu_\text{R}, 2 \mu_\text{F}) \text{,} \;
(0.5 \mu_\text{R}, 0.5 \mu_\text{F}) \text{,} \;
(2 \mu_\text{R}, \mu_\text{F}) \text{,} \;
(\mu_\text{R}, 2 \mu_\text{F}) \text{,} \;
(0.5 \mu_\text{R}, \mu_\text{F}) \text{,} \;
(\mu_\text{R}, 0.5 \mu_\text{F}) \text{.}
\label{eq:seven-point-scale-variation}
\end{equation}

\subsection{Phase-space volume}
\label{sec:phase-space-volumes}

We define the fiducial phase-space volume as follows, 
in agreement with the particle-level phase-space volume given in the ATLAS 
measurement~\cite{Aaboud:2019nmv}.
We first require at least two leading jets 
that are defined according to the
anti-$k_\text{T}$ algorithm~\cite{Cacciari:2008gp} with jet radius $R = 0.4$
and that fulfil the requirements
\begin{equation}
p_{\mathrm{T}} > \SI{35}{\giga\electronvolt} \text{,} \quad |y| < 4.5 \text{,} \quad \Delta R_{\mathrm{j} \ell} > 0.3 \text{.}
\end{equation}
The leading jet $\mathrm{j}_1$, \ie the one with largest transverse momentum, must fulfil the stronger cut
\begin{equation}
p_{\text{T},\mathrm{j}_1} > \SI{65}{\giga\electronvolt} \text{.}
\end{equation}
The next-to-leading jet $\mathrm{j}_2$ is the one with the second largest $p_{\mathrm{T}}$.
The two leading jets are then called \emph{tagging jets} and must furthermore fulfil the following 
invariant-mass and rapidity separation cuts (VBS cuts):
\begin{equation}
M_{\mathrm{j}_1 \mathrm{j}_2} > \SI{500}{\giga\electronvolt} \text{,} \quad |y_{\mathrm{j}_1} - y_{\mathrm{j}_2}| > 2 \text{.}
\label{eq:vbs-cuts}
\end{equation}
The two charged leptons must pass the cuts
\begin{equation}
p_{\mathrm{T}, \ell} > \SI{27}{\giga\electronvolt} \text{,} \quad |y_\ell| < 2.5 \text{.}
\end{equation}
Furthermore we require an invariant-mass cut and a distance separation of 
the two charged leptons:
\begin{equation}
M_{\ell \ell} > \SI{20}{\giga\electronvolt} \text{,} \quad \Delta R_{\ell \ell} > 0.3 \text{.}
\end{equation}
The missing transverse momentum carried away by 
the neutrinos has to obey
\begin{equation}
p_{\mathrm{T,miss}} \equiv
p_{\mathrm{T}, \nu_\mathrm{e} \nu_\mathrm{\mu}} > \SI{30}{\giga\electronvolt} \text{.}
\end{equation}

\subsection{Integrated cross sections}
\label{sec:full_results}

In \citere{Aaboud:2019nmv} ATLAS presented their first observation of EW production of two same-sign W~bosons, including 
$\PWm\PWm$-scattering, each boson decaying into either an (anti-)electron or an (anti-)muon.
In \citere{Aaboud:2019nmv} (text above Fig.~3) two conflicting theoretical predictions have been presented, which differ by a substantial amount.
Private communication with some authors of \citere{Aaboud:2019nmv}
revealed that for the two theoretical predictions two different scale choices had been used.
With \powheg~\cite{Frixione:2007vw,Alioli:2010xd,Jager:2011ms}
the factorization and renormalization scale was set to the W-boson mass, whereas with 
\sherpa~\cite{Schumann:2007mg,Gleisberg:2008ta,Hoeche:2009rj,Hoeche:2012yf}, 
a dynamical scale of $\mu = M_{\PW\PW}$ was chosen, which is the invariant mass of the two W~bosons.
Moreover, the used version of \sherpa included an error in the multi-jet merging which has been corrected in \citere{Buckley:2021gfw}.

Our results on the various LO and NLO contributions to the integrated cross section
for the like-sign W-boson
scattering process $\Pp \Pp \to \Pep \Pgne \, \Pgmp \Pgngm \, \mathrm{j} \mathrm{j} + \mathrm{X}$,
defined by the cuts of the previous section, are given in \refta{tab:integrated-cross-sections}.
\begin{table}
\centering
\begin{tabular}{rr|S[table-format=+1.8]S[table-format=+2.1,table-space-text-post=\si{\percent}]S[table-format=+2.1,table-space-text-post=\si{\percent}]S[table-format=2.1,table-space-text-post=\si{\percent}]}
\toprule
& Order & {Result [\si{\femto\barn}]} & $\delta$ [\si{\percent}] & \multicolumn{2}{c}{Scale uncertainty} \\
\midrule
 LO & $\mathcal{O}(\alpha^6 \alpha_\mathrm{s}^0)$ & 1.24597(5)   & &  -7.7\si{\percent} &  9.9\si{\percent} \\
    & $\mathcal{O}(\alpha^5 \alpha_\mathrm{s}^1)$ & 0.051133(3)  & & -14.0\si{\percent} & 17.7\si{\percent} \\
    & $\mathcal{O}(\alpha^4 \alpha_\mathrm{s}^2)$ & 0.18649(2)   & & -22.2\si{\percent} & 31.6\si{\percent} \\
    & sum                                         & 1.48359(5)   & &  -9.8\si{\percent} & 12.1\si{\percent} \\
\midrule
NLO & $\mathcal{O}(\alpha^7 \alpha_\mathrm{s}^0)$ & -0.1747(5)   & -11.8\si{\percent} \\
    & $\mathcal{O}(\alpha^6 \alpha_\mathrm{s}^1)$ & -0.0902(8)   &  -6.1\si{\percent} \\
    & $\mathcal{O}(\alpha^5 \alpha_\mathrm{s}^2)$ & -0.00017(19) &   0.0\si{\percent} \\
    & $\mathcal{O}(\alpha^4 \alpha_\mathrm{s}^3)$ & -0.0033(7)   &  -0.2\si{\percent} \\
    & sum                                         & -0.268(1)    & -18.1\si{\percent} \\
\midrule
LO+NLO & sum & 1.215(1) & & -4.0\si{\percent} & 1.5\si{\percent} \\
\bottomrule
\end{tabular}
\caption{LO and NLO contributions to the integrated cross section defined by setup
given in \refse{sec:setup}, including the respective uncertainties from scale variations
and PDF errors.
Note that in the case of the $\mathcal{O}(\alpha^5 \alpha_\mathrm{s}^2)$ correction the uncertainty is larger than the estimated value.
The quantity $\delta$, defined in Eq.~\eqref{eq:relative-correction}, gives the size of higher-order corrections relative to the sum of all leading-order contributions.}
\label{tab:integrated-cross-sections}
\end{table}
In addition to the results obtained with the central scale setting, 
\refta{tab:integrated-cross-sections} also shows the combined renormalization
and factorization scale uncertainty (see Eq.~\eqref{eq:seven-point-scale-variation}) for each LO contribution and the full LO and NLO cross sections (given as ``sum'').
Including the full tower of NLO corrections reduces the LO scale uncertainty of about
10--12\% to about 2--4\% at NLO.

\subsection{Differential cross sections} 
\label{sec:diff_cross_sections}

\begin{figure}
	\centering	
	\includegraphics[page=7, width=0.49\textwidth]{los}	
	\includegraphics[page=1, width=0.49\textwidth]{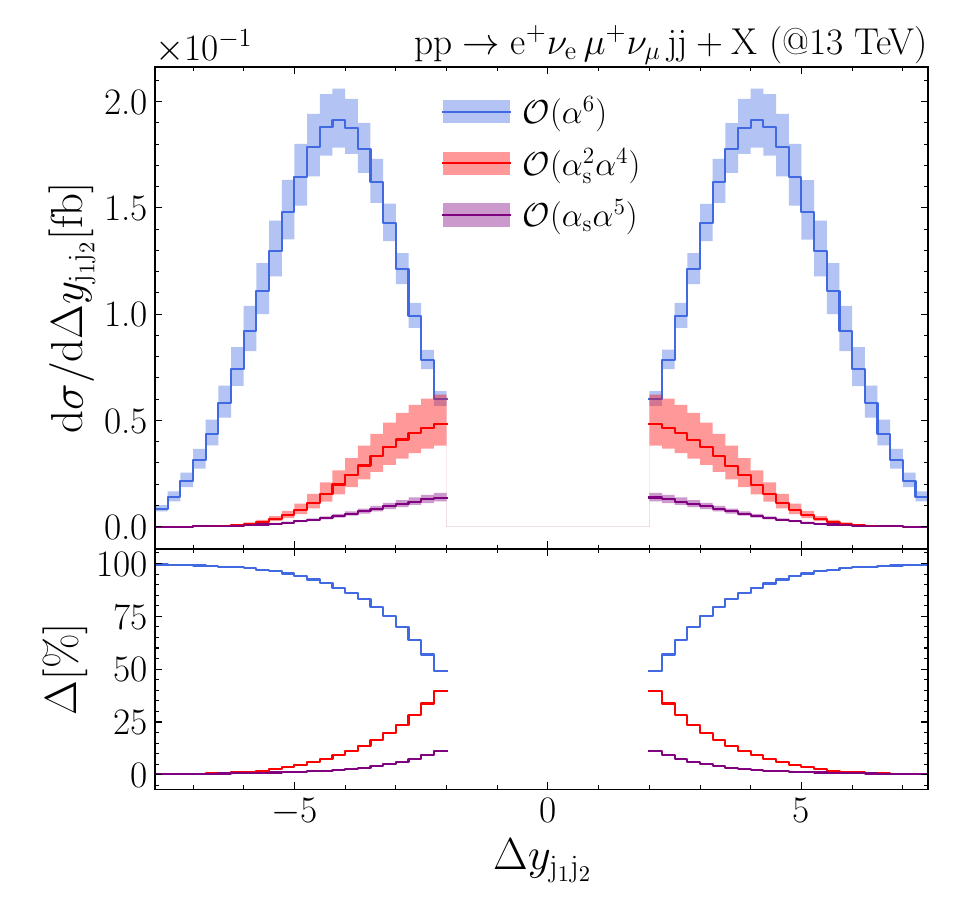}\\
	\includegraphics[page=6, width=0.49\textwidth]{los}	
	\includegraphics[page=10, width=0.49\textwidth]{los}
	\caption{LO differential distributions for $\Pp \Pp \to \Pep \Pgne \, \Pgmp \Pgngm \, \mathrm{j} \mathrm{j} + \mathrm{X}$ at the LHC with CM energy $13\TeV$:  invariant mass of the two jets (top left), rapidity difference of the two jets (top right), invariant mass of the two charged leptons (bottom left), and transverse momentum of the second hardest jet (bottom right). The upper panels show the absolute contributions of order $\mathcal{O}(\alpha^6)$ (EW),  $\mathcal{O}(\alpha_s\alpha^5)$  (interference), and $\mathcal{O}(\alpha_s^2\alpha^4)$ (QCD). The bands denote the envelope of the scale variation for each order. The lower panels show the relative LO contributions $\Delta$ to their sum in percent.}
	\label{fig:lo-distributions}
\end{figure}

In the following, we show differential distributions for 
$\Pp \Pp \to \Pep \Pgne \, \Pgmp \Pgngm \, \mathrm{j} \mathrm{j} + \mathrm{X}$ and
start with some LO predictions in \reffi{fig:lo-distributions}.
The upper panels show the absolute differential distributions of order $\mathcal{O}(\alpha^6)$ (EW),  $\mathcal{O}(\alpha_s\alpha^5)$ (interference), and $\mathcal{O}(\alpha_s^2\alpha^4)$ (QCD). 
In the lower panels, their relative contributions $\Delta$ to the full LO cross section are displayed which provide useful information for
the discussion of the relative NLO corrections presented below.
The two upper plots of \reffi{fig:lo-distributions} show the invariant-mass distribution and the rapidity-difference distribution of the two tagging jets. These observables are commonly used to separate EW and QCD contributions by appropriate cuts, \ie~$M_{\mathrm{j}_1\mathrm{j}_2}>500\GeV$ and $|\Delta y_{\mathrm{j}_1\mathrm{j}_2}|>2$.  It can be clearly seen in these distributions that for high rapidity differences as well as high invariant masses the EW contribution is dominant, justifying the cut selection.
In the lower left plot, the invariant mass of the two charged leptons, $M_{\Pep\mu^+}$, is shown. 
Around $M_{\Pep\mu^+}\sim800\GeV$, both the EW and QCD contributions are almost of the same size, 
because the EW contribution, which is dominant for low invariant masses, is falling more steeply than the QCD 
contribution for intermediate $M_{\Pep\mu^+}$ values up to $M_{\Pep\mu^+}\lsim600\GeV$.
In the transverse-momentum distribution of the subleading jet, shown in the lower right plot, the EW contribution is always dominant.
It should, however, be recalled in the whole discussion of $\Delta$ that,
although the
contribution of the VBS subprocess $\PW\PW\to\PW\PW$ to the cross section
is completely contained in the EW contribution,
the EW contribution is not identical with the contribution induced by $\PW\PW\to\PW\PW$.
The graphs shown in \reffis{fig:born_qq_www}--\ref{fig:born_qq_nonres} illustrate
various EW diagrams not of $\PW\PW\to\PW\PW$ type.
An unambiguous identification of the $\PW\PW\to\PW\PW$ part
of the cross section, called ``VBS part'' in the following,
is not possible on the basis of diagrams,
since the sum of all genuine VBS-type diagrams is not gauge independent.
A sensible way to define the VBS part is, for instance, via the effective
vector-boson approximation, as introduced in \refse{sec:evba} and numerically
discussed in \refse{sec:evba_results} below.

\begin{figure}
	\centering
	\includegraphics[page=9, width=0.49\textwidth]{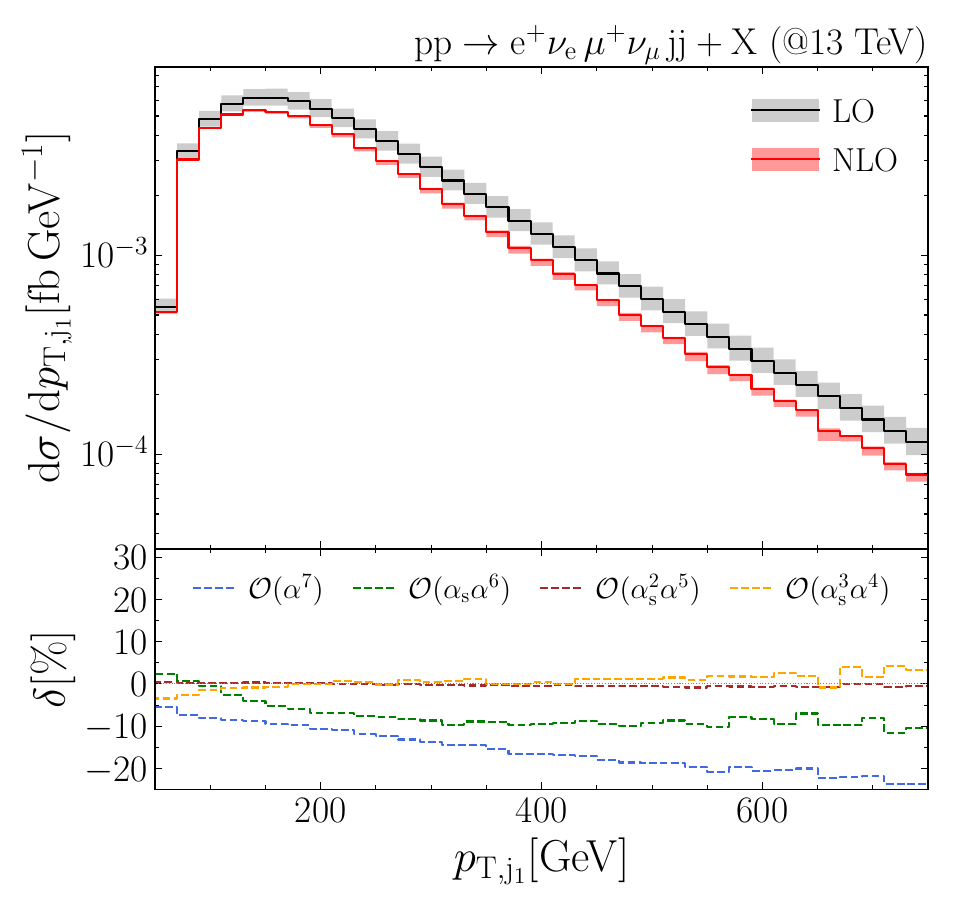}
	\includegraphics[page=10, width=0.49\textwidth]{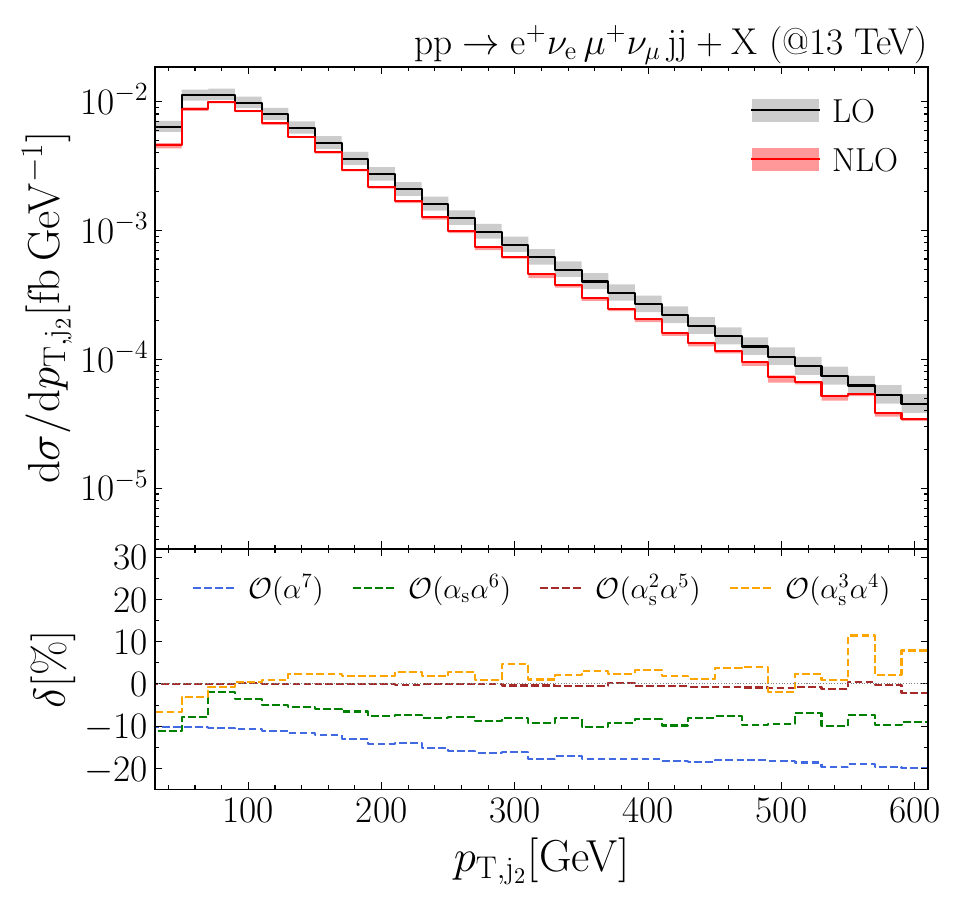}\\
	\includegraphics[page=11, width=0.49\textwidth]{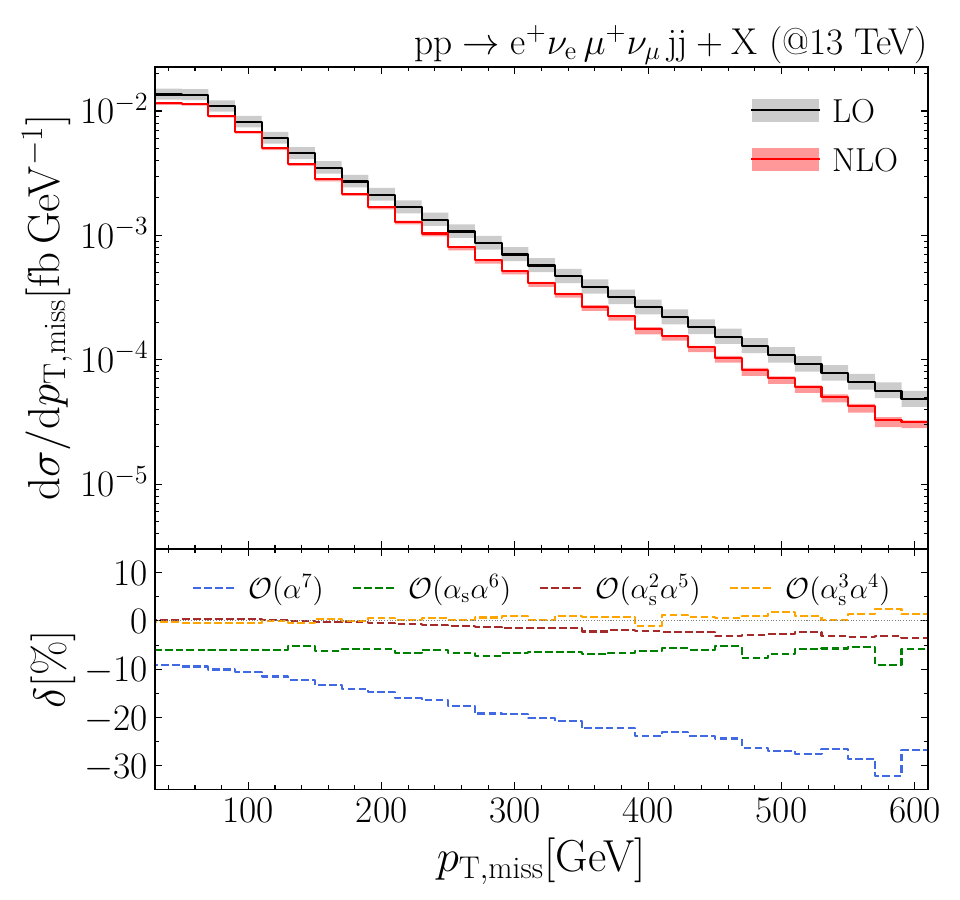}
	\includegraphics[page=8, width=0.49\textwidth]{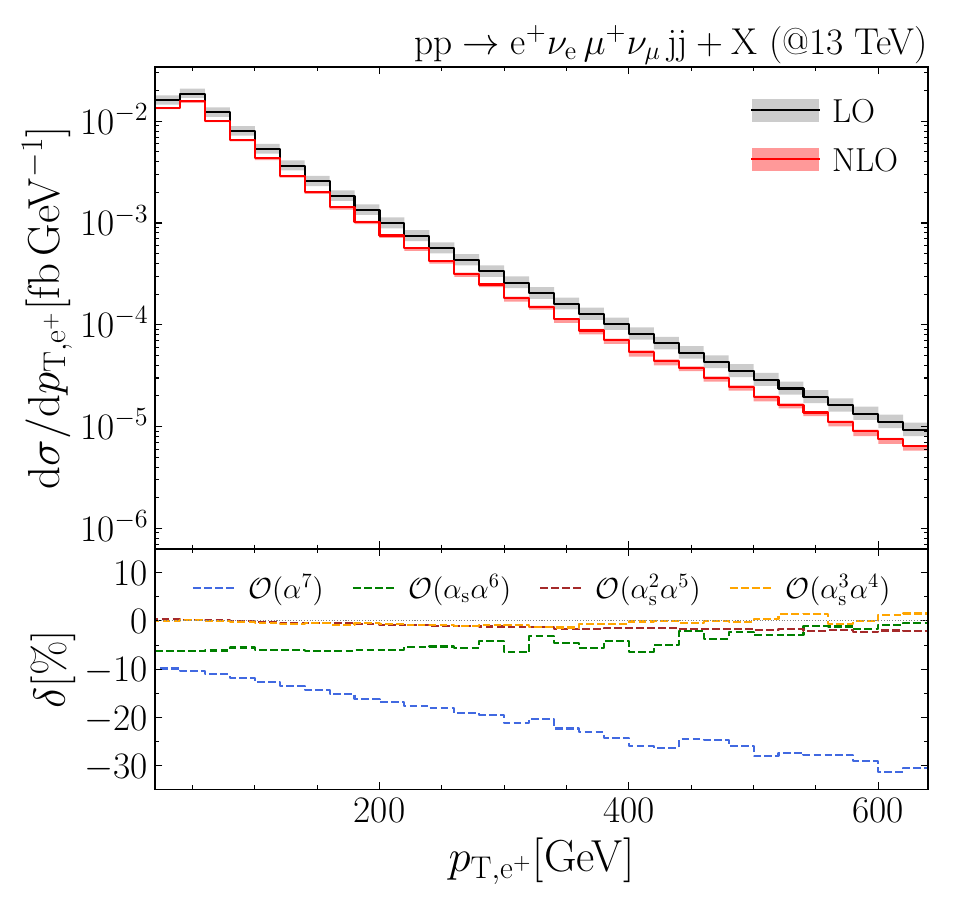}	
	\caption{Differential distributions for $\Pp \Pp \to \Pep \Pgne \, \Pgmp \Pgngm \, \mathrm{j} \mathrm{j} + \mathrm{X}$ at the LHC with CM energy $13\TeV$: transverse momentum of the hardest jet (top left), transverse momentum of the second hardest jet (top right), missing transverse energy (bottom left), and transverse momentum of the positron (bottom right). The upper panels show the LO and NLO contributions, the lower panels show the relative NLO corrections in percent.}
	\label{fig:nlo-distributions1}
\end{figure}

The subsequent figures show various NLO differential distributions. 
In the upper panels, the complete predictions at LO and NLO are shown. 
In the lower panels, we show the relative NLO contributions
\begin{align}
	\delta = \frac{\sigma_\text{NLO}}{\sigma_\text{LO}} -1
    \label{eq:relative-correction}
\end{align}
of $\mathcal{O}(\alpha^7)$, $\mathcal{O}(\alpha_s\alpha^6)$, $\mathcal{O}(\alpha_s^2\alpha^5)$, 
and $\mathcal{O}(\alpha_s^3\alpha^4)$ separately,
always normalized to the full LO cross section.
In \reffi{fig:nlo-distributions1}, various transverse-momentum distributions are shown. 
In the upper plots the transverse momentum of the leading and subleading jets are presented. 
In both observables, for moderate and large $p_{\mathrm{T}}$, the purely EW corrections of 
$\mathcal{O}(\alpha^7)$ are dominant, negative, and increasing in magnitude for 
increasing transverse momenta, reaching about $\sim-20\%$ at $p_{\mathrm{T}}=600\GeV$.
This behaviour is typical for EW corrections at momentum transfer much larger than
the EW scale $\MW$. We note, however, that the high-energy tails of the $p_{\mathrm{T}}$
distributions of the jets do not entirely zoom into the so-called Sudakov regime
of the WW$\to$WW subprocess, which demands large Mandelstam variables in the $2\to2$ subprocess and small virtualities of the incoming (off-shell) W~bosons.
For large $p_{\mathrm{T}}$ values of a jet, the virtuality of at least one of the incoming W~bosons is not small, and the $t$-channel momentum transfer in the WW$\to$WW subprocess is not forced to be large.
Therefore, the EW Sudakov double logarithms cannot really dominate the corrections to the
$p_{\mathrm{T}}$ of the jets, and all kinds of nominally subleading (singly-logarithmic)
EW high-energy corrections become relevant.
The mixed QCD--EW correction of $\mathcal{O}(\alpha_s\alpha^6)$ reduce the 
transverse-momentum distributions of the leading jet by 5--10\% above the maximum
in the distribution, which appears around $p_{\mathrm{T}, \mathrm{j}_1}\sim150\GeV$;
for smaller $p_{\mathrm{T}, \mathrm{j}_1}$ these corrections become smaller in size and
even turn slightly positive.
The $\mathcal{O}(\alpha_s\alpha^6)$ corrections to the $p_{\mathrm{T}, \mathrm{j}_2}$
distribution of the subleading jet are similar in size, but show a different behaviour
for small $p_{\mathrm{T}, \mathrm{j}_2}$ where they reach $-10\%$.
The suppression of the remaining corrections of $\mathcal{O}(\alpha_s^2\alpha^5)$
and $\mathcal{O}(\alpha_s^3\alpha^4)$ is in part due to the fact that the
cross-section contributions widely inherit the kinematic behaviour of the 
LO QCD contribution, which is small compared to the EW contribution over the
whole distribution, as shown explicitly for $\mathrm{j}_2$ in 
\reffi{fig:lo-distributions}. 
This special behaviour of the NLO corrections is, thus,
mainly enforced by the VBS cuts.
Similar features can be observed in the lower plots of \reffi{fig:nlo-distributions1},
where the distributions in missing transverse momentum and in the transverse momentum 
of the positron are depicted. 
Here, the purely EW correction is dominant for all values of $p_{\mathrm{T}}$,
reaching $\sim-30\%$ at $p_{\mathrm{T}}=600 \GeV$, while the other orders
remain at the level of few percent over the whole shown $p_{\mathrm{T}}$ range.
Note that the domain of large $p_{\mathrm{T}}$ of any of the decay leptons is 
in fact dominated by the Sudakov regime of the WW$\to$WW subprocess, 
because the preference of small jet transverse momenta leads to small
virtualities of the incoming W~bosons and the large transverse momentum of a decay
lepton requires both a large scattering energy (Mandelstam variable $s$) 
and large momentum transfer (Mandelstam variable $t$) of the subprocess.
This explains that the impact of the genuine EW corrections of $\mathcal{O}(\alpha^7)$
is larger in size for large leptonic $p_{\mathrm{T}}$ compared to 
large jet transverse momenta of the same value; for leptonic $p_{\mathrm{T}}=600\GeV$ 
an EW correction of $\sim-30\%$ is reached, compared to $\sim-20\%$ for jets.
The observed $\sim-30\%$ can be qualitatively reproduced by just taking into account
the EW Sudakov correction factor for the WW$\to$WW subprocess,
$\delta_{\mathrm{Sud}}=-\frac{2\alpha}{\pi s_{\mathrm{w}}^2}\ln^2(\hat s/\MW^2)$, 
where $\sqrt{\hat s}=\mathcal{O}(p_{\mathrm{T}})$ is the WW centre-of-mass energy and
$s_{\mathrm{w}}$ the sine of the weak mixing angle.
These features were already highlighted in 
\citeres{Biedermann:2016yds,Biedermann:2017bss,Chiesa:2019ulk}.
The next-to-largest corrections to the 
distributions of the transverse momentum of the charged leptons and in the missing
transverse momentum are the mixed QCD--EW corrections of $\mathcal{O}(\alpha_s\alpha^6)$,
which are almost uniformly $\sim -5\%$.
The remaining orders $\mathcal{O}(\alpha_s^2\alpha^5)$
and $\mathcal{O}(\alpha_s^3\alpha^4)$ hardly exceed the 1\% level, and only for large 
transverse momenta where the cross section is small.

\begin{figure}
	\centering
	\includegraphics[page=3, width=0.49\textwidth]{nlos}
	\includegraphics[page=4, width=0.49\textwidth]{nlos}\\
	\includegraphics[page=5, width=0.49\textwidth]{nlos}
	\includegraphics[page=1, width=0.49\textwidth]{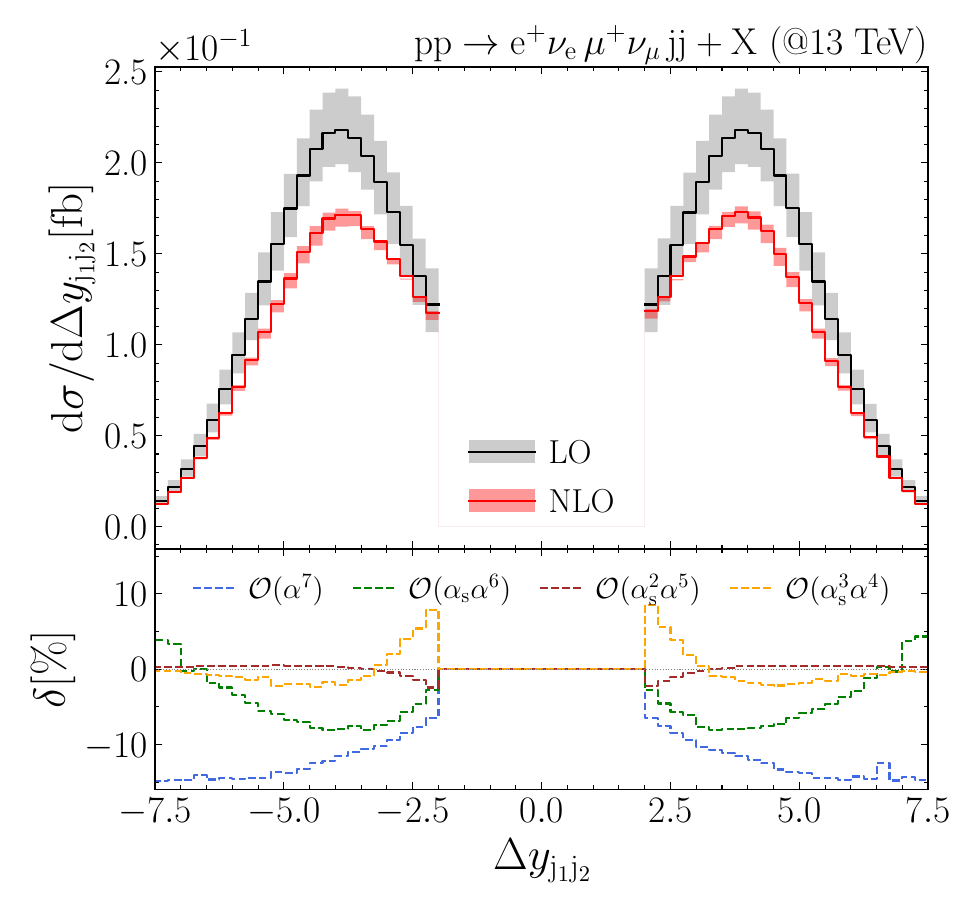}	
	\caption{Differential distributions for $\Pp \Pp \to \Pep \Pgne \, \Pgmp \Pgngm \, \mathrm{j} \mathrm{j} + \mathrm{X}$ at the LHC with CM energy $13\TeV$: rapidity of the hardest jet (top left), rapidity of the second hardest 
jets (top right), rapidity of the anti-muon (bottom left), and rapidity difference of the two tagging jets (bottom right). The upper panels show the LO and NLO contributions, the lower panels show the relative NLO corrections in percent.}
	\label{fig:nlo-distributions2}
\end{figure}

Figure~\ref{fig:nlo-distributions2} shows rapidity distributions. 
The upper plots show the rapidity distributions of the leading (left) and subleading (right) jet,
the lower plots show the rapidity distribution of a charged lepton and the distribution
in the jet rapidity difference.
The hierarchy among the various NLO contributions is similar as for the $p_{\mathrm{T}}$
distributions shown above, \ie the purely EW corrections of $\mathcal{O}(\alpha^7)$ are
the dominating ones, followed by the order $\mathcal{O}(\alpha_s\alpha^6)$, while
the other two orders with higher powers of $\alpha_s$ are widely suppressed.
This is again due to the global dominance of the EW LO contribution over its QCD 
counterpart.
The corrections of $\mathcal{O}(\alpha^7)$ show much less variations in shape 
than for the $p_{\mathrm{T}}$ distributions and are typically about $-10\%$ to
$-12\%$. This is due to the fact that the
large logarithmic EW high-energy
corrections uniformly contribute to all rapidities, in contrast
to the $p_{\mathrm{T}}$ distributions where they appear at high scales only.
The moderate variations in the $\mathcal{O}(\alpha^7)$ corrections mostly result
from the change in the LO normalization induced by the variation in its
composition from EW and QCD parts; normalizing the $\mathcal{O}(\alpha^7)$
contribution to the $\mathcal{O}(\alpha^6)$ LO part would produce a nearly flat relative 
$\mathcal{O}(\alpha)$ correction.
The overall second-largest corrections are again the ones of 
$\mathcal{O}(\alpha_s\alpha^6)$, which are dominated by the QCD corrections to the
EW LO channel. Their largest impact, growing even to $\sim20\%$, is on the leading jet
at high rapidities, an effect that also leaves an imprint for large jet rapidity differences,
but reduced by a factor of $\sim2$ in size. In the other rapidity regions
those corrections hardly exceed 5\%. 
The pure QCD corrections of $\mathcal{O}(\alpha_s^3\alpha^4)$ only exceed the 1\% level
significantly for central rapidities, and their maximal impact occurs for the smallest
jet rapidity differences, where the LO cross section almost equally consists of EW and QCD
parts (see \reffi{fig:lo-distributions}, upper right panel).
The mixed corrections of order $\mathcal{O}(\alpha_s^2\alpha^5)$ never exceed the
1\% level at all.

\begin{figure}
	\centering
	\includegraphics[page=7, width=0.49\textwidth]{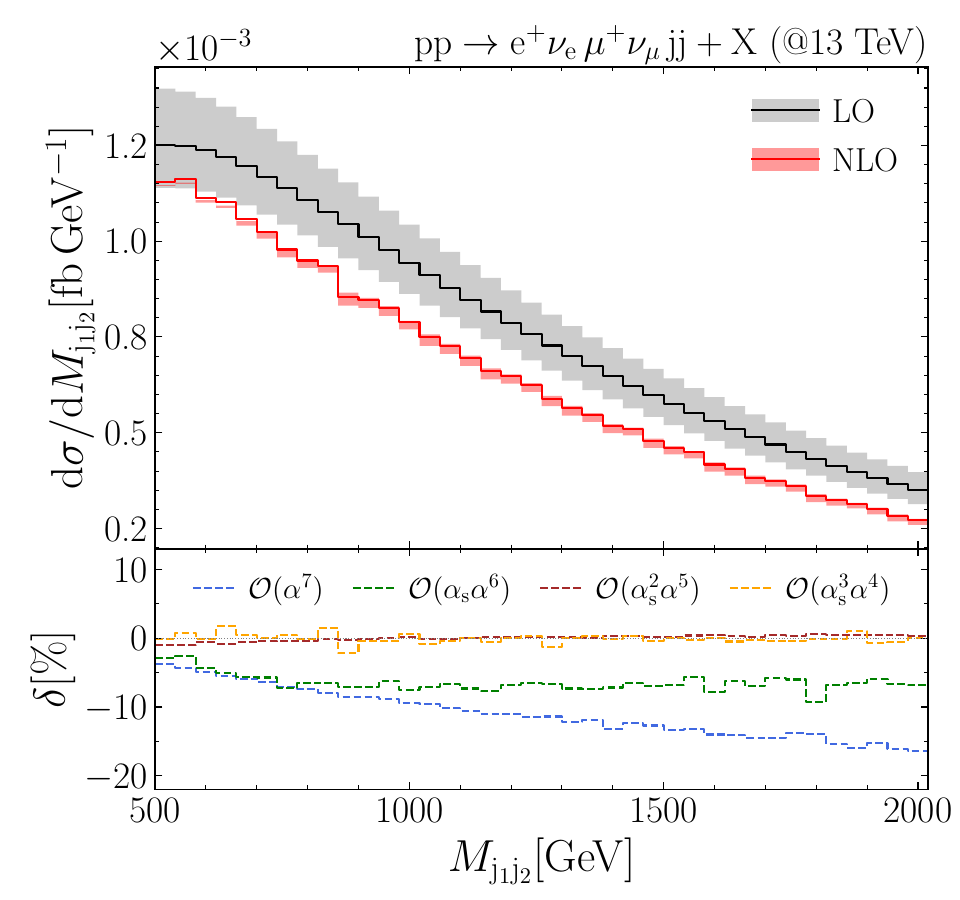}
	\includegraphics[page=6, width=0.49\textwidth]{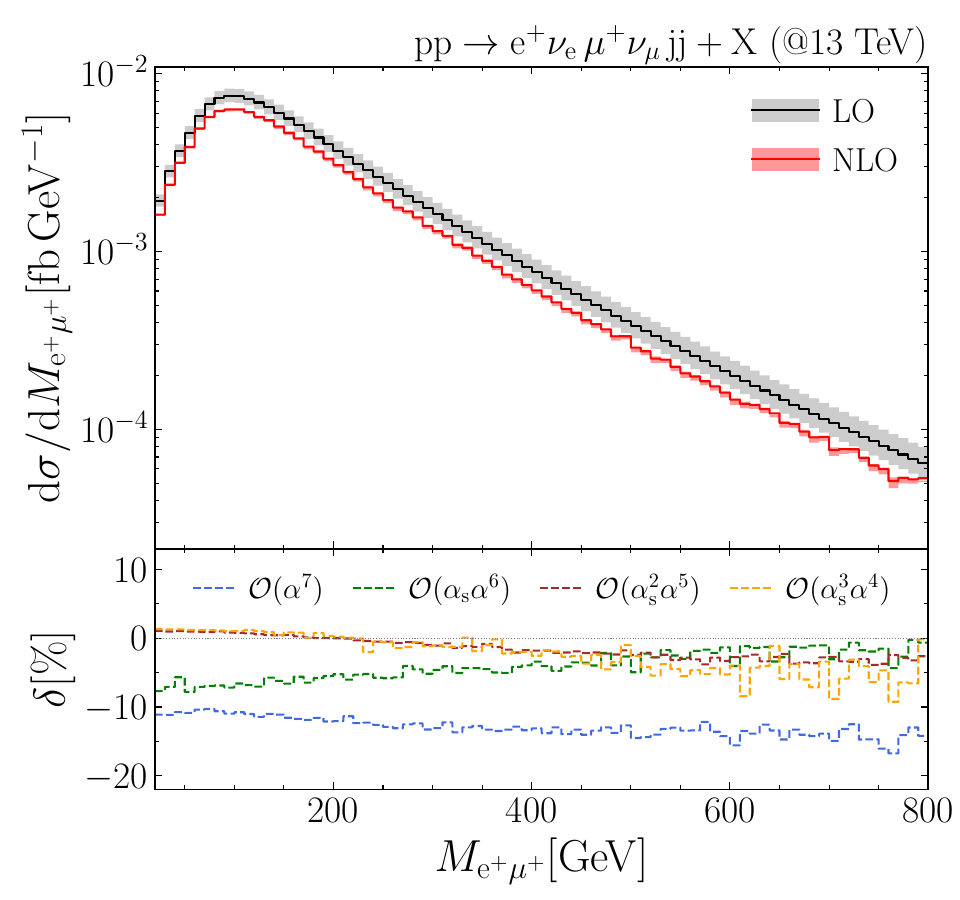}
	\caption{Differential distributions for $\Pp \Pp \to \Pep \Pgne \, \Pgmp \Pgngm \, \mathrm{j} \mathrm{j} + \mathrm{X}$ at the LHC with CM energy $13\TeV$:  invariant mass of the two tagging jets (left) and invariant mass of the four lepton system (right). The upper panels show the LO and NLO contributions, the lower panels show the relative NLO corrections in percent.}
	\label{fig:nlo-distributions3}
\end{figure}

Finally, we show the invariant-mass distribution of the 
two leading jets (left) and the invariant mass of the charged leptons (right) in \reffi{fig:nlo-distributions3}.
The hierarchy of the various corrections and their behaviour can be widely explained following
similar arguments as above.
The trend of the dominating genuine weak corrections of $\mathcal{O}(\alpha^7)$ towards
increasingly negative corrections for larger scales is visible as for the 
$p_{\mathrm{T}}$ distributions, but the increase in size to about $-15\%$ for the largest
considered scales is much less dramatic. This is due to the fact that the domain
of large invariant masses $M_{\mathrm{j}_1\mathrm{j}_2}$ or $M_{\Pep\mu^+}$
is not fully dominated by the Sudakov regime of the WW$\to$WW subprocess, 
because the $t$-channel-like momentum transfer in the subprocess is not forced to be large.
Thus, the impact of the leading EW Sudakov corrections is
damped to the size of the subleading singly-logarithmic EW high-energy corrections.
In the $M_{\mathrm{j}_1\mathrm{j}_2}$ distribution, which is mostly dominated by
EW LO contributions,
the corrections of $\mathcal{O}(\alpha_s\alpha^6)$ typically have an impact at the
5\%~level, while the remaining two orders with higher powers of $\alpha_s$ hardly
reach 1\%.
The mixed EW--QCD and the pure QCD corrections show, however, an interesting crossover
in the $M_{\Pep\mu^+}$ distribution at $M_{\Pep\mu^+}\sim400\GeV$, which we attribute
to the increasing influence of the LO QCD contribution 
(see \reffi{fig:lo-distributions}, lower left panel).
For $M_{\Pep\mu^+}<400\GeV$, where the EW part strongly dominates the LO cross section,
the $\mathcal{O}(\alpha_s\alpha^6)$ correction is the second largest
after the genuine EW correction, and the remaining NLO orders are at the 1\%~level.
For $M_{\Pep\mu^+}>400\GeV$, where the LO QCD part competes in size with the EW LO part, 
the corrections of
$\mathcal{O}(\alpha_s^2\alpha^5)$ and $\mathcal{O}(\alpha_s^3\alpha^4)$ dominate over
$\mathcal{O}(\alpha_s\alpha^6)$ and reach $\sim-5\%$ for large $M_{\Pep\mu^+}$.

\subsection{Quality of the DPA/VBS approximation}
\label{sec:dpa_vbs_results}

\begin{figure}
\centering
\includegraphics[page=9,width=0.49\textwidth]{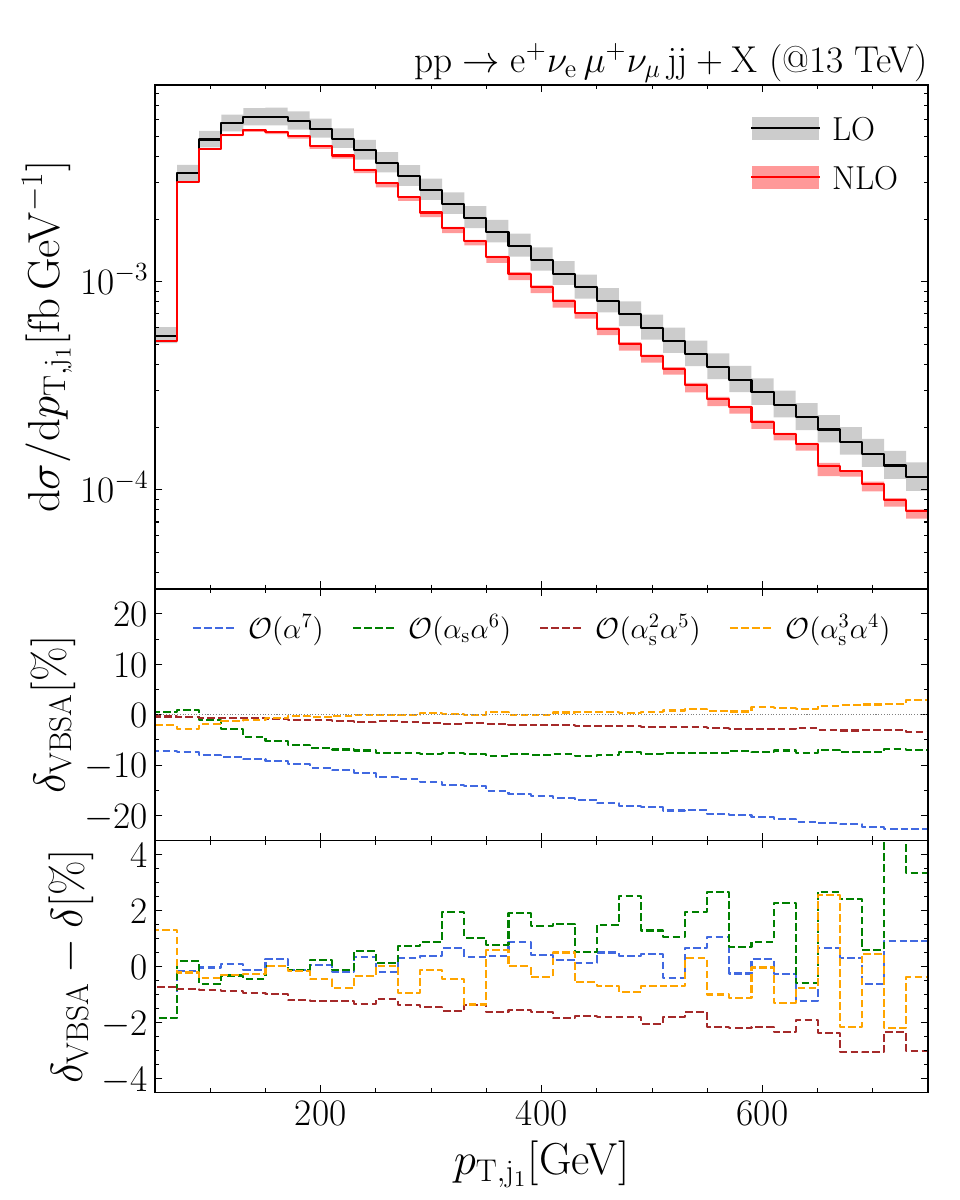}%
\includegraphics[page=10,width=0.49\textwidth]{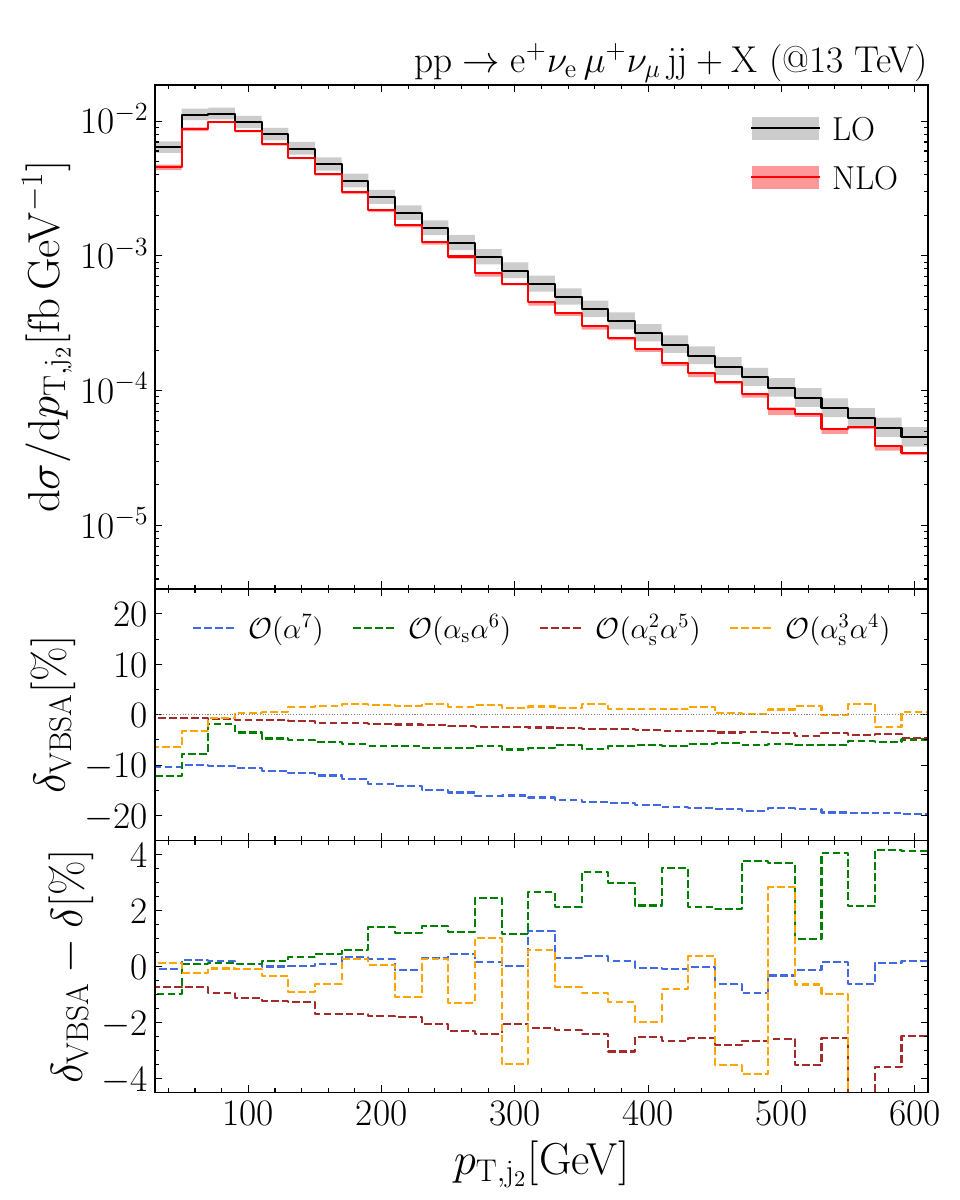}\\
\includegraphics[page=11,width=0.49\textwidth]{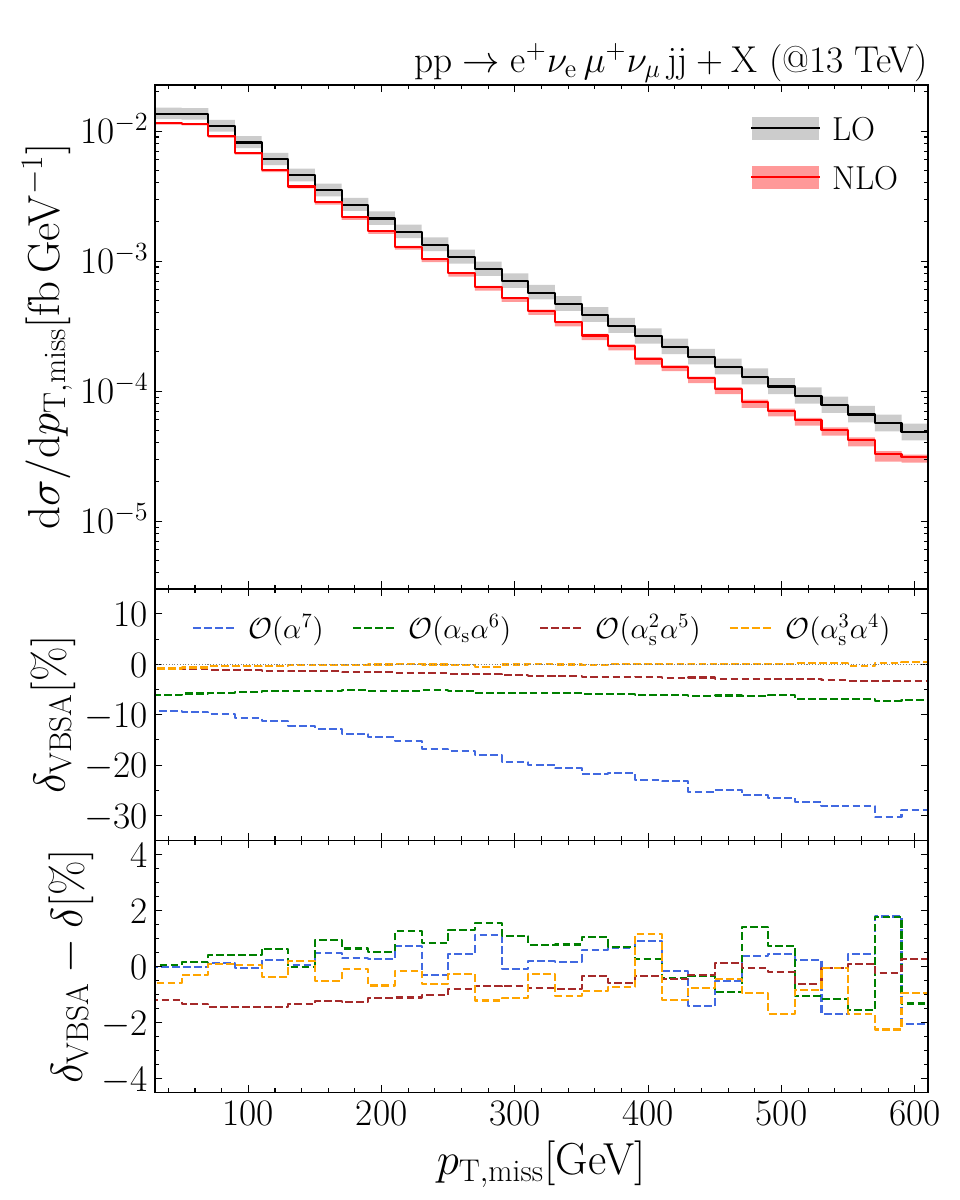}%
\includegraphics[page=8,width=0.49\textwidth]{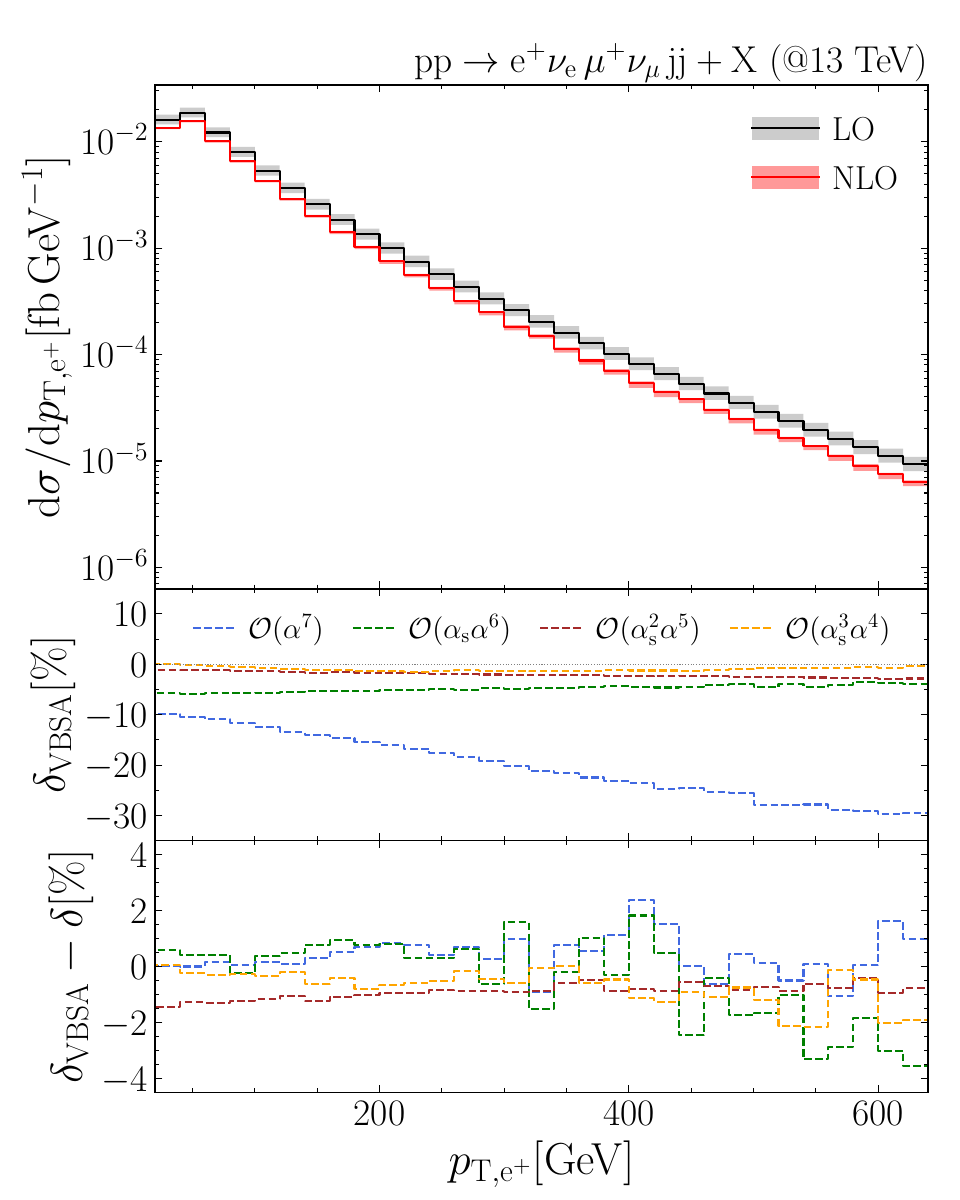}
\caption{Differential distributions for $\Pp \Pp \to \Pep \Pgne \, \Pgmp \Pgngm \, \mathrm{j} \mathrm{j} + \mathrm{X}$ at the LHC with CM energy $13\TeV$:  
transverse momentum of the hardest jet (top left), transverse momentum of the second hardest jet (top right), missing transverse energy (bottom left), and transverse momentum of the positron (bottom right).
The upper panels show the LO and full NLO contributions, 
the middle panels show the relative VBSA corrections in percent,
and the lower panels show the difference between VBSA and full relative corrections.}
\label{fig:nlo-distributions-vbsa1}
\end{figure}

\begin{figure}
\centering
\includegraphics[page=3,width=0.49\textwidth]{nlos_vs_vbsa}%
\includegraphics[page=4,width=0.49\textwidth]{nlos_vs_vbsa}\\
\includegraphics[page=5,width=0.49\textwidth]{nlos_vs_vbsa}%
\includegraphics[page=1,width=0.49\textwidth]{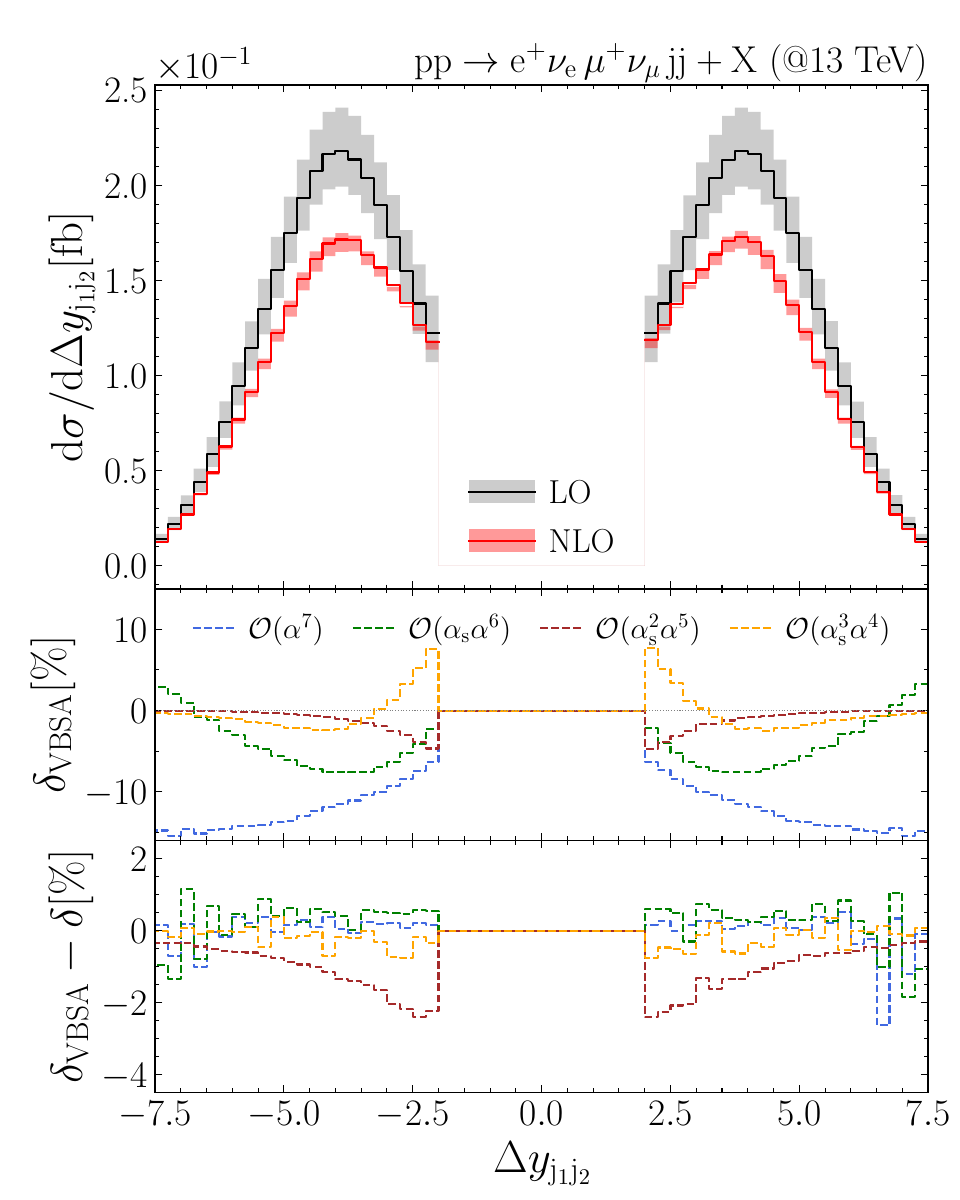}
\caption{Differential distributions for $\Pp \Pp \to \Pep \Pgne \, \Pgmp \Pgngm \, \mathrm{j} \mathrm{j} + \mathrm{X}$ at the LHC with CM energy $13\TeV$:  
rapidity of the hardest jet (top left), rapidity of the second hardest jet (top right), rapidity of the anti-muon (bottom left), and rapidity difference of the two tagging jets (bottom right). 
The upper panels show the LO and full NLO contributions, 
the middle panels show the relative VBSA corrections in percent,
and the lower panels show the difference between VBSA and full relative corrections.}
\label{fig:nlo-distributions-vbsa2}
\end{figure}

\begin{figure}
\centering
\includegraphics[page=7,width=0.49\textwidth]{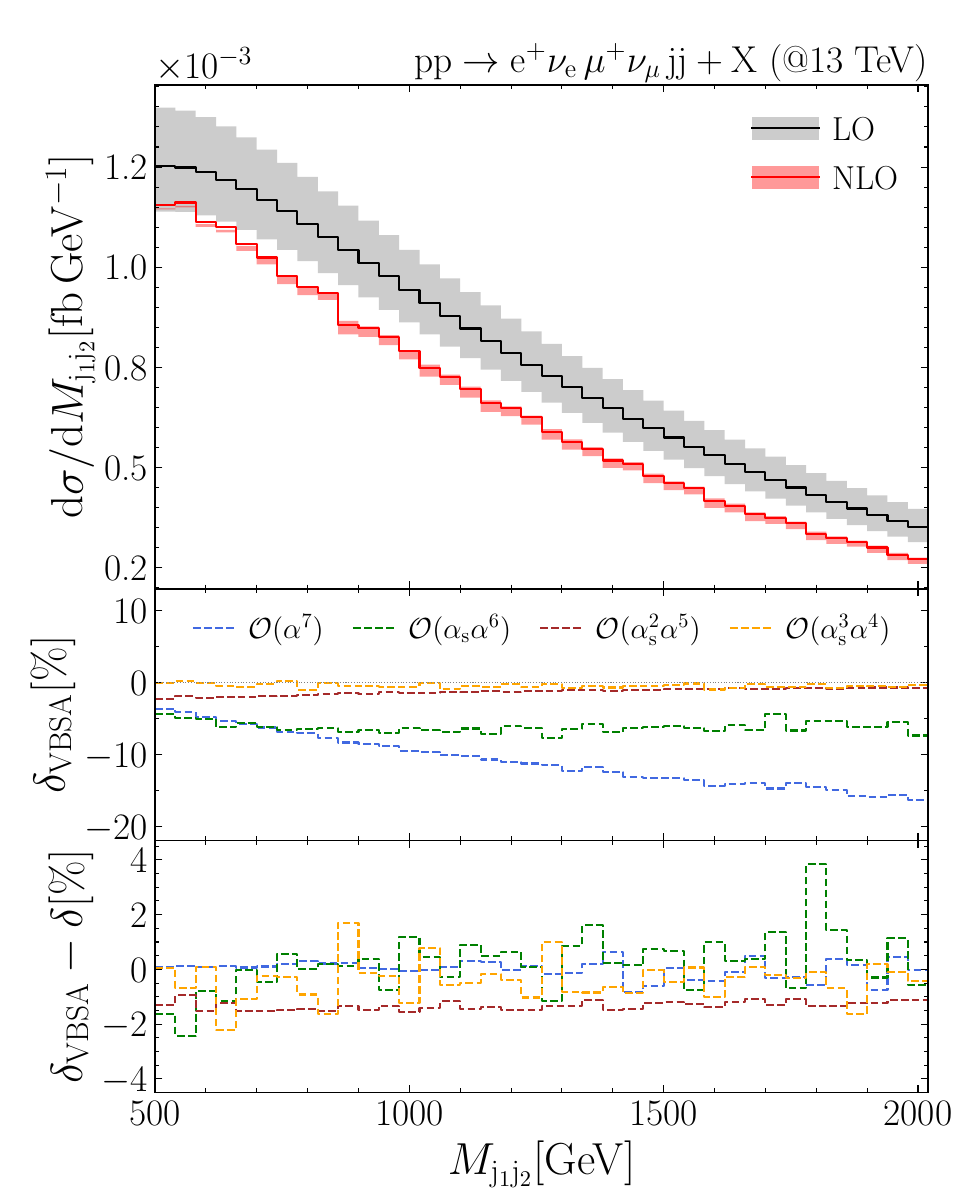}%
\includegraphics[page=6,width=0.49\textwidth]{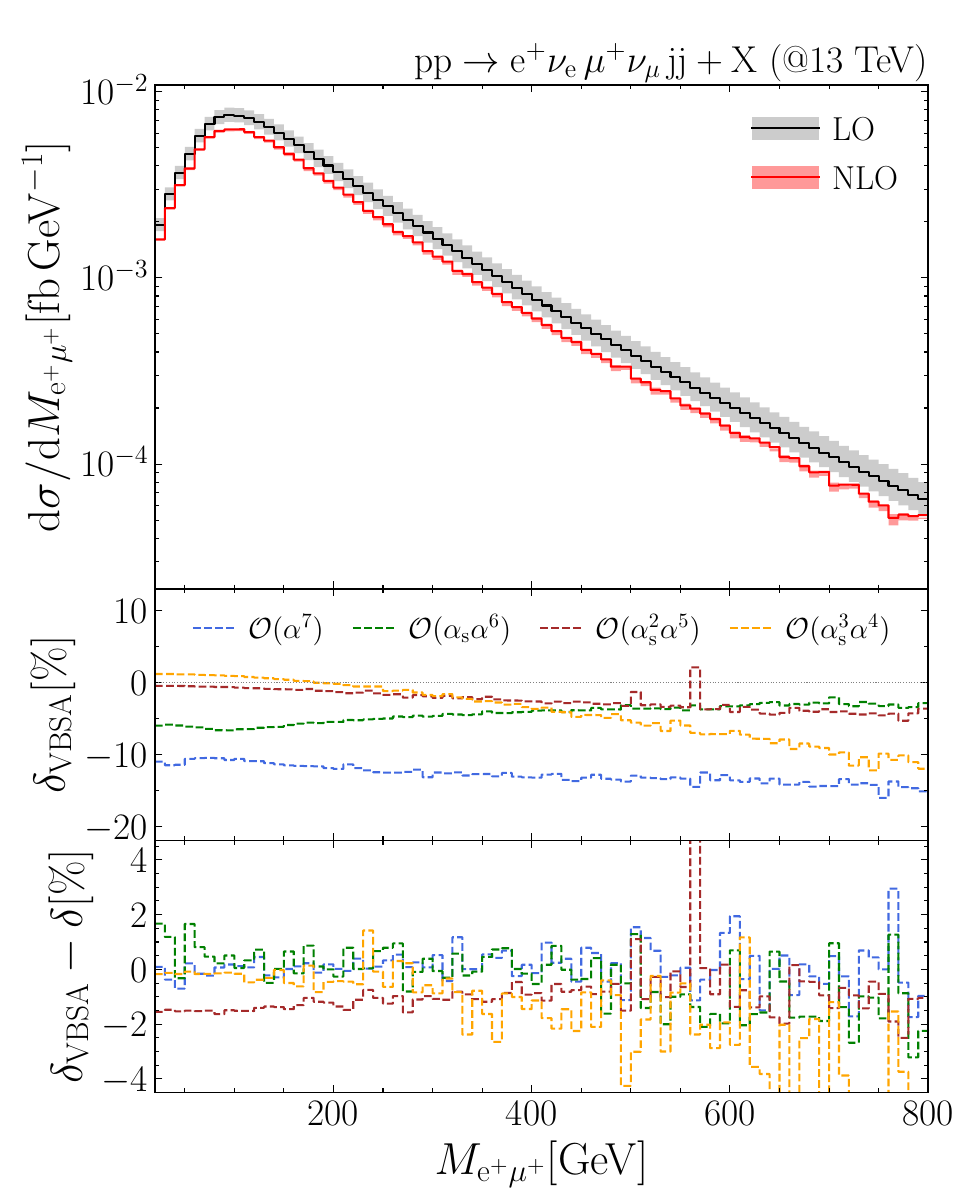}
\caption{Differential distributions for $\Pp \Pp \to \Pep \Pgne \, \Pgmp \Pgngm \, \mathrm{j} \mathrm{j} + \mathrm{X}$ at the LHC with CM energy $13\TeV$:  
invariant mass of the two tagging jets (left) and invariant mass of the four lepton system (right). 
The upper panels show the LO and full NLO contributions, 
the middle panels show the relative VBSA corrections in percent,
and the lower panels show the difference between VBSA and full relative corrections.}
\label{fig:nlo-distributions-vbsa3}
\end{figure}

In Figs.~\ref{fig:nlo-distributions-vbsa1}, \ref{fig:nlo-distributions-vbsa2}, 
and \ref{fig:nlo-distributions-vbsa3} we compare the
full NLO results discussed in \refse{sec:diff_cross_sections} with the combined DPA/VBS approximation
described in \refses{sec:vbsa} and \ref{sec:dpa}, respectively.
For brevity we denote the approximation with \enquote{VBSA}, but it also includes the DPA.
In the upper panels, the figures show the full off-shell LO and NLO results, 
in the middle panels the approximated relative corrections, 
and in the lower panels the differences between the relative corrections of 
1) the full off-shell calculation and 2) the approximated calculation, in detail defined as
\begin{equation}
\delta_\text{VBSA} - \delta = 
\frac{\sigma^\text{NLO, VBSA} - \sigma^\text{NLO}}{\sigma^\text{LO}} \text{.}
\end{equation}
In both calculations the LO matrix elements are those from the full off-shell calculation.
In phase-space regions that dominate the cross section the difference between full off-shell
calculation and the approximative VBSA result hardly exceed the 1\% level, nicely establishing
the validity of the approximation. 
In the tails of distributions, where the cross section is suppressed, the differences sometimes
reach a few percent. 
We emphasize that the Monte Carlo integration uncertainties, which can be inferred from the 
fluctuations of neighbouring bins, is by far dominated by the full off-shell calculation,
where the evaluation of loop amplitudes are the limiting factor in CPU time.

Moreover, we point out that the DPA part of the overall approximation
(\ie without the VBS approximation at all) typically causes
only small deviations from the full result, \ie not more than $0.5\%$.
This observation is consistent with the comparison made in \citere{Schwan:2018nhl} at LO. 
The good quality of pole expansions was also observed in related processes,
such as 
$\Pep\Pem\to\PW\PW\to4\,$fermions~\cite{Denner:2005es,Denner:2005fg,Denner:2019vbn},
$\Pp\Pp\to\PW\PW\to4\,$leptons~\cite{Biedermann:2016guo},
and $\Pp\Pp\to\PW\PW\PW\to6\,$leptons~\cite{Dittmaier:2019twg}.
We recall that we apply the DPA only to the virtual corrections, rendered
infrared-finite by adding the endpoint terms from dipole subtraction. 
A degradation of the DPA quality is only expected when loop diagrams with less resonances
start to play a role, which for instance happens in high-energy tails of
transverse-momentum distributions, as observed for diboson production in
\citere{Biedermann:2016guo}.

The amount of CPU time required to evaluate the DPA loop amplitudes is roughly a factor 10 faster than fully off-shell loop amplitudes.
The evaluation of the virtual $\mathcal{O} (\alpha^7)$ corrections are the bottleneck in terms of the required statistics, making it the dominant source of Monte Carlo uncertainty per CPU hour.
The real corrections yield a much larger uncertainty per generated event.
However, since they are much faster to evaluate, we can increase the statistics in the phase-space integration of the real corrections relative to statistics used by the virtuals, so that the Monte Carlo uncertainties of the real corrections are subdominant.
The fact that the impact of the (dipole-subtracted) real corrections is quite small further adds to this damping of the corresponding statistical error.
The possibility to evaluate the NLO corrections roughly ten times faster than the full calculation renders the DPA an interesting tool for quick analyses.

The differences between full NLO and VBSA results are, thus, dominated by the VBS
approximations described in \refses{sec:vbsa}.
This feature is particularly prominent in the $\mathcal{O} (\alpha_\mathrm{s}^2 \alpha^5)$
corrections of mixed QCD--EW type, where the approximative quality of the VBSA is only 
at the 1.5\% level in the most important phase-space regions.
For this particular order, the VBS approximation
typically predicts a smaller correction than the full off-shell calculation
($\delta_\text{VBSA} - \delta < 0$).

\subsection{Numerical study of the EVA for LO predictions}
\label{sec:evba_results}

The effective vector-boson approximation (EVA) (see Sec.~\ref{sec:evba}) has the
computational advantage of involving only a drastically reduced number of diagrams 
that have to be evaluated and provides some knowledge about the VBS part of the 
underlying scattering process.
However, the subset of Feynman diagrams defining the EVA 
does not automatically define a gauge-independent amplitude.
Special measures are required to define a gauge-independent EVA amplitude, such as an on-shell projection of momenta that are involved in the $\PW\PW\to\PW\PW$ scattering subprocess and the restoration of transversality conditions of polarization vectors or of the currents describing vector-boson production or decays, as described in detail in \refapp{sec:evba-const}.
These modifications comprise some freedom
in practice, leading to different EVA variants, and it is not clear a priori
which variants are better than others. To investigate this, we have
implemented several different EVA variants, as defined in 
\refapp{sec:evba-variants} and summarized in \refta{tab:vba_modes}.
\begin{table}
	\centering
\renewcommand{\arraystretch}{0.8}
	\begin{tabular}{c|c|ccccc}
		\toprule
		 \multicolumn{2}{c|}{Method} &  $c_1$ & $c_2$ & $c_3$ & $c_4$ & $c_5$\\
		\midrule
		\multicolumn{2}{c|}{EVA KS~\cite{Kuss:1995yv}} & 0  & 1 & 1 & 1 & 1 \\
		\midrule
		\multirow{4}{*}{EVA 1} & a &  1 & 1 & 1 & 0 & 0  \\
		& b  & 1 & 1 & 0 & 0 & 0  \\
		& c & 1 & 0 & 0 & 0 & 0 \\
		& d & 1 & 0 & 1 & 0 & 0 \\
		\midrule
		\multirow{4}{*}{EVA 2} & a &  0 & 1 & 1 & 0 & 0  \\
		& b  & 0 & 1 & 0 & 0 & 0  \\
		& c & 0 & 0 & 0 & 0 & 0 \\
		& d & 0 & 0 & 1 & 0 & 0 \\
		\midrule
		\multirow{4}{*}{EVA 3} & a &  1 & 1 & 1 & 1 & 0  \\
		& b  & 1 & 1 & 0 & 1 & 0  \\
		& c & 1 & 0 & 0 & 1 & 0 \\
		& d & 1 & 0 & 1 & 1 & 0 \\
		\midrule
		\multirow{4}{*}{EVA 4} & a &  0 & 1 & 1 & 1 & 0  \\
		& b  & 0 & 1 & 0 & 1 & 0  \\
		& c & 0 & 0 & 0 & 1 & 0 \\
		& d & 0 & 0 & 1 & 1 & 0 \\
		\bottomrule
	\end{tabular}
	\caption{Survey of different EVA versions implemented in our Monte Carlo integrator,
as defined by the flags $c_1,\dots,c_5$ described in App.~\ref{sec:evba-variants} in detail.
In short terms, the $c_k=0/1$ means no/yes for:
$c_1$:~restoration of transversality for initial-state 
W~bosons in VBS amplitude; 
$c_2$:~restoration of transversality for final-state 
W~bosons in VBS amplitude; 
$c_3$:~restoration of transversality for final-state lepton currents;
$c_4$:~restoration of relative sign factor between transverse and longitudinal polarization
vectors in the completeness relation of the incoming W-boson propagators that gets
lost by on-shell projection of 
longitudinal initial-state W~bosons;
$c_5$:~``Kuss--Spiesberger'' (KS) factor to restore explicit $1/\sqrt{-k_i^2}$ 
factors in the polarization vectors for longitudinal initial-state W~bosons.}
	\label{tab:vba_modes}
\end{table}
In the following, we systematically compare numerical results from these EVA variants for integrated and differential cross sections to explore the reliability of EVA versions and to see which properties are the most important to maximize the quality of the EVA.

\subsubsection*{Integrated cross section}

We start by comparing the integrated cross sections obtained with the EVA 
implementations, $\sigma_\text{EVA}$, with the full LO calculation $\sigma_\text{LO}$ 
at $\mathcal{O}(\alpha^6\alpha_s^0)$. In order to quantify the quality of the EVA, we define the relative difference
\begin{align}
	\Delta_\text{EVA}=\frac{\sigma_\text{EVA}-\sigma_\text{LO}}{\sigma_\text{LO}},
	\label{eq:relativediff}
\end{align}
where $\sigma_\text{LO}$ is the LO cross section based on the full off-shell matrix elements.
As the EVA is initially motivated in the collinear limit of the (anti)quarks radiating the vector bosons that are initiating the VBS process, we choose different kinematic setups that are based on experimental analyses on the one hand but vary the selection of forward/backward scattered jets on the other:
\begin{enumerate}[(I)]
\item 
The standard setup from Sec.~\ref{sec:phase-space-volumes}, where
the leading (subleading) jet has to fulfil 
$p_{\text{T},\mathrm{j}} > \SI{65(35)}{\giga\electronvolt}$.
Note that this requirement actually excludes the phase-space region of very small
$p_{\text{T},\mathrm{j}}$ where the EVA is best motivated.
\item 
We replace the cut on the jet transverse momentum with an inverse cut,
\begin{align}
	p_{\text{T},\mathrm{j}} < \SI{150}{\giga\electronvolt},
\end{align} 
without demanding a lower cut on $p_{\text{T},\mathrm{j}}$,
to restrict the tagged jets to the region where the EVA is motivated.
\item 
Finally, we apply an even stronger inverse cut on the jet transverse momentum,
\begin{align}
	p_{\text{T},\mathrm{j}} < \SI{100}{\giga\electronvolt},
\end{align} 
to zoom deeper into the EVA region.
\end{enumerate}
The results for the integrated cross sections are given in \refta{tab:EVAdefault}.
\begin{table}
	\centering
	\begin{tabular}{c|c| S[table-format=2.5e+1]@{} c | S[table-format=1.5] c| S[table-format=1.6] c}
		\toprule
		\multicolumn{2}{c}{ } & \multicolumn{2}{c}{Setup I} &  \multicolumn{2}{c}{Setup II} & \multicolumn{2}{c}{Setup III}\\
		\cmidrule(r){3-4} \cmidrule(r){5-6} \cmidrule(r){7-8}
		\multicolumn{2}{c}{ } &  {$\sigma$ [fb]} &  {$\Delta_\text{EVA}$ [\%]}  &  {$\sigma$ [fb]} & {$\Delta_\text{EVA}$ [\%]}  & {$\sigma$ [fb]} & {$\Delta_\text{EVA}$ [\%]} \\
		\midrule
		\multicolumn{2}{c|}{LO} & 1.2455(2) & - & 0.4559(2) & - & 0.17191(9) & - \\
		\midrule
		\multicolumn{2}{c|}{EVA KS} & \color{mygood} 1.8739(3) & \color{mygood} 50.5 & 0.7521(3) & 65.0 & 0.3364(2) & 95.7  \\
		\midrule
		\multirow{4}{*}{EVA 1} & a & 5.603(1) & 350 & 0.6011(2) & 31.8 &  0.17236(9) & 0.26 \\
		& b & \color{mybad}5.86(5)e+4 &  \color{mybad} $>\!10^6$ & 0.6338(2) & 39.0 & 0.17649(9) & 2.7\\
		& c & 6.876(3) & 452 & 0.58311(2) & 27.9 & \color{mygood} 0.17211(9) & \color{mygood}0.12 \\
		& d & 15.74(5) & $>\!10^3$ & 0.6311(2) & 38.4 & 0.17931(9) & 4.3\\
		\midrule
		\multirow{4}{*}{EVA 2} & a & 2.7453(5) & 120 & 0.9350(3) & 105 & 0.3220(2) & 87.3 \\
		& b & 3.864(3) & 210 & 0.9795(4) & 115 & 0.3355(2) & 95.2\\
		& c & 2.4017(5) & 92.8 & 0.9131(3) & 100 & 0.3196(2) & 85.9 \\
		& d & 3.0143(6) & 142 & 0.9672(4) & 112 & 0.3279(2) & 90.7\\
		\midrule
		\multirow{4}{*}{EVA 3} & a & 5.133(1) & 312 & 0.5357(2) & 17.5 & 0.16511(8) & $-4.0$ \\
		& b  & \color{mybad}5.83(3)e+4 & \color{mybad} $>\!10^6$ & 0.5630(2) & 23.5 & 0.16841(9) & $-2.0$\\
		& c & 6.273(3) & 404 & \color{mygood} 0.5135(2) & \color{mygood} 12.6 & 0.16360(8) & $-4.8$ \\
		& d & 15.62(6) & $>\!10^3$ & 0.5633(2) & 23.6 & 0.17113(9) & $-0.45$\\
		\midrule
		\multirow{4}{*}{EVA 4} & a & 2.2918(4) & 84.0 & 0.7277(2) & 59.6 & 0.2517(2) & 46.4 \\
		& b & 3.346(3) & 169 & 0.7630(3) & 67.4 & 0.2639(2) & 53.5\\
		& c & 1.9536(3) & 56.9 & 0.7203(2) & 58.0 & 0.2574(2) & 49.7 \\
		& d & 2.5877(5) & 108 & 0.7748(3) & 69.9 & 0.2656(2) & 54.5\\
		\bottomrule
	\end{tabular}
	\caption{Integrated cross sections at order $\mathcal{O}(\alpha^6)$ for the process 
	$\Pp \Pp \to \Pep \Pgne \, \Pgmp \Pgngm \, \mathrm{j} \mathrm{j} + \mathrm{X}$
	using the full LO matrix element, and various EVA implementations. For the EVA, we also present the relative difference $\Delta_\text{EVA}$ defined in \eqref{eq:relativediff}. The statistical uncertainties from the Monte Carlo integration are given in parentheses, and the best EVA results are highlighted in green. Particularly bad results are shown in red.}
	\label{tab:EVAdefault}
\end{table}
Apart from the EVA~KS version,
the EVA seems to perform best in regions of low $p_{\text{T},\mathrm{j}}$, as expected.
The deviation $\Delta_\text{EVA}$ globally decreases in the sequence of setups~I, II, III. 
Within a fixed setup, there are, however, huge discrepancies in the quality of the EVA depending on the particular choice of implementation. 
The EVA~1 and EVA~3 implementations, which both guarantee that the polarization vectors $\varepsilon_{\lambda,i}^*$ of the incoming W~bosons are orthogonal to the on-shell-projected W-boson momenta $\hat{k}_i$, outperform the other implementations
in setups~II and III.
In contrast, the predicted integrated cross section does not change much 
inside a given set of EVA~$n$ variants (\ie keeping $n=1,2,3,4$ fixed),
in which the members vary the other transversality conditions connected to final-state W~bosons and their decays.
While all EVA versions 1--4 show the expected trend of improving agreement with the full LO calculation, 
in changing setup according to $\mathrm{I}\to\mathrm{II}\to\mathrm{III}$
the EVA KS implementation~\cite{Kuss:1995yv, Kuss:1996ww} shows the 
opposite behaviour and appears to get worse in the collinear region, where any EVA version should become reliable by construction.

Inspecting the actual size of $\Delta_\text{EVA}$ in setup~I, which is inspired by an ATLAS analysis, we find that almost all EVA variants except for EVA~KS fail by $100\%$ (or more), even though their results show the correct trend in the sequence $\mathrm{I}\to\mathrm{II}\to\mathrm{III}$.
On the other hand, EVA~KS at least offers an approximative quality of $\sim50\%$ in setup~I, but shows a wrong trend in $\mathrm{I}\to\mathrm{II}\to\mathrm{III}$.
This surprising behaviour can be understood better after inspecting distributions.

\subsubsection*{Distributions}

\begin{figure}
	\centering
	\includegraphics[page=5,width=0.5\textwidth]{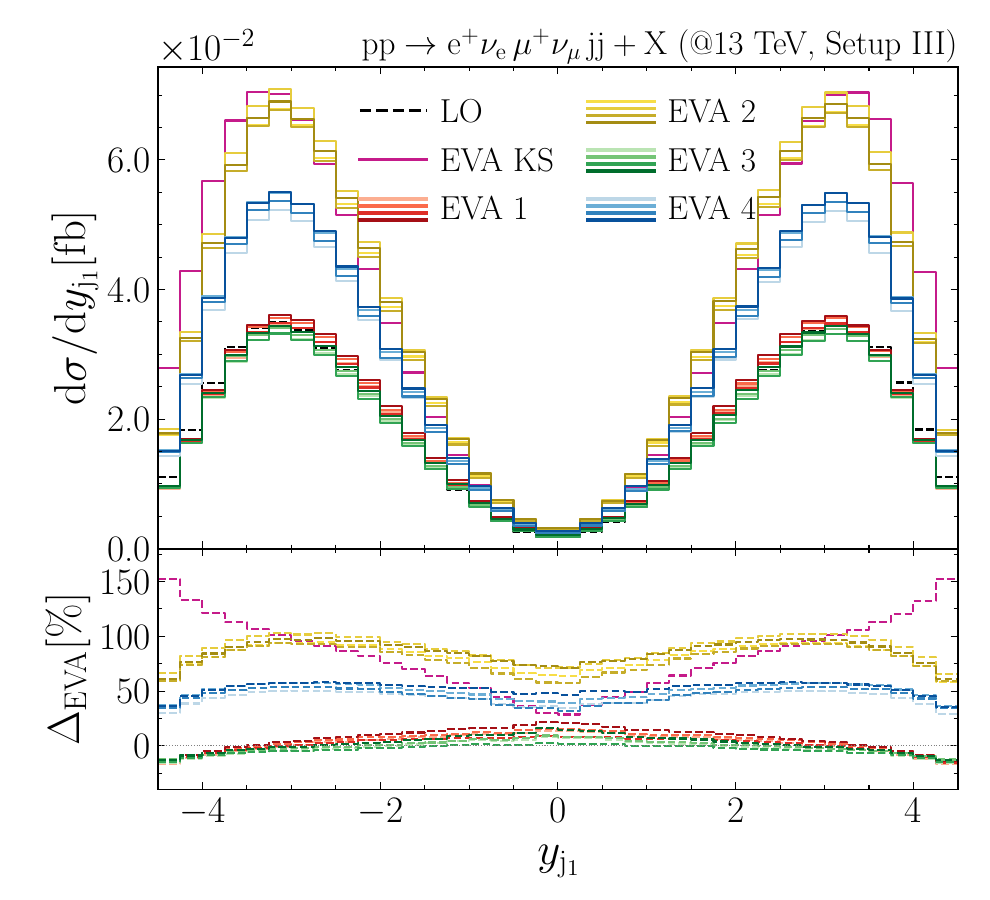}\hfill
	\includegraphics[page=3,width=0.5\textwidth]{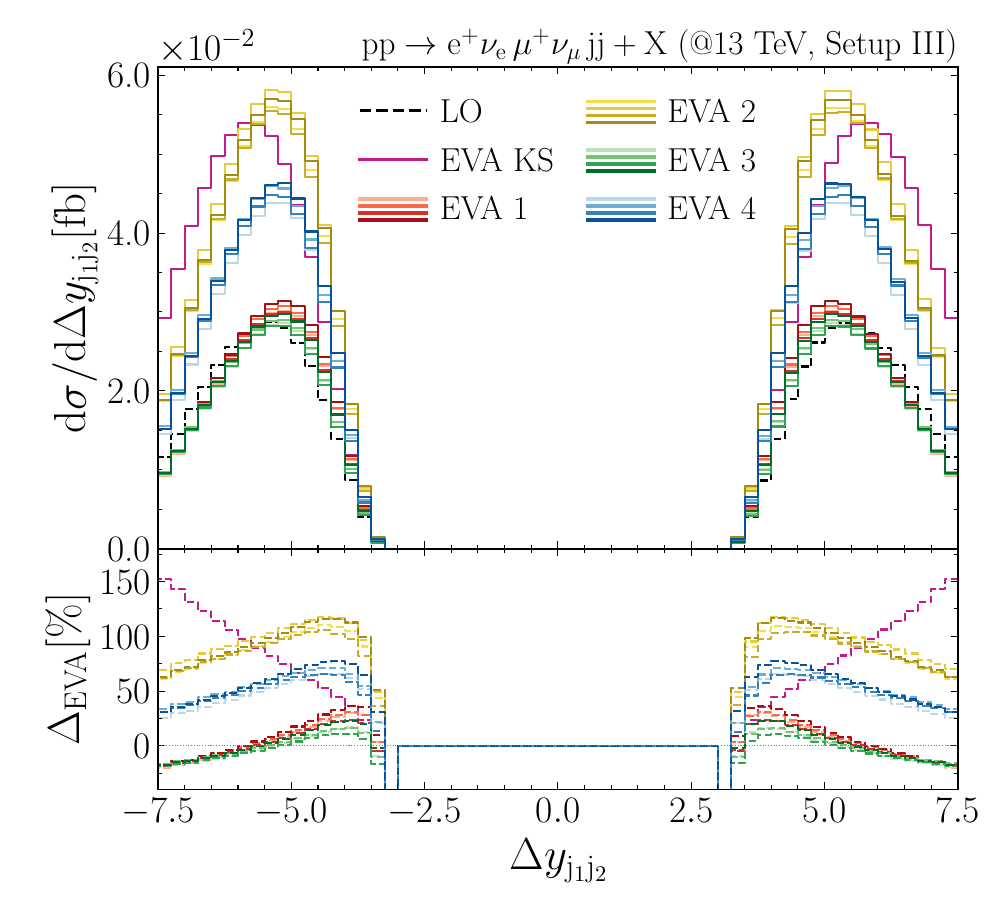}\\
	\includegraphics[page=4,width=0.5\textwidth]{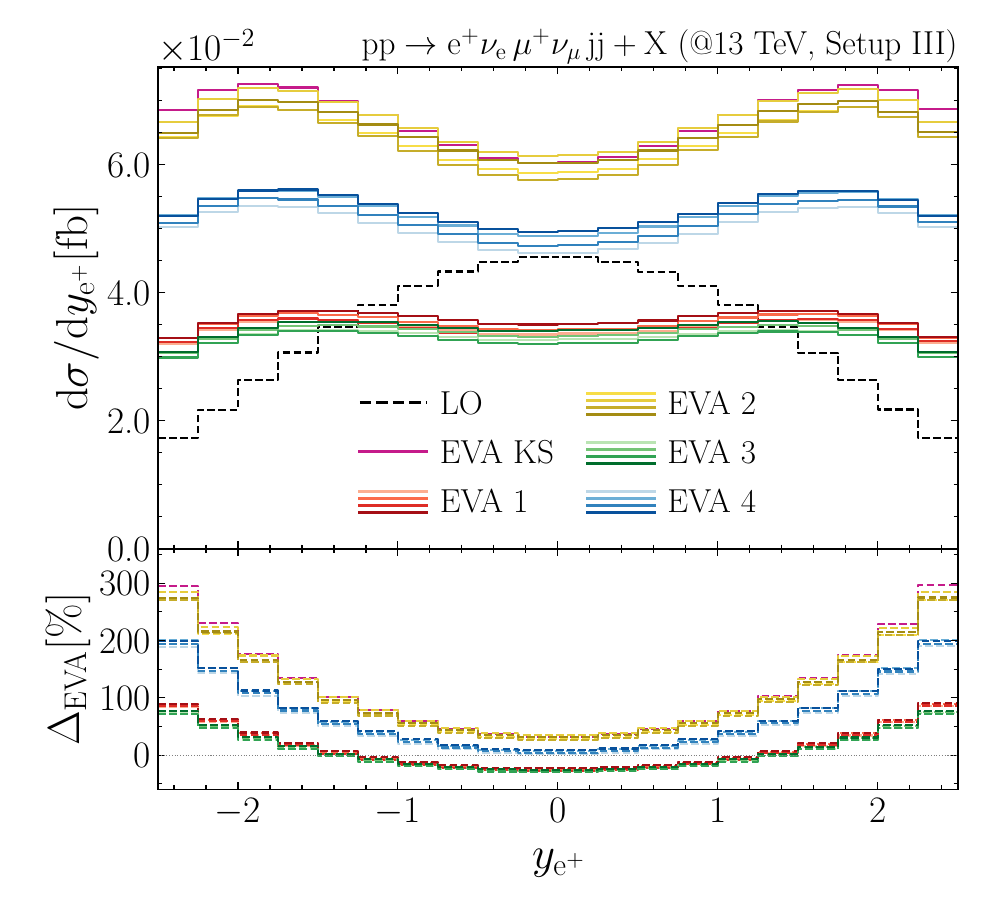}\hfill
	\includegraphics[page=2,width=0.5\textwidth]{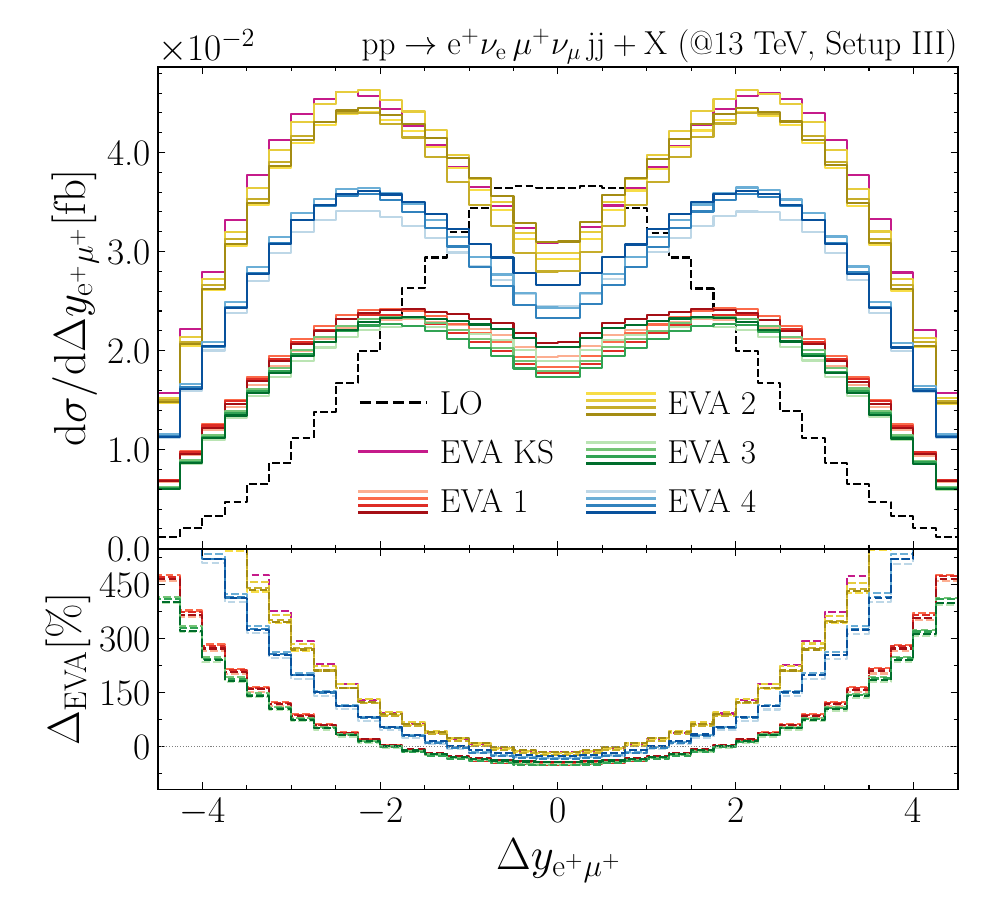}
	\caption{Differential distributions in rapidities $y_i$ and rapidity differences
$\Delta y_{ij}=y_i-y_j$
for $\Pp \Pp \to \Pep \Pgne \, \Pgmp \Pgngm \, \mathrm{j} \mathrm{j} + \mathrm{X}$ 
at the LHC with CM energy $13\TeV$ in setup~III. The upper panels show the LO cross-section contributions of
$\mathcal{O}(\alpha^6)$ and the corresponding EVA predictions, 
the lower panels show the relative deviation $\Delta_\text{EVA}$ in percent.}
\label{fig:EVAdist1}
\end{figure}

\begin{figure}
	\centering
	\includegraphics[page=10,width=0.5\textwidth]{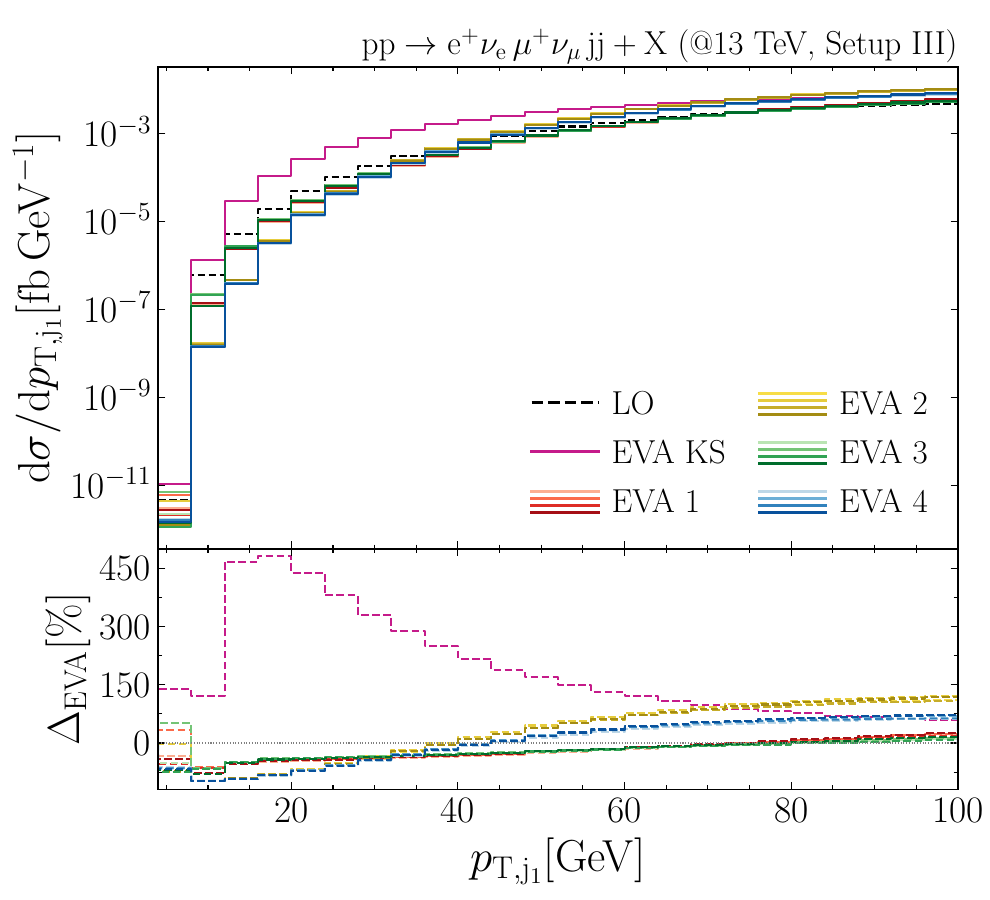}\hfill
	\includegraphics[page=8,width=0.5\textwidth]{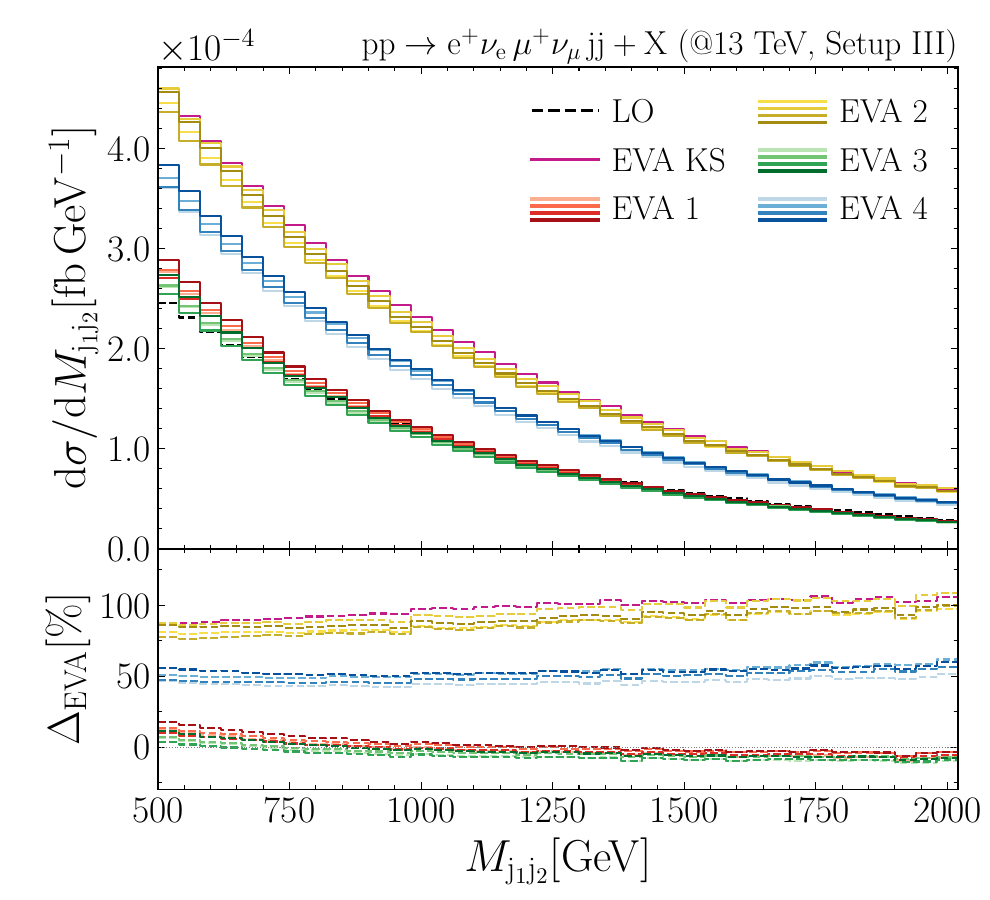}\\
	\includegraphics[page=9,width=0.5\textwidth]{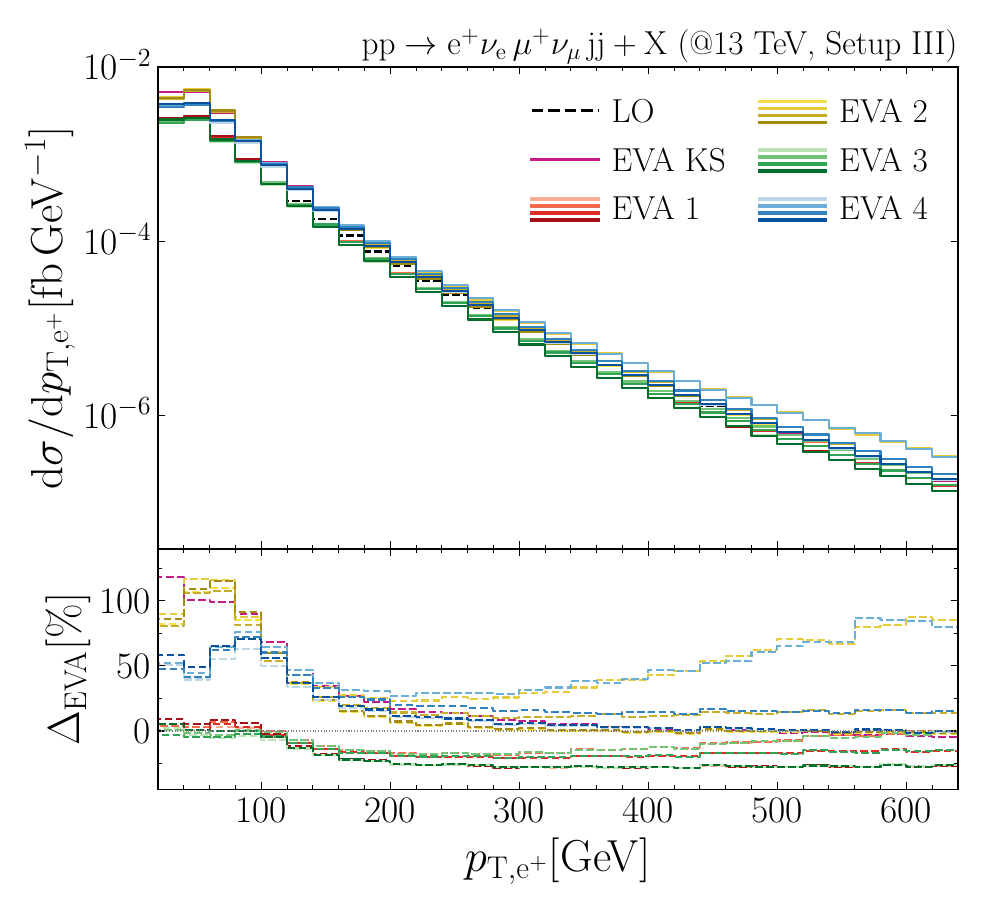}\hfill
	\includegraphics[page=7,width=0.5\textwidth]{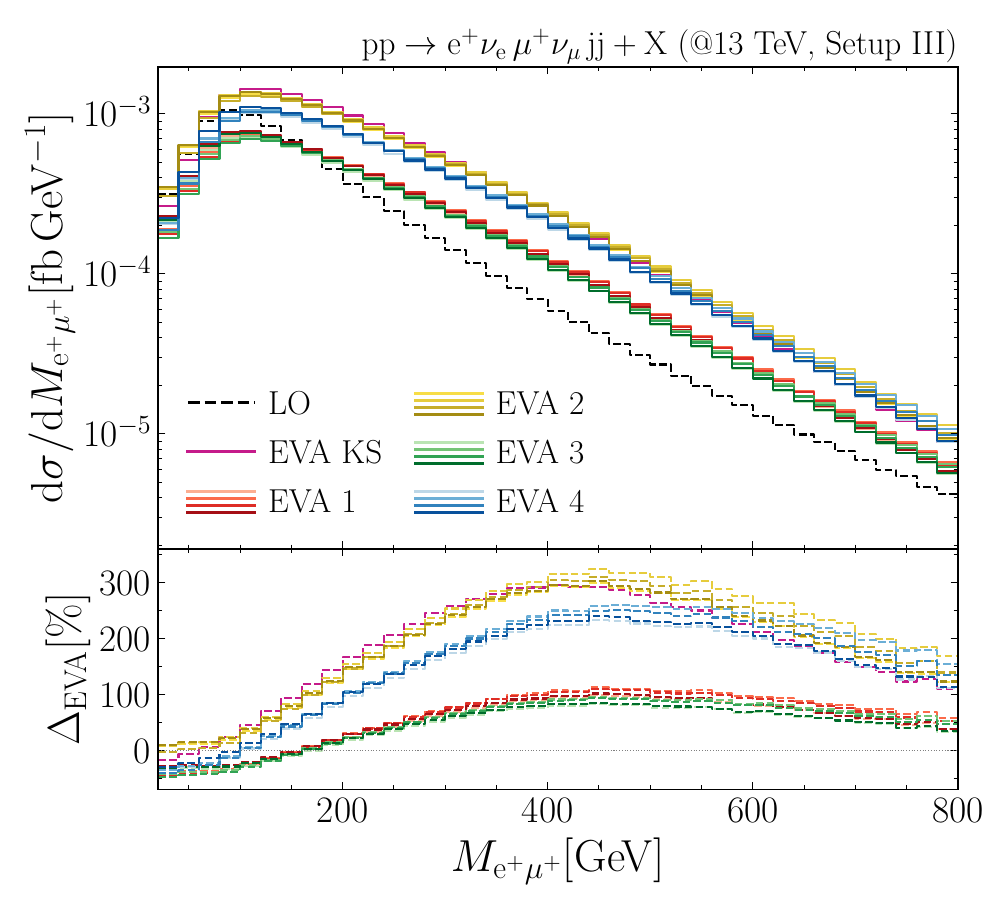}
	\caption{Differential distributions in transverse momenta $p_{\mathrm{T},i}$ 
and invariant masses $M_{ij}$ of the jet and charged-lepton pairs
for $\Pp \Pp \to \Pep \Pgne \, \Pgmp \Pgngm \, \mathrm{j} \mathrm{j} + \mathrm{X}$ 
at the LHC with CM energy $13\TeV$ in setup~III. The upper panels show the LO cross-section contributions of
$\mathcal{O}(\alpha^6)$ and the corresponding EVA predictions, 
the lower panels show the relative deviation $\Delta_\text{EVA}$ in percent.}
	\label{fig:EVAdist2}
\end{figure}

\begin{figure}
	\centering
	\includegraphics[page=1,width=0.5\textwidth]{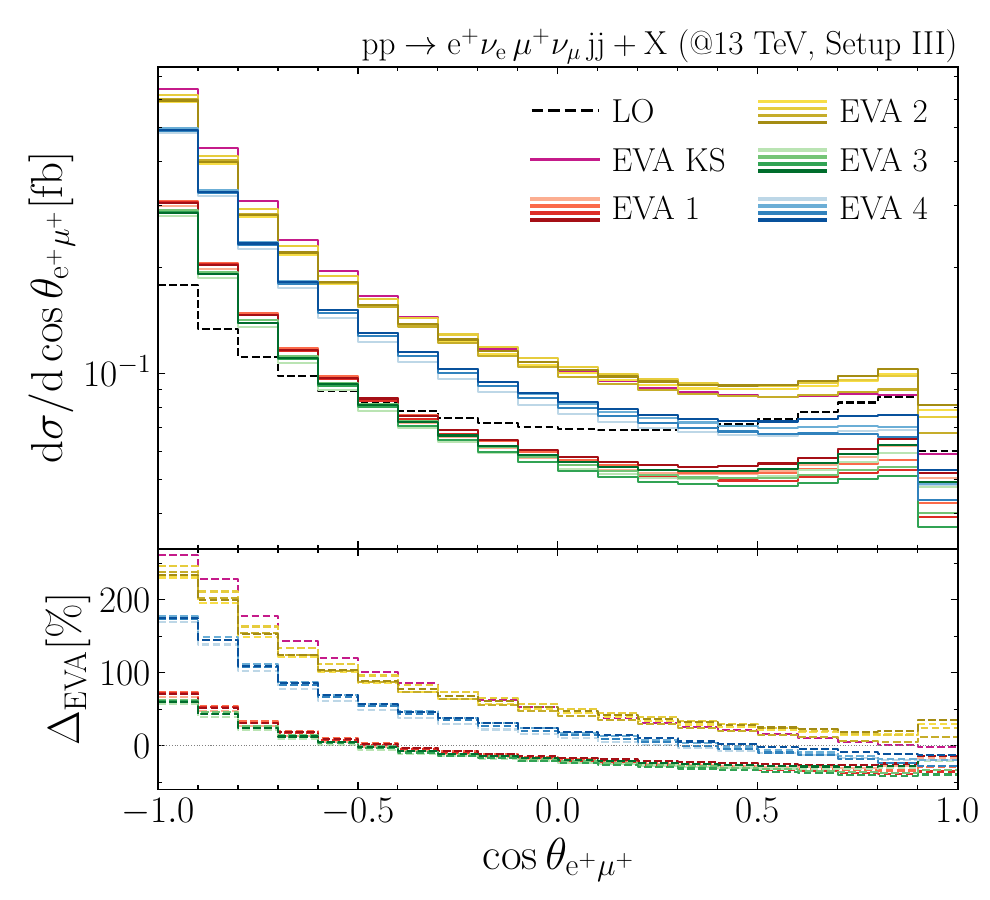}\hfill
	\includegraphics[page=6,width=0.5\textwidth]{wbas_ptj100.pdf}
	\caption{Differential distributions in the cosine of the angle $\theta_{\Pep\mu^+}$
between the charged leptons and in the four-lepton invariant mass $M_{4\ell}$
for $\Pp \Pp \to \Pep \Pgne \, \Pgmp \Pgngm \, \mathrm{j} \mathrm{j} + \mathrm{X}$ 
at the LHC with CM energy $13\TeV$ in setup~III. The upper panels show the LO cross-section contributions of
$\mathcal{O}(\alpha^6)$ and the corresponding EVA predictions, 
the lower panels show the relative deviation $\Delta_\text{EVA}$ in percent.}
	\label{fig:EVAdist3}
\end{figure}

We start our discussion by showing differential cross sections in setup~III, 
which zooms into the low-$p_{\text{T}}$ region for the two jets, to
check the validity of the EVA versions in the region where they should
provide a reasonable approximation.
In detail, \reffi{fig:EVAdist1} shows distributions in rapidities,
\reffi{fig:EVAdist2} in transverse momenta and invariant masses of jet and
charged-lepton pairs,
and \reffi{fig:EVAdist3} in the angle between the charged leptons and
in the four-lepton invariant mass. 

%
%

Similar to the situation for integrated cross sections described above,
the EVA predictions show up in four groups of curves called EVA~$n$ ($n=1,2,3,4$) plus the single curve for EVA~KS. 
This confirms the observation already made for the integrated cross section that details on handling the polarization vectors for the produced W bosons and the corresponding decay currents are of minor importance. 
At the same time, the results depend on the details of handling the initial-state W bosons quite sensitively.
The shapes of all jet observables are reproduced very well by the EVA, but only EVA~1 and EVA~3 can also reproduce their normalization at the 10--20\% level. 
In contrast, the EVA~KS prediction generally fails to reproduce the shapes of the jet observables, in particular
the $p_{\mathrm{T,j_1}}$ distribution is not reproduced at all.
This failure can be attributed to the KS factors $1/\sqrt{-k_i^2}$ for longitudinally polarized incoming W~bosons, because EVA~KS and EVA~4a differ only in these factors. 
We conclude that the KS factors do not systematically improve the
quality of the EVA, as might have been conjectured from the results on integrated cross sections given in \refta{tab:EVAdefault} alone. 
It seems that the application of the KS factors leads to a drastic overestimation of the cross section for small $p_{\mathrm{T,j_1}}$, where the virtualities $k_i^2$ of the incoming off-shell W~bosons are small, and to some damping of the typical overestimation of EVA predictions at large $p_{\mathrm{T,j_1}}$, where these virtualities are large in size. The resulting EVA~KS cross-section prediction in the ATLAS-inspired setup~I,
which deviates from the full LO result by only $\sim50\%$ (see \refta{tab:EVAdefault}), appears widely accidental.

In contrast to the relatively good approximation of the shapes of the jet distributions by the various EVA versions, the EVA generally cannot reproduce the shapes of leptonic observables well. 
For instance, the EVA approximations of the leptonic rapidity distributions shown in \reffi{fig:EVAdist1} seem to be systematically shifted towards larger rapidities.
We attribute this behaviour to some extent to the on-shell projection of the
W~momenta, which is designed to preserve the Mandelstam variables
$t_{\Pw\Pw}$ and $u_{\Pw\Pw}$ of the VBS subprocess (see \refapp{sec:evba-variants}).
Keeping $t_{\Pw\Pw}$ and $u_{\Pw\Pw}$ fixed in the on-shell projection, leads to a systematic increase of the WW CM energy in the transition 
$s_{\Pw\Pw}\to\hat s_{\Pw\Pw}$ because of the Mandelstam relations~\refeq{sec:Mandelstam-rel} and the fact that $k^2_i<0$ for the incoming W~bosons (for the outgoing W~bosons $k^2_i\sim\MW^2$ hardly changes during the projection).
For higher WW scattering energies, the VBS matrix element
systematically prefers scattering angles closer to the forward and backward directions.

\begin{figure}
	\centering
	\includegraphics[page=10,width=0.5\textwidth]{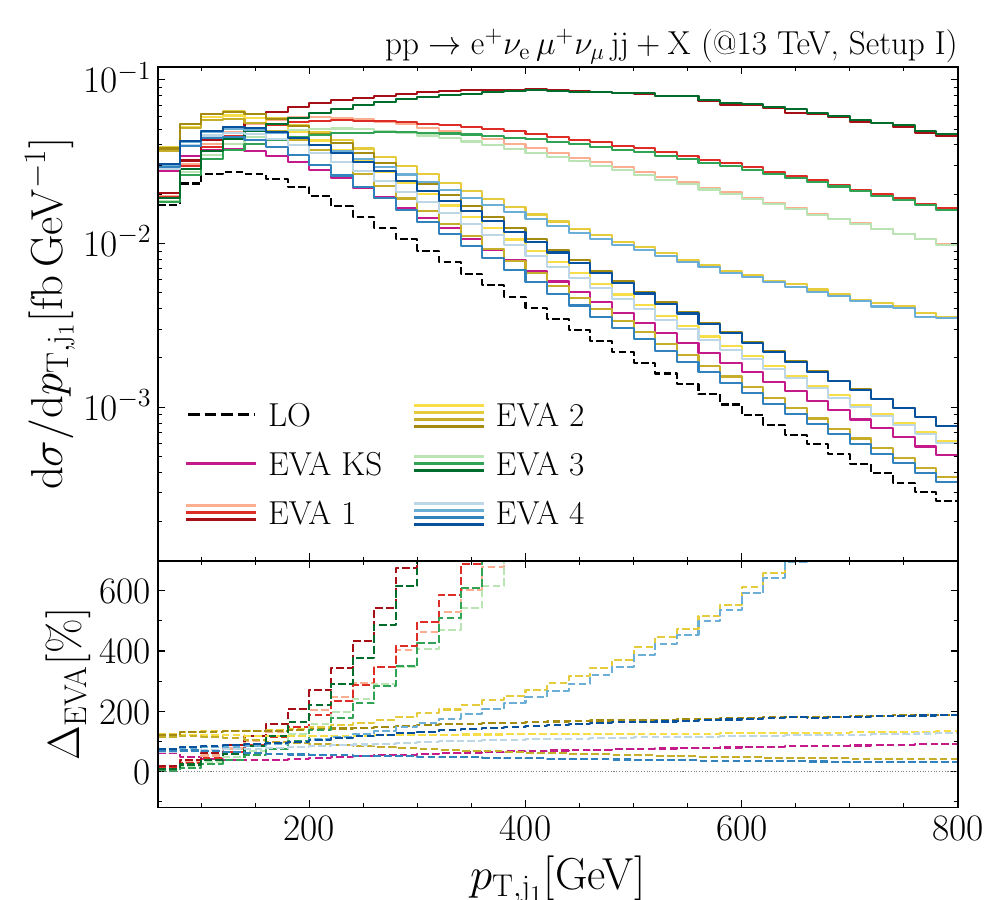}\hfill
	\includegraphics[page=3,width=0.5\textwidth]{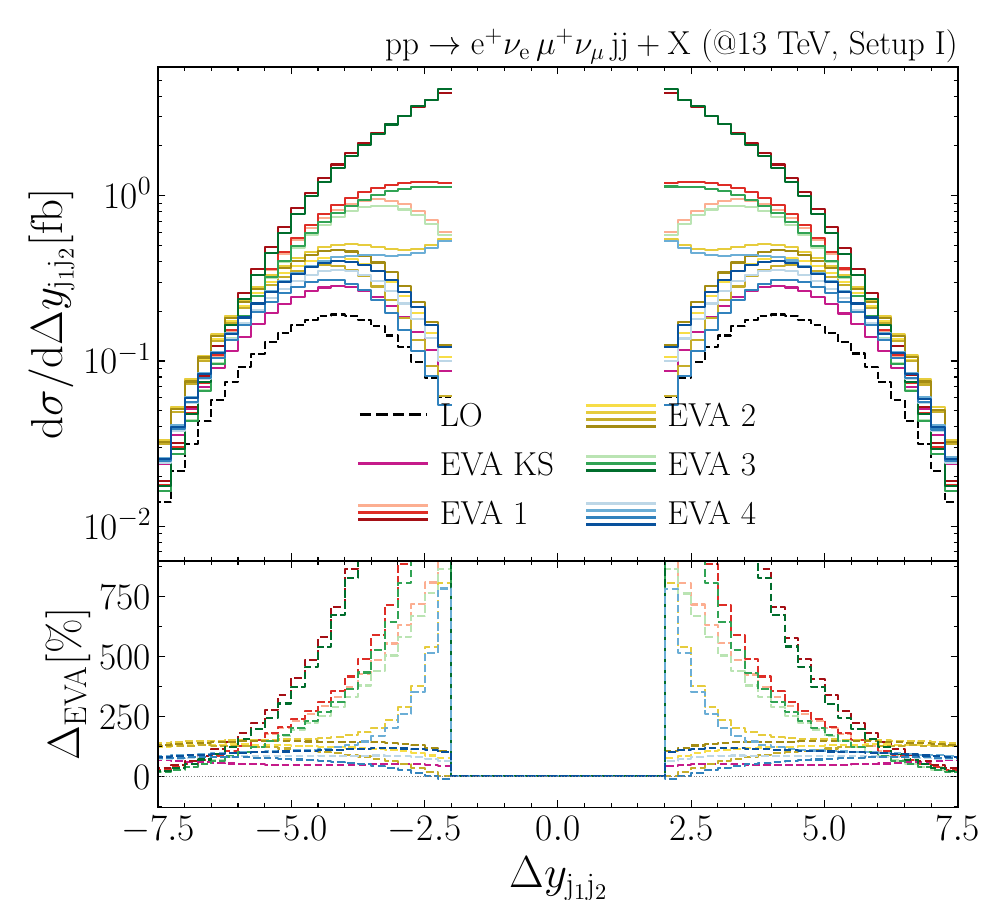}\\
	\includegraphics[page=9,width=0.5\textwidth]{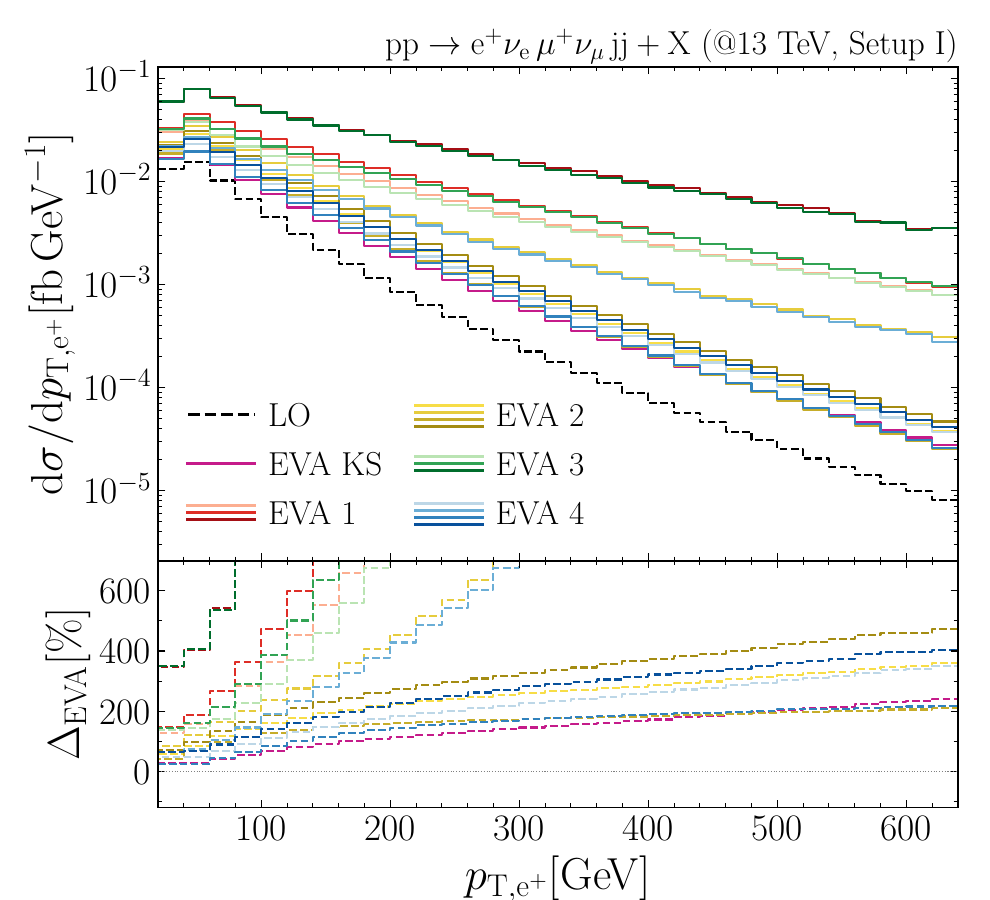}\hfill
	\includegraphics[page=2,width=0.5\textwidth]{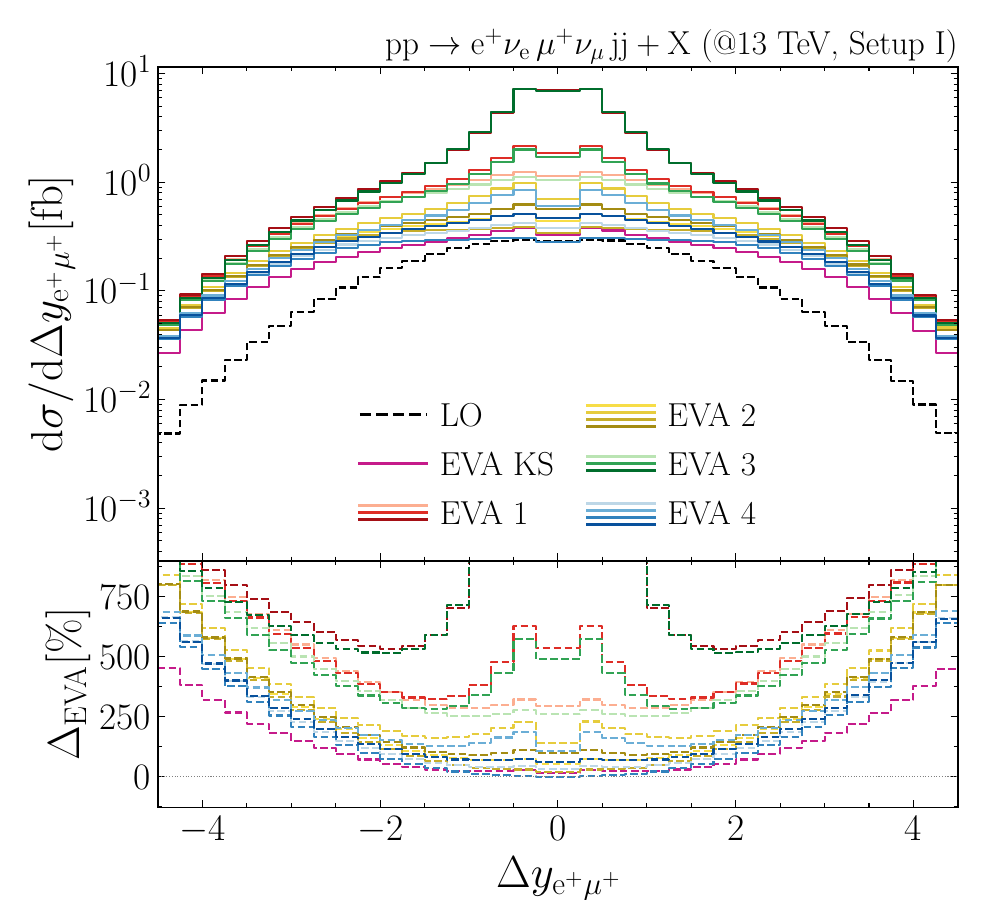}
	\caption{Differential distributions for $\Pp \Pp \to \Pep \Pgne \, \Pgmp \Pgngm \, \mathrm{j} \mathrm{j} + \mathrm{X}$ at the LHC with CM energy $13\TeV$ in setup I. Due to the size of deviation, we leave out the distributions for the EVA 1b and 3b variants. The upper panels show the  $\mathcal{O}(\alpha^6)$ and EVA contributions, the lower panels show the relative deviation in percent. }
	\label{fig:EVA_ptjfull}
\end{figure}

The results shown in \reffis{fig:EVAdist1}--\ref{fig:EVAdist3} for setup~III
demonstrate the validity of the EVA in general, in particular by the good 
approximation of the jet observables, which are most directly connected to the kinematic approximations for incoming W~bosons in the VBS process.
Going over to setup~II, which increases the range of jet transverse momenta, the features discussed above remain widely valid, although a loss in approximative quality is already observed.
In the ATLAS-inspired setup~I, which excludes the lowest jet transverse
momenta and includes the corresponding large-$p_{\text{T}}$ tails of the jets, all EVA variants fail to properly approximate the LO cross sections.
This is illustrated in \reffi{fig:EVA_ptjfull}.

\subsubsection*{Further investigations}

The unsatisfactory quality of the EVA in the realistic, ATLAS-inspired setup~I raises the question if any motivated changes in the EVA variant can improve this situation. 
Of course, in hindsight, we could always motivate any damping or fudge factors that improve the quality of the EVA because we know the full LO cross section.
However, such a procedure would not be useful when applying the EVA or related approximations to new situations. 
Instead of trying any fine-tuning, we have performed several additional investigations of the EVA by changing details in the EVA construction, such as the on-shell projection, or by changing parts of the experimental setup to understand the showstoppers better.
For instance, one of the major limitations of the EVA at LHC energies originates from the fact that the relevant energies in the process are not much larger than the W-boson mass.
Consequently, we have also varied the experimental 
setup~I to increase the energies in the subprocesses for radiating the W~bosons off the (anti)quarks and/or in the actual VBS scattering process.
The most important findings are:
\begin{itemize}
\item 
Increasing the pp CM energy from $13\TeV$ to $100\TeV$: \\
As expected, only increasing the hadronic scattering energy, does not change the picture at all. 
The change effectively only shifts the relevant momentum fractions~$x_i$ of the initial-state partons, but the whole partonic scattering process is evaluated with the same scattering energies and the same event selection.
The only difference originates from the fact that the PDFs are evaluated at lower 
$x_i$~values.


\item 
Tighter cuts on the leptonic invariant mass,
$M_{\Pep\Pgmp}> \{100,200, 1000\}\GeV$: \\
The EVA shows the same features as before, but the EVA cross sections increase 
by some non-trivial factor compared to the respective full LO calculation.
Higher cuts on $M_{\Pep\Pgmp}$ did not improve the overall performance for any of the EVA implementations either.
\item  
Stronger cut on transverse momentum of the leptons, 
$p_{\mathrm{T},\ell}>  \{100,200\}\GeV$: \\
Without combining this with the inverse $p_{\mathrm{T},\mathrm{j}}$ cut as in setups~II and III, 
this also does not show any improvement of the EVA quality
and only reduces the overall cross section.
However, in combination with, for instance, $p_{\mathrm{T},\mathrm{j}}<150$ GeV, 
the tighter cut on $p_{\mathrm{T},\ell}$
improves the quality of the EVA, see \reffi{fig:EVAptl}. 
Of course, this tight event selection drastically reduces the cross section.
 
\begin{figure}
	\centering
	\includegraphics[page=1,width=0.5\textwidth]{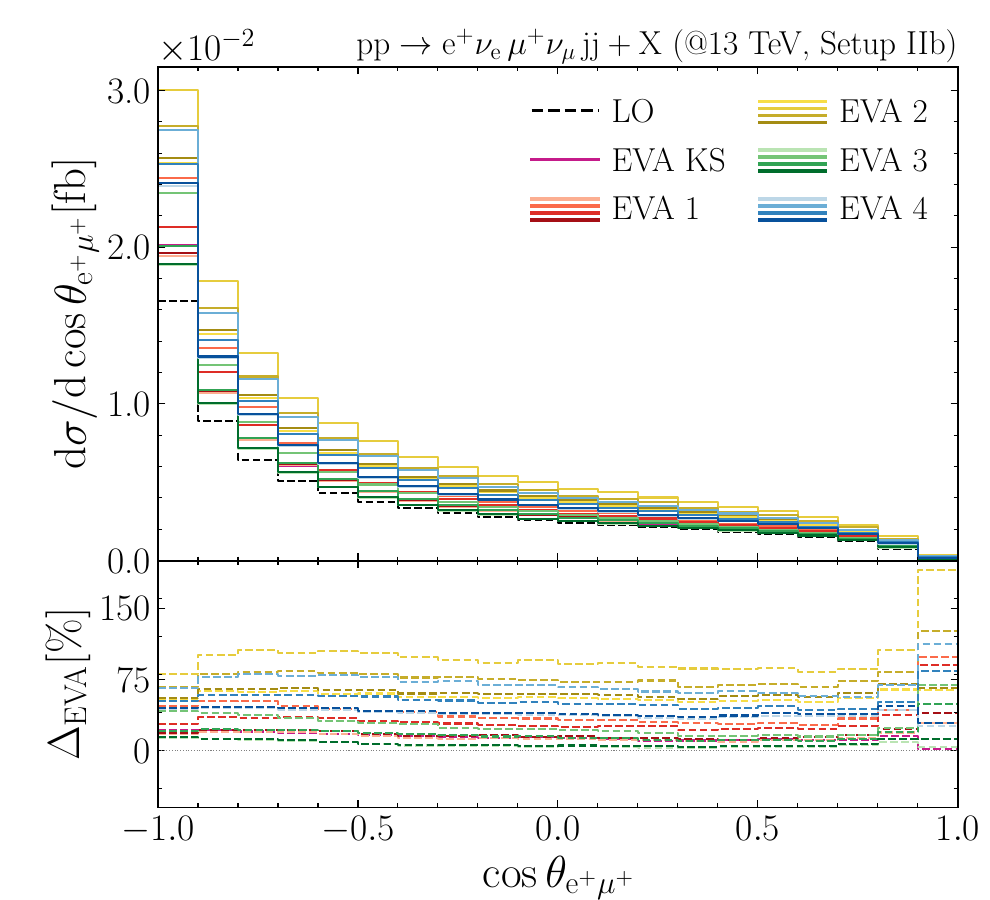}\hfill
	\includegraphics[page=2,width=0.5\textwidth]{wbas_ptj150_ptl100.pdf}\\
	\includegraphics[page=6,width=0.5\textwidth]{wbas_ptj150_ptl100.pdf}\hfill
	\includegraphics[page=7,width=0.5\textwidth]{wbas_ptj150_ptl100.pdf}
	\caption{Differential distributions for $\Pp \Pp \to \Pep \Pgne \, \Pgmp \Pgngm \, \mathrm{j} \mathrm{j} + \mathrm{X}$ at the LHC with CM energy $13\TeV$ in setup II and an additional $p_{\mathrm{T},\ell}>100\GeV$ cut. The upper panels show the  $\mathcal{O}(\alpha^6)$ and EVA contributions, the lower panels show the relative deviation in percent.}
	\label{fig:EVAptl}
\end{figure}

\item 
Damping factor: \\
The discussion of EVA results presented above has shown that the KS factor
$f^\text{KS}_{\lambda_i,1}$ defined in \refeq{eq:KSfactor}
positively affects the EVA's quality for large virtualities $|k_i^2|$
of the incoming W~bosons, but ruins the EVA for small $|k_i^2|\ll\MW^2$.
Therefore, we have investigated the effects of a damping factor
\begin{align}
\zeta_{i}(n)=
\begin{cases}
	\sqrt{\frac{n\MW^2}{\vert k_i^2\vert }}, & \text{if}\;\;\vert k_i^2\vert >n\MW^2,\\
	1, & \text{otherwise},
\end{cases}
\qquad i=1,2,
\end{align}
which is applied to all helicity contributions for each of the two incoming W~bosons
of momentum $k_i$.
Specifically, we have employed factors $\zeta_{i}(n)$ with $n=2,4$.
However, this only scales down the EVA predictions without improving the approximative 
quality in the distributions.

\item 
Changing the on-shell projection: \\
In \refapp{sec:evba-const}
we have argued that it is vital to take care of the positions of the
Coulomb poles in the EVA amplitude, \ie to the poles $\propto 1/t_\text{ww}$ 
or $1/u_\text{ww}$ for small $t_\text{ww}$ or $u_\text{ww}$,
respectively. Allowing for significant deformations of $t_\text{ww}$ for
$t_\text{ww}\to0$ and likewise for $u_\text{ww}$, easliy leads to diverging
cross-section predictions. The solution employed in our EVA formulation
preserved $t_\text{ww}$ and $u_\text{ww}$ during the on-shell projection,
however, at the price of increasing $s_\text{ww}$, as mentioned above. 
As an alternative, we have employed a modified on-shell projection
which keeps $s_\text{ww}$ fixed, but
modifies $t_\text{ww}$ and $u_\text{ww}$ in such a way that $t_\text{ww}$ is
unchanged in the limit $t_\text{ww}\to0$ and likewise for $u_\text{ww}$.
For instance, this can be achieved by the variants described in \eqref{eq:tu_limit},
and choosing 
$a=15$ to make sure that $\Delta K \cdot f_B(\cos\theta;a)$, with $\Delta K$  being defined in \eqref{eq:delta_K},
is small against $\MW^2$, for $\cos \theta \to -1$.

It turns out that these variants only lead to finite predictions
if events are discarded in which the hierarchy 
$t_\text{ww}\,\parbox{1em}{$> \\[-.9em] <$}\,u_\text{ww}$ is broken, \ie
if $\hat t_\text{ww} \,\parbox{1em}{$> \\[-.9em] <$}\,\hat u_\text{ww}$ is not fulfilled.
But even with this restriction, the approximative quality of the EVA is not
improved.

As another alternative, we have used the projection defined in \eqref{eq:tu_step}
which does not lead to finite cross sections even when
the preservation of the hierarchy
$t_\text{ww}\,\parbox{1em}{$> \\[-.9em] <$}\,u_\text{ww}$ is enforced by
discarding events. This shortcoming can be attributed to the non-smooth nature of this prescription which is a consequence of its discontinuity at $\vert t_\text{ww}\vert = \vert u_\text{ww}\vert$.
\end{itemize}

From the above analysis of different EVA variants we can only draw the conclusion
that it is not possible to make statements of the approximative quality of any EVA
variant without knowing the full off-shell cross section. 
There are two types of reasons for this ``failure'' of the EVA for LHC physics:
the fact that LHC analyses do not zoom into the phase-space region where the EVA
is designed for and internal difficulties in setting up the EVA.

As for the phase-space problem, the EVA requires that the jets that radiate off
the W~bosons should be very high energetic and extremely forward/backward pointing,
but the experimental analyses exclude the low-$p_{\mathrm{T}, \mathrm{j}}$ region
and include large $p_{\mathrm{T}, \mathrm{j}}$; moreover, the relevant energies
are not very much larger than $\MW$ at the LHC.
In the analysis of \citere{Accomando:2006hq} for VBS in $\Pep\Pem$ collisions, on the other hand,
the approximate quality of an EVA appeared more promising. We attribute this to a great deal
to the different event selection: For $\PW\PW\to\PW\PW$ in $\Pep\Pem$ collisions,
the forward/backward jets of our analysis are replaced by outgoing neutrinos which are
not constrained by any cuts, \ie the very forward/backward regions are fully integrated over.
Moreover, the incoming $\Pep\Pem$ always have a high energy and
in \citere{Accomando:2006hq} the VBS kinematics was even enhanced by cuts on the W~bosons
rather than by imposing cuts on their decay products.

As for internal difficulties in setting up an EVA,
our study has clearly shown that the EVA results very sensitively depend on the
treatment of the off-shellness, the on-shell projection
of the momenta, and the intermediate polarizations of the incoming W~bosons, while
the treatment of the nearly resonant outgoing W~bosons is not critical at all.
Moreover, the issue of photon poles in $t$- and $u$-channel diagrams for
the $\PW\PW\to\PW\PW$ subamplitude sets tight constraints on the kinematic
treatment of the VBS subprocess.
Certainly it would be possible to fine-tune the setting of such an approximation
by exploiting its internal freedom to obtain some acceptable approximative quality,
but this was not the incentive of our investigation. The true value of a working
EVA would have been to provide a reasonable approximation without knowing the
full result.

\section{Conclusion}
\label{sec:conclusion}

The experimental analysis of the scattering of massive electroweak gauge bosons
at the LHC
is an important milestone in the investigation of electroweak interactions,
in at least two different respects.
Firstly, quartic non-abelian gauge interactions, and thus gauge symmetry,
are probed directly. Secondly, massive gauge-boson scattering provides
an indirect window to electroweak symmetry breaking that is complementary to the
direct investigation of Higgs bosons.
Like-sign W-boson scattering is among the most important VBS channels
owing to its relatively clean experimental signature of two like-sign charged leptons
in the final state.

Theoretical predictions for VBS have been refined over the decades
several times, but full NLO calculations for scattering processes
with more than four particles in the final state have only become feasible since a couple of years ago owing to the
great progress in automation of multi-leg NLO calculations.
In this article we have presented a new calculation of the full tower of NLO
corrections to the process
$\Pp \Pp \to \Pep \Pgne \Pgmp \Pgngm \mathrm{j} \mathrm{j} + \mathrm{X}$,
which contains like-sign W-boson scattering as a subprocess.
Before our calculation, results for the full set of NLO corrections have been known only from a single calculation.
Particularly noteworthy is the size of the genuine electroweak
corrections, which have been found to be very large.
Our independent calculation confirms those corrections and thus put the theoretical
predictions for like-sign W-boson scattering on a more solid basis.
In detail, the older results are based on the Monte Carlo program \mocanlo using 
amplitudes from the generator \recola, while we have extended the
independent Monte Carlo integrator \bonsay,\footnote{The code \bonsay is partially publicly available at \url{https://github.com/cschwan/hep-mc} (Monte Carlo integrator) and \url{https://github.com/cschwan/hep-ps} (phase-space integration maps, implementation of the subtraction procedure, and interface to \openloops).
The remaining parts can be obtained from the authors upon request.} which uses amplitudes from
\openloops by default.
This Monte Carlo program has been used before in calculations for similar $2 \to 6$ multi-leg processes~\cite{Ballestrero:2018anz,Denner:2019tmn,Dittmaier:2019twg}.
Our tuned comparison to the existing NLO results reveals 
mutual agreement
between the two calculations within Monte Carlo errors
for all contributions but a single one that is, however, numerically unimportant.
While fully confirming the phenomenologically relevant NLO corrections that were
previously known, our new calculation underlines the importance of different independent
calculations for predictions for processes as complex as VBS, in order to 
guarantee full reliability in predictions.

Moreover, we have presented a detailed survey of NLO predictions for integrated and
differential cross sections in 
a setup with VBS selection cuts inspired by the ATLAS analysis.
The overall dominance of the genuine weak corrections of $\mathcal{O}(\alpha^7)$, which are
typically of the order of $\sim-12\%$ 
and even larger in high-energy tails of
distributions, is clearly visible. 
This overall dominance is enforced by the
dedicated VBS cuts, which widely suppress the QCD contribution at LO.
The corrections of $\mathcal{O}(\alpha_s\alpha^6)$,
which are mainly QCD corrections to the EW LO contribution, are subleading with
a typical impact at the 5\%~level.
The remaining two types of corrections of
$\mathcal{O}(\alpha_s^2\alpha^5)$ and $\mathcal{O}(\alpha_s^3\alpha^4)$ are kinematically
related to the QCD LO cross section and grow beyond the 1\%~level only where 
EW and QCD LO parts are of similar size.

In addition to the very complex calculation of the full NLO corrections we have worked
out a much simpler approximation, dubbed ``VBSA'',
based on a combination of the so-called VBS approximation,
which neglects suppressed partonic channels and interference effects,
and a double-pole approximation, which performs expansions about the produced W-boson resonances.
Since the approximation quality of this simplified calculation is about 
$\lsim1.5\%$ for integrated cross
sections and for the dominating parts of differential distributions,
it may serve as a basis for the efficient inclusion of new-physics effects beyond
the SM, for which a full NLO calculation might be too expensive.

Finally, for LO cross sections we have constructed and numerically tested
different versions of effective W-boson approximations, which are based on the
picture of W~bosons as partons in the proton. Since, however, the underlying 
assumption that collinear W-boson emission off quarks strongly dominates the cross
sections is not perfectly fulfilled for LHC energies, the approximation
quality of this approach is more qualitative, so that this approximation cannot serve
as basis for serious predictions at the LHC without fine-tuning the approximation
to a full off-shell prediction, in line with previous findings in the
literature. 
Our study shows that the effective W-boson approximation critically
requires to zoom into the high-energetic forward/backward region of the jets that radiate off
the scattering W~bosons, a requirement that is not really fulfilled in the VBS event selection
at the LHC.
Moreover, we find that the approximative quality very sensitively depends on
the treatment of off-shell and polarization effects of the incoming (but not of the outgoing) 
W~bosons. The appearance of photon poles in the $\PW\PW\to\PW\PW$ scattering amplitudes for
forward or backward scattered W~bosons sets further consistency constraints in the 
construction of the approximation. 


The overall message of this article is that in spite of the enormous complexity
of NLO predictions for $2\to6$ particle processes at hadron colliders
theoretical predictions for like-sign W-boson scattering are very well under control
in the SM. The construction of the technically much simpler approximation
for the NLO corrections, called VBSA above, was successful as well.
A logical next step is, thus, to extend the existing precision calculations
beyond the SM, either by including widely model-independent 
non-standard couplings within effective theories or by switching to full SM extensions.
This step beyond the SM might also be made by employing approximations similar to
the VBSA presented in this paper.

\section*{Acknowledgements}

We would like to thank Ansgar Denner, Sandro Uccirati, and Mathieu Pellen for their help in the comparison to results from
\mocanlo/\recola and Olivier Mattelaer for fruitful discussions on the EVA.
S.D.\ and C.S.\ acknowledge support by the DFG through grant DI 784/3.
Moreover, C.S.\ is supported by the German Research Foundation (DFG) under reference number DE 623/6-2.
R.W.\ acknowledges support by FRS-FNRS (Belgian National Scientific Research Fund) IISN projects 4.4503.16.
The authors acknowledge support by the state of Baden-Württemberg through bwHPC and the DFG through grant no INST 39/963-1 FUGG. Additional computational resources have been provided by the supercomputing facilities of the Université catholique de Louvain (CISM/UCL) and the Consortium des Équipements de Calcul Intensif en Fédération Wallonie Bruxelles (CÉCI) funded by the Fond de la Recherche Scientifique de Belgique (F.R.S.-FNRS) under convention 2.5020.11 and by the Walloon Region.

\appendix
\section{Detailed construction of the effective vector-boson approximation}
\label{sec:evba-const}

We consider the partonic subprocess
\begin{align}
	q_1(p_1)\,q_2(p_2)\to  q_1'(p'_1)\,q_2'(p'_2) +\nu_{\Pe}(p_3)\,\Pe^+(p_3^\prime)\; \nu_\mu(p_4)\,\mu^+(p_4^\prime),
	\label{eq:wbaprocess}
\end{align}
where $q_i$ ($q_i'$) with $i=1,2$ are the quarks or antiquarks in the initial (final) state with their corresponding 
momenta $p_i$ ($p_i^\prime$). The final-state leptons (neutrinos) have momenta 
$p_f^\prime$ ($p_f$) with $f=3,4$.
The EVA takes into account only diagrams containing the actual VBS subprocess
$\PW\PW\to\PW\PW$. 
The generic diagram representing the process \eqref{eq:wbaprocess}, in which the genuine VBS  subprocess occurs, is given in the first diagram of \reffi{fig:vbsEWA1}. We recall that this subset of Feynman diagrams does not define a gauge-invariant contribution to the full amplitude if no further changes to this contribution are made.

\subsection{Factorization of the amplitude}

\begin{figure}[!htbp]
	\centering
	\includegraphics[width=0.95\textwidth]{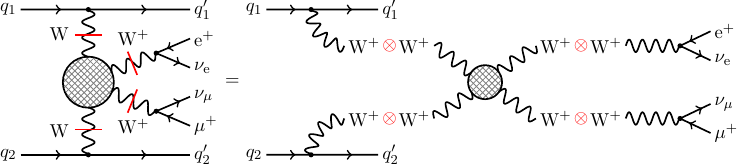}
	\caption{Generic VBS diagram split into its building blocks: 
	two $\PW$-boson production processes, $\PW^+\PW^+\to\PW^+\PW^+$ subprocess, and two $\PW$-boson decay processes.}
	\label{fig:vbsEWA1}
\end{figure}

Figure~\ref{fig:vbsEWA1} shows diagrammatically how the amplitude is split into building 
blocks to decompose the full $2\to6$ particle process into two W-boson production processes, 
the $2\to2$ VBS process $\PW\PW\to\PW\PW$, and the subsequent leptonic W-boson decays.

The corresponding amplitude can be written as
\begin{align}
\mathcal{M}_{qq}=J_1^{\mu}\,J_2^{\nu}\,\frac{\ci P_{\mu\mu'}(k_1)}{K_1}\frac{\ci P_{\nu\nu'}(k_2)}{K_2}\,\mathcal{M}_{\PW\PW\to \PW\PW}^{\mu^\prime\nu^\prime\sigma^\prime\rho^\prime}\,\frac{\ci P_{\sigma\sigma^\prime}(k_3)}{K_3}\frac{\ci P_{\rho\rho^\prime}(k_4)}{K_4}\,J_3^\sigma J_4^\rho,
\label{eq:amplitude}
\end{align}
where
\begin{align}
K_i&=k_i^2-M_\text{W}^2, &k_i&=p_i-p^\prime_i,& k_i^2&<0,& i&=1,2,\notag\\
K_f&=k_f^2-M_\text{W}^2+\ci M_\text{W}\Gamma_\text{W}, & k_f&=p_f+p^\prime_f,& k_f^2&>0,& f&=3,4,\notag\\
P^{\alpha\beta}(k_j)&=-g^{\alpha\beta}+k_j^{\alpha}k_j^{\beta}/k_j^2, &&&& & j&=1,2,3,4,
\end{align}
with the W-boson momenta $k_j$ and the momenta $p^{(\prime)}_i$ and $p^{(\prime)}_f$ of the massless
fermions in the initial and final states, respectively.
The four-vectors $J_j^\mu$ denote the fermion currents, and the amplitude of the $\PW\PW\to \PW \PW$ 
subprocess is denoted by $\mathcal{M}_{\PW\PW\to \PW\PW}^{\mu^\prime\nu^\prime\sigma^\prime\rho^\prime}$. 
Note the appearance of complex W-boson masses in the propagators of the outgoing W~bosons to properly describe the W~resonances.
The fermion currents $J_j^\mu$ can be written in the usual Dirac formalism as
\begin{align}
J_i^\mu&\equiv J_i^\mu(p_i,p_i^\prime)=\begin{cases}
\frac{\ci e}{\sqrt{2}s_\text{w}}\,\overline{u}(p'_i)\gamma^\mu \omega_-u(p_i),& \text{if}\,q_i,\,q'_i\,\text{are quarks},\\
\frac{\ci e}{\sqrt{2}s_\text{w}}\,\overline{v}(p_i)\gamma^\mu \omega_-v(p'_i),& \text{if}\,q_i,\,q'_i\,\text{are antiquarks},
\end{cases} & i&=1,2,\notag\\
J_f^\mu&\equiv J_f^\mu(p_f,p_f^\prime)=
\frac{\ci e}{\sqrt{2}s_\text{w}}\,\overline{u}(p_f)\gamma^\mu \omega_-v(p'_f),& f&=3,4,
\end{align}
where $\omega_-=(1-\gamma_5)/2$ is the left-handed chirality projector.
Owing to the Dirac equation, the currents fulfil the transversality conditions
\begin{align}
k_{i,\mu} J_i^\mu =
k_{f,\mu} J_f^\mu = 0.
\label{eq:Jtrans}
\end{align}
In \refeq{eq:amplitude} we made already use of this transversality when inserting 
the transversal projectors $P^{\alpha\beta}(k_j)$ instead of $-g^{\alpha\beta}$ in the numerators of the W-boson propagators.
In order to split the VBS contribution to the full amplitude into building blocks, 
we decompose the transversal projectors $P^{\alpha\beta}(k_j)$ into contributions from individual
W~polarizations described by 
four sets of polarization vectors $\varepsilon_{\lambda_j}^\mu\equiv\varepsilon_{\lambda_j}^\mu(k_j)$ 
($j=1,2,3,4$), \ie one set for each W~boson.
For each $j$, the $\varepsilon_{\lambda_j}^\mu$
are orthogonal to each other and to $k_j^\mu$, and are normalized according to
\begin{align}
\varepsilon_{\lambda_i}\cdot k_i &=0, & \varepsilon_{\lambda_i}\cdot\varepsilon_{\lambda_i^\prime}^*&=(-1)^{\lambda_i}\delta_{\lambda_i,\lambda_i'},&\lambda_i&=0,\pm1& i&=1,2,\notag\\
\varepsilon_{\lambda_f}\cdot k_f &=0, &\varepsilon_{\lambda_f}\cdot\varepsilon_{\lambda_f^\prime}^*&=-\delta_{\lambda_f,\lambda_f'},&\lambda_f&=0,\pm1, & f&=3,4.
\label{eq:epstrans}
\end{align}
As we introduce off-shell polarization vectors for the space-like momenta $k_i$ ($i=1,2$),
the norms of the longitudinal polarization vectors for the incoming W~bosons inherit $+$ signs according to the 
sign of $k_i^2$.
These four sets of polarization vectors satisfy the completeness relations
\begin{align}
\sum_{\lambda_i=-1}^{1}(-1)^{\lambda_i+1}\varepsilon_{\lambda_i}^\mu\varepsilon_{\lambda_i}^{*\nu}&=-g^{\mu\nu}+\frac{k_i^\mu k_i^\nu}{k_i^2}=P^{\mu\nu}(k_i),& i&=1,2,\notag\\
\sum_{\lambda_f=-1}^{1}\varepsilon_{\lambda_f}^\mu\varepsilon_{\lambda_f}^{*\nu}&=-g^{\mu\nu}+\frac{k_f^\mu k_f^\nu}{k_f^2}=P^{\mu\nu}(k_f),& f&=3,4.
\label{eq:epscompleteness}
\end{align}
Thus, the VBS amplitude can be written as
\begin{align}
\mathcal{M}_{qq}=\sum_{\lambda_1,\lambda_2,\lambda_3,\lambda_4}\frac{(-1)^{\lambda_1+\lambda_2}}{K_1K_2K_3K_4}\,\mathcal{M}_{\lambda_1}^{\text{prod}}\mathcal{M}_{\lambda_2}^{\text{prod}}\times\mathcal{M}^\text{VBS}_{\lambda_1\lambda_2\lambda_3\lambda_4}\times\mathcal{M}_{\lambda_3}^{\text{decay}}\mathcal{M}_{\lambda_4}^{\text{decay}},
\label{eq:evba_helicity}
\end{align}
where we have introduced the following shorthand for the subamplitudes
\begin{align}
\ci\mathcal{M}_{\lambda_i}^{\text{prod}}&=J_i^\mu\,\varepsilon^*_{\lambda_i,\mu},  
\qquad
\ci\mathcal{M}_{\lambda_f}^{\text{decay}}=J_f^\mu\,\varepsilon_{\lambda_f,\mu}, \notag\\
\mathcal{M}^\text{VBS}_{\lambda_1\lambda_2\lambda_3\lambda_4} &= \mathcal{M}_{\PW\PW\to \PW\PW}^{\mu^\prime\nu^\prime\sigma^\prime\rho^\prime} \,\varepsilon_{\lambda_1\mu^\prime} 
\,\varepsilon_{\lambda_2,\nu^\prime} \,\varepsilon^*_{\lambda_3,\rho^\prime} 
\,\varepsilon^*_{\lambda_4,\sigma^\prime}.
\label{eq:sub_helicity}
\end{align}
%

\subsubsection*{Explicit parametrization of off-shell momenta  and polarization vectors}

We construct the kinematical quantities for the
matrix elements in the centre-of-mass (CM) frame of the VBS subprocess, \ie
\begin{align}
	(k_1+k_2)^\mu=(k_3+k_4)^\mu=(\sqrt{s_{\Pw\Pw}},0,0,0),\qquad (k_1+k_2)^2=(k_3+k_4)^2=s_{\Pw\Pw},
\end{align}
where $\sqrt{s_{\Pw\Pw}}$ is the CM energy of the VBS subprocess. 
Within this frame we can choose the off-shell W-boson momenta as follows,
\begin{align}
	k_1^\mu&=(k_{0,1},0,0,k), & k_2^\mu&=(k_{0,2},0,0,-k),\nonumber\\
	k_3^\mu&=(k_{0,3},k^\prime\sin\theta,0,k^\prime\cos\theta), & k_4^\mu&=(k_{0,4},-k^\prime\sin\theta,0,-k^\prime\cos\theta),
	\label{eq:offshellmomenta}
\end{align}
with $\theta$ denoting the scattering angle,
\begin{align}
	k_{0,1}&=\frac{s_{\Pw\Pw}+k_1^2-k_2^2}{2\sqrt{s_{\Pw\Pw}}},& k_{0,2}&=\frac{s_{\Pw\Pw}-k_1^2+k_2^2}{2\sqrt{s_{\Pw\Pw}}}, 
	& k&=\frac{\sqrt{\lambda(s_{\Pw\Pw},k_1^2,k_2^2)}}{2\sqrt{s_{\Pw\Pw}}}, \qquad
\notag\\
	k_{0,3}&=\frac{s_{\Pw\Pw}+k_3^2-k_4^2}{2\sqrt{s_{\Pw\Pw}}},& k_{0,4}&=\frac{s_{\Pw\Pw}-k_3^2+k_4^2}{2\sqrt{s_{\Pw\Pw}}}
	&k' &=\frac{\sqrt{\lambda(s_{\Pw\Pw},k_3^2,k_4^2)}}{2\sqrt{s_{\Pw\Pw}}},
\end{align}
%
%
and the K\"all\'en function
\begin{align}
\lambda(x,y,z) = x^2+y^2+z^2-2xy-2xz-2yz.
\end{align}
Using this, we can further give a set of off-shell polarization vectors as
\begin{align}
	\varepsilon^\mu_{\pm,1}&=\frac{1}{\sqrt{2}}(0,-1,\mp \ci,0), &
	\varepsilon_{0,1}^\mu&=\frac{1}{\sqrt{-k_1^2}}(k,0,0,k_{0,1}),\nonumber\\
	\varepsilon^\mu_{\pm,2}&=\frac{1}{\sqrt{2}}(0,1,\mp \ci,0), &
	\varepsilon_{0,2}^\mu&=\frac{1}{\sqrt{-k_2^2}}(k,0,0,-k_{0,2}),\nonumber\\
	\varepsilon^\mu_{\pm,3}&=\frac{1}{\sqrt{2}}(0,-\cos\theta,\mp \ci,\sin\theta), &
	\varepsilon_{0,3}^\mu&=\frac{1}{\sqrt{ k_3^2}}(k^\prime,k_{0,3}\sin\theta,0,k_{0,3}\cos\theta),\nonumber\\
	\varepsilon^\mu_{\pm,4}&=\frac{1}{\sqrt{2}}(0,\cos\theta,\mp \ci,-\sin\theta), &
	\varepsilon_{0,4}^\mu&=\frac{1}{\sqrt{ k_4^2}}(k^\prime,-k_{0,4}\sin\theta,0,-k_{0,4}\cos\theta).
	\label{eq:eps_off}
\end{align}

\subsection{Helicity amplitudes}
\label{sec:helamplitudes}

Here we give explicit expressions for the helicity amplitudes appearing in 
\eqref{eq:evba_helicity} and \eqref{eq:sub_helicity}, using the Weyl--van-der-Waerden (WvdW) formalism for helicity
amplitudes in the conventions of \citere{Dittmaier:1998nn}. All amplitudes have been derived by hand and cross-validated with amplitudes generated using our standard in-house \mathematica routines before implementing them into \bonsay.

\subsubsection*{Helicity amplitudes of the emission process}

As the external fermions are considered massless, the helicities of the incoming quarks and 
antiquarks are conserved. Since W~bosons couple only to left-handed chirality, only one helicity configuration contributes to the
matrix element $\mathcal{M}^\text{prod}_{\lambda_i}$ in each initial-state splitting
process initiating the VBS process. This non-vanishing splitting amplitude is denoted as
$\mathcal{M}^\text{prod}_{\lambda_i}\equiv\mathcal{M}(\text{L},\text{L},\lambda_i)$ for quarks and as $\mathcal{M}^\text{prod}_{\lambda_i}\equiv\mathcal{M}(\text{R},\text{R},\lambda_i)$ for antiquarks, where L/R corresponds to the 
chirality of the incoming (anti)quark.
In terms of WvdW spinors, the non-vanishing amplitude of the emission process can be written as
\begin{align}
	\parbox[c][2.5cm]{0.25\textwidth}{\includegraphics[width=0.25\textwidth]{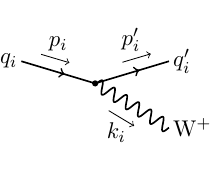}}\equiv\mathcal{M}^\text{prod}_{\lambda_i}(p_i,p_i^\prime,k_i)
=\frac{e}{\sqrt{2}s_\text{w}}p_i^{\dot{A}}p_i^{\prime B}\varepsilon^*_{\lambda_i,\dot{A}B},\qquad i=1,2,
	\label{eq:emission1}
\end{align}
where the spinors $p_i^{\dot{A}}$ and $p_i^{\prime B}$ are related to the four-vectors $p_i^\mu$ and $p_i^{\prime\mu}$ according to (2.9) and (2.20) of Ref.~\cite{Dittmaier:1998nn}.
The transition from the polarization vector $\varepsilon_{\lambda_i}^{*\mu}$ to the $2\times2$ matrix $\varepsilon^*_{\lambda_i,\dot{A}B}$ is given by
\begin{align}
	\varepsilon^*_{\lambda_i,\dot{A}B}=\varepsilon_{\lambda_i,\mu}^{*}\sigma^\mu_{\dot{A}B},\qquad i=1,2,
\end{align}
where the basic definitions for the $\sigma^\mu$ matrices and the spinorial
objects are given in Sect.~2 of Ref.~\cite{Dittmaier:1998nn}.
If the fermion line is replaced by an antifermion line the amplitude reads
\begin{align}
	\parbox[c][2.5cm]{0.25\textwidth}{\includegraphics[width=0.25\textwidth]{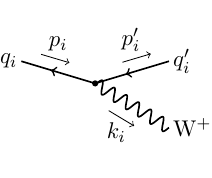}}\equiv\mathcal{M}^\text{prod}_{\lambda_i}(p_i,p_i^\prime,k_i)
=\frac{ e}{\sqrt{2}s_\text{w}}p_i^{\prime\dot{A}}p_i^{ B}\varepsilon^*_{\lambda_i,\dot{A}B},\qquad i=1,2,
\end{align}
where we have consistently replaced $p_i\leftrightarrow p_i^\prime$ compared to \refeq{eq:emission1}.

\subsubsection*{Helicity amplitudes of the decay process}

Analogously, we can derive the amplitude for the W-boson decay. In this case, only the helicity configuration $\pmb{\sigma}=(\sigma_f,\sigma_f^\prime)=(\text{L},\text{R})$ gives a non-vanishing contribution. The corresponding non-vanishing amplitude $\mathcal{M}^\text{decay}_{\lambda_f}\equiv\mathcal{M}(\text{L},\text{R},\lambda_f)$ can be expressed in the following way
\begin{align}
	\parbox[c][2.5cm]{0.25\textwidth}{\includegraphics[width=0.25\textwidth]{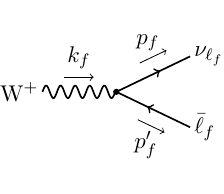}}\equiv\mathcal{M}^\text{decay}_{\lambda_f}(p_f,p_f^\prime,k_f)
=\frac{e}{\sqrt{2}s_\text{w}}p_f^{\prime\dot{A}}p_f^{ B}\varepsilon_{\lambda_f,\dot{A}B},\qquad f=3,4.
\end{align}
%

\subsubsection*{Helicity amplitudes of the VBS subprocess}

The Born matrix element $\mathcal{M}^\text{VBS}_{\lambda_1\lambda_2\lambda_3\lambda_4}$
of the VBS subprocess $\PW^+\PW^+\to \PW^+\PW^+$ 
receives contributions from seven diagrams.
We write the full amplitude as a sum over all contributions,
\begin{align}
	\mathcal{M}^\text{VBS}=\mathcal{M}^\text{QGC}+\mathcal{M}^{\gamma,t}+\mathcal{M}^{\mathrm{Z},t}+\mathcal{M}^{\mathrm{H},t}+\mathcal{M}^{\gamma,u}+\mathcal{M}^{\mathrm{Z},u}+\mathcal{M}^{\mathrm{H},u},
\end{align}
where $\mathcal{M}^\text{QGC}$ denotes the diagram including the quartic W-boson coupling, $\mathcal{M}^{N,t}$ and $\mathcal{M}^{N,u}$ denote the diagrams in which a neutral boson $N=\gamma,\PZ,\PH$ 
is exchanged in the $t$- and $u$-channel, respectively.
The corresponding analytical expressions are given in the following, using the conventions and Feynman rules of \citere{Denner:2019vbn}.
For simplicity, we give explicit expressions only for $t$-channel diagrams as
the corresponding $u$-channel diagrams can be obtained by consistently interchanging the outgoing bosons, \ie $k_3\leftrightarrow k_4$, $t\leftrightarrow u$, and $\lambda_3\leftrightarrow \lambda_4$. 
\begin{enumerate}[(a)]
	\item \textbf{QGC diagram}\\
	The contribution from the Feynman diagram including a quartic-gauge coupling reads
	\begin{align}
			&\parbox[c][2.75cm]{0.25\textwidth}{\includegraphics[width=0.25\textwidth]{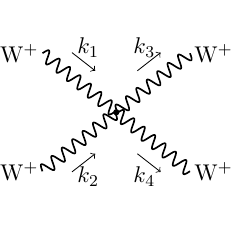}}\equiv\mathcal{M}^\text{QGC}_{\lambda_1\lambda_2\lambda_3\lambda_4}\notag\\
			&=\frac{e^2}{s_\text{w}^2}\Big\{2\,(\varepsilon_{\lambda_1}\cdot\varepsilon_{\lambda_2})(\varepsilon^*_{\lambda_3}\cdot\varepsilon^*_{\lambda_4})-(\varepsilon_{\lambda_1}\cdot\varepsilon^*_{\lambda_4})(\varepsilon_{\lambda_2}\cdot\varepsilon^*_{\lambda_3}) 
			- (\varepsilon_{\lambda_1}\cdot\varepsilon^*_{\lambda_3})(\varepsilon_{\lambda_2}\cdot\varepsilon^*_{\lambda_4})\Big\}.
	\end{align}
	\item \textbf{TGC diagrams}\\
	The contribution from $t$-channel diagrams with an intermediate boson $V=\mathrm{Z},\gamma$ is given by
	\begin{align}
			&\parbox[c][3cm]{0.25\textwidth}{\includegraphics[width=0.25\textwidth]{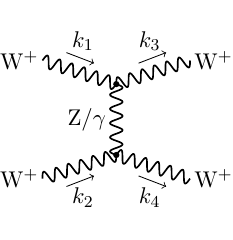}}\equiv\mathcal{M}^{V,t}_{\lambda_1\lambda_2\lambda_3\lambda_4}\notag\\
			&=-\frac{e^2\,C^2_{V\PW \PW}}{t-M^2_V}\Big\{4\,(\varepsilon_{\lambda_2}\cdot\varepsilon^*_{\lambda_4})\left[(\varepsilon_{\lambda_1}\cdot k_2)(\varepsilon^*_{\lambda_3}\cdot k_1)+(\varepsilon_{\lambda_1}\cdot k_3)(\varepsilon^*_{\lambda_3}\cdot k_2)\right]\notag\\
			&\hphantom{=-\frac{e^2}{t}}-4\,(\varepsilon^*_{\lambda_3}\cdot k_1)\left[(\varepsilon_{\lambda_1}\cdot \varepsilon^*_{\lambda_4})(\varepsilon_{\lambda_2}\cdot k_4)+(\varepsilon_{\lambda_1}\cdot \varepsilon_{\lambda_2})(\varepsilon^*_{\lambda_4}\cdot k_2)\right]\notag\\
			&\hphantom{=-\frac{e^2}{t}}-4\,(\varepsilon_{\lambda_1}\cdot k_3)\left[(\varepsilon_{\lambda_2}\cdot \varepsilon^*_{\lambda_3})(\varepsilon^*_{\lambda_4}\cdot k_2)+(\varepsilon^*_{\lambda_3}\cdot \varepsilon^*_{\lambda_4})(\varepsilon_{\lambda_2}\cdot k_4)\right]\notag\\
			&\hphantom{=-\frac{e^2}{t}}+4\,(\varepsilon_{\lambda_1}\cdot \varepsilon^*_{\lambda_3})\left[(\varepsilon_{\lambda_2}\cdot k_4)(\varepsilon^*_{\lambda_4}\cdot k_1)+(\varepsilon^*_{\lambda_4}\cdot k_2)(\varepsilon_{\lambda_2}\cdot k_1)\right]\notag\\
			&\hphantom{=-\frac{e^2}{t}}+(\varepsilon_{\lambda_1}\cdot \varepsilon^*_{\lambda_3})(\varepsilon_{\lambda_2}\cdot\varepsilon^*_{\lambda_4})(u-s)\Big\},
		\label{eq:Vtdiagram}
	\end{align}
	where $C_{\gamma \PW \PW}=1$ and $C_{\PZ\PW \PW}=-c_\text{w}/s_\text{w}$ are the triple-gauge couplings, and $M_V$ is the vector-boson mass, with $M_\gamma=0$.
	\item \textbf{Higgs-boson diagrams}\\
		The contribution from the $t$-channel diagram with an intermediate Higgs boson reads
	\begin{align}
			\parbox[c][3.3cm]{0.25\textwidth}{\includegraphics[width=0.25\textwidth]{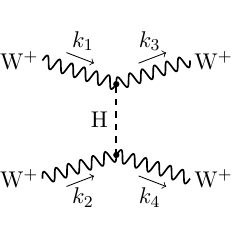}}&\equiv\mathcal{M}^{\mathrm{H},t}_{\lambda_1\lambda_2\lambda_3\lambda_4}
			=-\frac{e^2}{s_\text{w}^2}\frac{M_\text{W}^2}{t-M_\text{H}^2}\,(\varepsilon_{\lambda_1}\cdot\varepsilon^*_{\lambda_3})(\varepsilon_{\lambda_2}\cdot\varepsilon^*_{\lambda_4}).
	\end{align}
\end{enumerate}

\subsection{On-shell projection}

Up to this point, the treatment of the kinematics in the amplitudes and sub-amplitudes is exact.
However, an important step in the EVA is the projection onto on-shell W bosons, \ie $k_j\to\hat{k}_j$ with $\hat{k}_j^2=M_\text{W}^2$.
This projection is crucial as the diagrams considered in \reffi{fig:vbsEWA1} do not form a gauge-invariant subset of the complete $2\to6$ process.
In order to guarantee gauge independence 
of the amplitude and thus of the resulting cross section, we need to consider on-shell W-boson scattering.
Consequently, we evaluate the VBS amplitude $\mathcal{M}^{\text{VBS}}_{\lambda_1\lambda_2\lambda_3\lambda_4}$ with on-shell-projected momenta $\hat{k}_j$.

\subsubsection*{Issues with photon poles}

When performing the on-shell projection, we have to be especially careful due to the 
presence of massless photons in the VBS subprocesses (see \reffis{fig:vbs_t} and \ref{fig:vbs_u}), 
because the exchange of photons between the W~bosons can lead to poles in the matrix elements 
in or at the boundary of the phase space.
For the full $2\to6$ process \eqref{eq:VBSprocess}, the photon 
poles lie outside the physical phase space,
or are least sufficiently suppressed at the boundary of the phase space,
and do not hamper the phase-space integration, 
so that the total cross section exists without any kinematic cuts. However, in general, after on-shell projection $k_j\to\hat{k}_j$, the kinematic invariants $s_{\Pw\Pw},t_{\Pw\Pw}$, and $u_{\Pw\Pw}$  relevant for the appearance of these poles are deformed,
\begin{align}
	\hat{s}_{\Pw\Pw}&=(\hat{k}_1+\hat{k}_2)^2\ne(k_1+k_2)^2=s_{\Pw\Pw},\nonumber\\
	\hat{t}_{\Pw\Pw}&=(\hat{k}_1-\hat{k}_3)^2\ne(k_1-k_3)^2=t_{\Pw\Pw},\nonumber\\
	\hat{u}_{\Pw\Pw}&=(\hat{k}_1-\hat{k}_4)^2\ne(k_1-k_4)^2=u_{\Pw\Pw}.
\end{align}
Indeed, it is possible to keep up to two invariants fixed in the on-shell projection, 
for instance $\hat{s}_{\Pw\Pw}=s_{\Pw\Pw}$ and $\hat{u}_{\Pw\Pw}=u_{\Pw\Pw}$, but the third invariant will unavoidably change, as the invariants are related to each other according to the Mandelstam relations
%
\begin{align}
	s_{\Pw\Pw}+t_{\Pw\Pw}+u_{\Pw\Pw}=\sum_{j=1}^{4}k_j^2, \qquad \hat{s}_{\Pw\Pw}+\hat{t}_{\Pw\Pw}+\hat{u}_{\Pw\Pw}=4M_\text{W}^2.
\label{sec:Mandelstam-rel}
\end{align}
Hence, in the example
$\hat{s}_{\Pw\Pw}=s_{\Pw\Pw}$ and $\hat{u}_{\Pw\Pw}=u_{\Pw\Pw}$, the modified invariant $\hat{t}_{\Pw\Pw}$
is given by
\begin{align}
	\hat{t}_{\Pw\Pw}&=t_{\Pw\Pw}+\sum_{j=1}^{4}\Delta k_j^2,\qquad\Delta k_j^2=M_\text{W}^2-k_j^2.
\end{align}
For all energies $\sqrt{s_{\Pw\Pw}}>2M_\text{W}+n\Gamma_\text{W}$
sufficiently above the WW threshold ($n\gsim3$), 
the kinematic configuration in which the outgoing W-boson momenta are near the resonance, 
\ie $k_f^2\simeq M_\text{W}^2$, are dominant, and hence $|\Delta k_f^2|\ll \MW^2$.
In contrast, the incoming W-boson momenta $k_i$ are always space-like, 
\ie $k_i^2<0$, which implies $\Delta k_i^2>0$ and $\Delta k_i^2={\cal O}(\MW^2)$. 
Consequently, in the above-mentioned variant of an on-shell projection
the invariant $t_{\Pw\Pw}<0$ typically gets bigger after the projection:
\begin{align}
	\hat{t}_{\Pw\Pw}>t_{\Pw\Pw}, \quad\text{if }\;\hat{s}_{\Pw\Pw}=s_{\Pw\Pw},\quad\text{and }\;\hat{u}_{\Pw\Pw}=u_{\Pw\Pw}.
	\label{eq:biggeru}
\end{align}
Now consider the diagram of the VBS process in which a photon is exchanged in the $t$-channel in 
(\ref{eq:Vtdiagram}). Following Eq.\,(\ref{eq:biggeru}) we obtain $\hat{t}_{\Pw\Pw}\to0$ for certain kinematic regions, and subsequently
$\mathcal{M}^{\gamma,t}\propto 1/\hat{t}_{\Pw\Pw}\longrightarrow\infty.$
Indeed, the same considerations lead to an enhancement of the $u$-channel diagrams if we keep $s_{\Pw\Pw}$ and $t_{\Pw\Pw}$ fixed instead. This artifact renders the EVA-approximated cross section non-integrable for forward/backward scattering of the W bosons, in analogy to the situation in Rutherford forward scattering, as the invariants are related to the
on-shell-projected scattering angle $\hat{\theta}$ according to
\begin{align}
	\hat{t}_{\Pw\Pw}&=-\frac{\hat{s}_{\Pw\Pw}\hat{\beta}^2}{2}(1-\cos\hat{\theta})\xrightarrow{\hat{\theta}\to0}0,\nonumber\\
	\hat{u}_{\Pw\Pw}&=-\frac{\hat{s}_{\Pw\Pw}\hat{\beta}^2}{2}(1+\cos\hat{\theta})\xrightarrow{\hat{\theta}\to\pi}0,
\end{align}
where $\hat{\beta}=\sqrt{1-4M_\text{W}^2/\hat{s}_{\Pw\Pw}}$ is the velocity of the W bosons for on-shell-projected momenta in the VBS CM frame. 

\subsubsection*{Variants of on-shell projections}
\label{sec:evba-variants}

While the on-shell projection (OSP) is needed to obtain a gauge-independent
prediction, its numerical implementation is not unique and allows 
to explore different variants. 
The investigated variants are constructed to conserve certain quantities
of the process. However, owing to kinematic constraints such as the 
Mandelstam relations, each variant can conserve only a subset of the kinematic quantities and is forced to alter the rest.

\myparagraph{OSP 0 -- A variant conserving scattering energy and angle in the WW rest frame}

In order to avoid problems with the photon poles in the EVA-approximated cross section, we have to be careful with the chosen on-shell projection. A possible on-shell projection, as for instance proposed in Ref.~\cite{Accomando:2006hq},
requires that the angles of incoming and outgoing W~bosons are fixed while performing their on-shell limit. In our chosen VBS CM frame, this would translate into
\begin{align}
	\hat{k}_1^{\mu}&=\frac{\sqrt{s_{\Pw\Pw}}}{2} (1,0,0,\beta), & \hat{k}_2^\mu&=\frac{\sqrt{s_{\Pw\Pw}}}{2}(1,0,0,-\beta),\nonumber\\
	\hat{k}_3^\mu&=\frac{\sqrt{s_{\Pw\Pw}}}{2}(1,\beta\sin\theta,0,\beta\cos\theta), & \hat{k}_4^\mu&=\frac{\sqrt{s_{\Pw\Pw}}}{2}(1,-\beta\sin\theta,0,-\beta\cos\theta),
\end{align}
where $\hat{s}_{\Pw\Pw}=s_{\Pw\Pw}$ and thus $\hat{\beta}=\beta=\sqrt{1-4M_\text{W}^2/s_{\Pw\Pw}}$,
and $\hat{\theta}=\theta$ is the angle between incoming and outgoing W~boson directions in the WW CM frame.
In this case, the invariants $t_{\Pw\Pw}$ and $u_{\Pw\Pw}$ are altered, 
but our numerical results (not all presented in this article)
have shown that the EVA cross section indeed diverges. A possible way to avoid the poles while keeping this on-shell projection could be achieved by imposing a phase-space cut
on the scattering angle $\theta$ of the outgoing W bosons
(as, \eg, done in \citere{Accomando:2006hq}). 
However, these cuts are not practical, since we cannot reconstruct the full W-boson momenta. Furthermore, the EVA should maintain the existence of the total cross section without applying any cuts. 

\myparagraph{OSP 1 -- A variant conserving $t_{\Pw\Pw}$ and $u_{\Pw\Pw}$}

A possible solution is to choose the on-shell projection such that the invariants $t_{\Pw\Pw}$ and $u_{\Pw\Pw}$ are fixed in order to avoid the photon poles, 
and $s_{\Pw\Pw}$ is modified.
Then, the on-shell-projected momenta are given by
\begin{align}
	\hat{k}_1^\mu&=\frac{\sqrt{\hat{s}_{\Pw\Pw}}}{2}(1,0,0,\hat{\beta}), & \hat{k}_2^\mu&=\frac{\sqrt{\hat{s}_{\Pw\Pw}}}{2}(1,0,0,-\hat{\beta}),\nonumber\\
	\hat{k}_3^\mu&=\frac{\sqrt{\hat{s}_{\Pw\Pw}}}{2}(1,\hat{\beta}\sin\hat{\theta},0,\hat{\beta}\cos\hat{\theta}), & \hat{k}_4^\mu&=\frac{\sqrt{\hat{s}_{\Pw\Pw}}}{2}(1,-\hat{\beta}\sin\hat{\theta},0,-\hat{\beta}\cos\hat{\theta}),
	\label{eq:momenta_on}
\end{align}
where 
\begin{align}
	\hat{s}_{\Pw\Pw}=4\,M_\text{W}^2-t_{\Pw\Pw}-u_{\Pw\Pw} = {s}_{\Pw\Pw}+\sum_{j=1}^4 \Delta k_j^2,
\end{align}
follows from the Mandelstam relation, and $\hat{\beta}=\sqrt{1-4M_\text{W}^2/\hat{s}_{\Pw\Pw}}$. 
Since $\sum_{j=1}^4 \Delta k_j^2$ is systematically $>0$ for all relevant events,
$\hat{s}_{\Pw\Pw}$ is typically increased over ${s}_{\Pw\Pw}$.
The scattering angle $\hat{\theta}$ is determined by the condition
\begin{align}
	t_{\Pw\Pw}=\hat{t}_{\Pw\Pw}=(\hat{k}_1-\hat{k}_3)^2=-\frac{\hat{s}_{\Pw\Pw}\hat{\beta}^2}{2}(1-\cos\hat{\theta}).
\end{align}
Consequently, the scattering angle is given by
\begin{align}
	\cos\hat{\theta}=1+\frac{2t_{\Pw\Pw}}{\hat{s}_{\Pw\Pw}\hat{\beta}^2}.
	\label{eq:theta_on}
\end{align}
The corresponding on-shell polarization vectors are then 
\begin{align}
	\hat{\varepsilon}^\mu_{\pm,1}&=\frac{1}{\sqrt{2}}(0,-1,\mp \ci,0), &
	\hat{\varepsilon}_{0,1}^\mu&=\frac{1}{M_\text{W}}\sqrt{\frac{\hat{s}_{\Pw\Pw}}{4}}\,(\hat{\beta},0,0,1),\nonumber\\
	\hat{\varepsilon}^\mu_{\pm,2}&=\frac{1}{\sqrt{2}}(0,1,\mp \ci,0), &
	\hat{\varepsilon}_{0,2}^\mu&=\frac{1}{M_\text{W}}\sqrt{\frac{\hat{s}_{\Pw\Pw}}{4}}\,(\hat{\beta},0,0,-1),\nonumber\\
	\hat{\varepsilon}^\mu_{\pm,3}&=\frac{1}{\sqrt{2}}(0,-\cos\hat{\theta},\mp \ci,\sin\hat{\theta}), &
	\hat{\varepsilon}_{0,3}^\mu&=\frac{1}{M_\text{W}}\sqrt{\frac{\hat{s}_{\Pw\Pw}}{4}}\,(\hat{\beta},\sin\hat{\theta},0,\cos\hat{\theta}),\nonumber\\
	\hat{\varepsilon}^\mu_{\pm,4}&=\frac{1}{\sqrt{2}}(0,\cos\hat{\theta},\mp \ci,-\sin\hat{\theta}), &
	\hat{\varepsilon}_{0,4}^\mu&=\frac{1}{M_\text{W}}\sqrt{\frac{\hat{s}_{\Pw\Pw}}{4}\,}(\hat{\beta},-\sin\hat{\theta},0,-\cos\hat{\theta}).
	\label{eq:eps_on}
\end{align}
%

\myparagraph{OSP 2 -- A variant conserving $s_{\Pw\Pw}$ and the forward/backward limits of
$t_{\Pw\Pw}$ and $u_{\Pw\Pw}$}

An alternative on-shell projection is given by keeping $s_{\Pw\Pw}$ fixed and restoring the correct limits $t_{\Pw\Pw}, u_{\Pw\Pw}\to0$. For this we define
\begin{align}
\hat{t}_{\Pw\Pw}&= t_{\Pw\Pw} + \Delta K \,f(\cos\theta), \notag\\
\hat{u}_{\Pw\Pw}&= u_{\Pw\Pw} + \Delta K \,(1-f(\cos\theta)),
\label{eq:tu_limit}
\end{align}
with
\begin{align}
 \Delta K = \sum_{j=1}^4 M_\text{W}^2 - k_j^2,
 \label{eq:delta_K}
\end{align}
and where $\theta$ is the scattering angle of the off-shell momenta
in the WW rest frame and $f$ is some appropriate 
function. For $f$ we require that $f(\cos\theta)$ tends to $1(0)$ if $\cos\theta$ approaches $-1(+1)$.

In our applications we have tested the following smooth functions
\begin{align}
	f_A(\cos \theta)=\frac{1-\cos \theta}{2}\equiv\sin^2 \frac{\theta}{2},\qquad 
	f_B(\cos \theta;a)=\frac{1}{1+e^{a\cos \theta}},
\end{align}
with some parameter $a>0$ that controls the steepness of the function $f_B$ at $\cos \theta=0$, \ie  
$f'_B(0;a)=a/4$.%
\footnote{The parameter has to be sufficiently large to guarantee $f(1)\approx0$ and
$f(-1)\approx 1$ to sufficient accuracy.}
Additionally, we also tested some step function
\begin{align}
	f_C(\cos\theta)&=\begin{cases}
		1, & \text{if } \vert t_{\Pw\Pw}\vert> \vert u_{\Pw\Pw}\vert,\\
		0, & \text{otherwise}.
	\end{cases}
\label{eq:tu_step}
\end{align}
Then, the on-shell projected momenta are given by
\begin{align}
	\hat{k}_1^\mu&=\frac{\sqrt{s_{\Pw\Pw}}}{2}(1,0,0,\beta), & \hat{k}_2^\mu&=\frac{\sqrt{s_{\Pw\Pw}}}{2}(1,0,0,-\beta),\nonumber\\
	\hat{k}_3^\mu&=\frac{\sqrt{s_{\Pw\Pw}}}{2}(1,\beta\sin\hat{\theta},0,\beta\cos\hat{\theta}), & \hat{k}_4^\mu&=\frac{\sqrt{s_{\Pw\Pw}}}{2}(1,-\beta\sin\hat{\theta},0,-\beta\cos\hat{\theta}),
	\label{eq:momenta_on_alt}
\end{align}
where  $\beta=\sqrt{1-4M_\text{W}^2/\hat{s}_{\Pw\Pw}}$ and in analogy to \refeq{eq:theta_on} the scattering angle $\hat{\theta}$ is given by
\begin{align}
	\cos\hat{\theta}=1+\frac{2\hat{t}_{\Pw\Pw}}{s_{\Pw\Pw}\beta^2}
	= 1+\frac{2\left(t_{\Pw\Pw} + \Delta K f(\cos\theta)\right)}{s_{\Pw\Pw}\beta^2}.
	\label{eq:theta_on_alt}
\end{align}
%

\subsubsection*{Transversality and normalization conditions}

Before on-shell projection, the external fermion currents $J_j(p_j,p_j')$ and the W-boson polarization vectors $\varepsilon_{\lambda_j}$ are transversal to the off-shell W-boson momenta and obey normalization and completeness relations, \ie~\eqref{eq:Jtrans} to \eqref{eq:epscompleteness}.
Instead, after on-shell projection, all polarization vectors are normalized according to the standard signature,
\begin{align}
	\hat{\varepsilon}_{\lambda_j}\cdot	\hat{\varepsilon}^*_{\lambda'_j}=-\delta_{\lambda_j,\lambda_j'},\qquad\lambda_j=0,\pm1,\quad j=1,2,3,4,
	\label{eq:epstrans_on}
\end{align}
and the completeness relations \refeq{eq:epscompleteness}
changes to
\begin{align}
	\sum_{\lambda_j=-1}^{1}\hat{\varepsilon}_{\lambda_j}^\mu\hat{\varepsilon}_{\lambda_j}^{*\nu}=-g^{\mu\nu}+\frac{\hat{k}_j^\mu \hat{k}_j^\nu}{M_W^2},\qquad j=1,2,3,4.
	\label{eq:epscompleteness_on}
\end{align}
Furthermore, the transversality of the on-shell momenta $\hat{k}_j$ to the external fermion currents $J_j$ is violated in general, if the momenta $p_j$ and $p_j'$ appearing in these currents are not simultaneously transformed as well. A possible projection of the outgoing fermion momenta is given by the following parametrization
\begin{alignat}{2}
	\hat{p}_3^{\prime\mu}&=p_3^{\prime\mu}\frac{M_\text{W}^2}{2\hat{k}_3\cdot p_3^\prime}, 
	\qquad & \hat{p}_3^\mu&=\hat{k}_3^\mu-\hat{p}_3^{\prime\mu},\notag\\
	\hat{p}_4^{\prime\mu}&=p_4^{\prime\mu}\frac{M_\text{W}^2}{2\hat{k}_4\cdot p_4^\prime}, 
	&  \hat{p}_4^\mu&=\hat{k}_4^\mu-\hat{p}_4^{\prime\mu},
	\label{eq:out_projection}
\end{alignat}
where we fix the angles of the momenta $p_3^\prime$ and $p_4^\prime$ 
corresponding to the observable (charged) final-state leptons. By this transformation, 
the transversality condition is preserved, 
\begin{align}
J_f(\hat{p}_f,\hat{p}_f')\cdot\hat{k}_f=0,\qquad f=3,4.
	\label{eq:trans_in}
\end{align}
However, as the external quarks $q_i$ and $q_i^\prime$ are massless, the incoming W-boson momenta $k_i=p_i-p_i^\prime$ are always space-like ($k_i^2<0$) and never get on shell within the physical phase-space region. This implies that there is no deformation $p_i\to\hat p_i$ of the incoming quark momenta in the physical phase space leading to $(\hat p_i-\hat p_i')^2=\MW^2$, \ie
\begin{align}
	p_i-p_i'=k_i
	\centernot{\longrightarrow}
	\hat{p}_i-\hat{p}_i'=\hat{k}_i,\qquad i=1,2,
\end{align}
and hence, the transversality condition $J_i\cdot \hat{k}_i\ne0$ is always broken.

\subsection{Implemented EVA variants}

In the following, we introduce different versions of the EVA amplitude, 
conserving or violating various transversality conditions, 
the results of which are investigated in \refse{sec:evba_results}. Our starting point is the fully decomposed off-shell amplitude \refeq{eq:evba_helicity} with its different helicity subamplitudes:
\begin{alignat}{3}
	\text{production:}&	
	&\quad  &\mathcal{M}_{\lambda_i}^\text{prod}(\varepsilon_{\lambda_i}^*,p_i,p_i'),
	& \qquad i&=1,2,\notag\\
	\text{decay:}&  
	& \quad
	&\mathcal{M}_{\lambda_f}^\text{decay}(\varepsilon_{\lambda_f},p_f,p_f'),
	&  f&=3,4,\notag\\
	\text{VBS:}& 
	& &\mathcal{M}_{\lambda_1\lambda_2\lambda_3\lambda_4}^\text{VBS}
	(\varepsilon_{\lambda_i}, \varepsilon_{\lambda_f}^*,k_j),
	& \qquad j&=1,2,3,4.
	\label{eq:hel_subamps}
\end{alignat}
For all EVA versions, we employ the on-shell transitions in the VBS subprocess,
\begin{alignat}{3}
	\varepsilon_{\lambda_i}&\xrightarrow{\eqref{eq:eps_on}}\hat{\varepsilon}_{\lambda_i},
	&\qquad \quad
	\varepsilon^{*}_{\lambda_f}&\xrightarrow{\eqref{eq:eps_on}}\hat{\varepsilon}^{*}_{\lambda_f},
	&\qquad \quad
	k_j&\xrightarrow{\eqref{eq:momenta_on}}\hat{k}_j,\notag\\
	i&=1,2, &\quad f&=3,4, &\quad j&=1,2,3,4,
	\label{eq:vbs_transition}
\end{alignat}
which defines the on-shell-projected VBS subamplitude
\begin{align}
	\mathcal{M}_{\lambda_1\lambda_2\lambda_3\lambda_4}^\text{VBS}
	\xrightarrow{\eqref{eq:vbs_transition}}
	\widehat{\mathcal{M}}_{\lambda_1\lambda_2\lambda_3\lambda_4}^\text{VBS}\equiv\widehat{\mathcal{M}}_{\lambda_1\lambda_2\lambda_3\lambda_4}^\text{VBS}(\hat{\varepsilon}_{\lambda_i},\hat{\varepsilon}_{\lambda_f}^*,\hat{k}_j). 
	\label{eq:vbs_on}
\end{align}
These are crucial, since any inconsistencies in the VBS subamplitude would potentially
result in a wrong behaviour of $\mathcal{M}_{\lambda_1\lambda_2\lambda_3\lambda_4}^\text{VBS}$
owing to violations of gauge cancellations.
In other parts of the amplitude we have more freedom and can check different approximations.
To be precise, we introduce five independent modification categories, 
characterized by $c_k=0,1$ for $k=1,\dots,5$, which are summarized in \refta{tab:vba_modes}.
Further, to allow for a 
transparent presentation of the different versions, we introduce the subscript notation
\begin{align}
	\mathcal{M}^\text{EVA}_{c_1c_2c_3c_4c_5},\qquad c_k=0,1,\quad k=1,\dots,5
\end{align}
in the EVA amplitude.

\subsubsection*{\boldmath{Production process -- category $c_1$}}

In the first category, we consider the transversality of the polarization vectors 
$\varepsilon^*_{\lambda_i}$ 
of the incoming W~bosons and the on-shell-projected W-boson momenta $\hat{k}_i$ in the production process, as shown in the first column of \refta{tab:vba_modes}. In order to achieve this transversality, we consistently replace the off-shell polarization vectors with their corresponding on-shell versions
\begin{align}
\varepsilon^{*}_{\lambda_i}
\xrightarrow[c_1=1]{\eqref{eq:eps_on}}
\hat{\varepsilon}^{*}_{\lambda_i},\qquad i=1,2,
	\label{eq:prod_eps_on}
\end{align}
in the production amplitudes
\begin{align}
\mathcal{M}_{\lambda_i,0}^\text{prod}(\varepsilon_{\lambda_i}^*,p_i,p_i^\prime)
	\,\xrightarrow{\eqref{eq:prod_eps_on}}
	\mathcal{M}_{\lambda_i,1}^\text{prod}(\hat{\varepsilon}_{\lambda_i}^*,p_i,p_i^\prime),\qquad  i=1,2,
	\label{eq:evba_prod}
\end{align}
where we have added the additional subscript $c_1$ on the amplitude $\mathcal{M}_{\lambda_i,c_1}^\text{prod}$
to exemplify the on-shell transition \eqref{eq:prod_eps_on}.

\subsubsection*{\boldmath{Decay process -- categories $c_2,c_3$}}

In the second column of \refta{tab:vba_modes}, we display whether the polarization vectors $\varepsilon_f$ in the decay process are transversal to the on-shell projected W-boson momenta $\hat{k}_f$.  Like in \eqref{eq:prod_eps_on}, this transversality can be satisfied by the transition
\begin{align}
	\varepsilon_{\lambda_f}
	\xrightarrow[c_2=1]{\eqref{eq:eps_on}}
	\hat{\varepsilon}_{\lambda_f},\qquad f=3,4.
	\label{eq:dec_eps_on}
\end{align}
Further, it is possible to satisfy the transversality of the lepton currents $J_f(p_f,p_f')$ and the on-shell projected W-boson momenta $\hat{k}_f$ by performing
\begin{align}
	\left\{p_f,p_f'\right\}
	\xrightarrow[c_3=1]{\eqref{eq:out_projection}}
		\left\{\hat{p}_f, \hat{p}'_f\right\},\qquad f=3,4.
	\label{eq:lep_mom_on}
\end{align}
Using, for instance, both transitions, we obtain the fully on-shell projected decay amplitude according to
\begin{align}
\mathcal{M}_{\lambda_f,00}^\text{decay}(\varepsilon_{\lambda_f},p_f,p_f^\prime)
	\xrightarrow{\eqref{eq:dec_eps_on}\,\eqref{eq:lep_mom_on}}
	\mathcal{M}_{\lambda_f,11}^\text{decay}(\hat{\varepsilon}_{\lambda_f},\hat{p}_f,\hat{p}_f^\prime),\qquad  f=3,4.
\end{align}
In general, the transitions \refeq{eq:dec_eps_on} and \refeq{eq:lep_mom_on} allow for 
four different versions of the decay amplitudes.

\subsubsection*{\boldmath{Helicity factor -- $c_4$}}

In the introduction of the off-shell polarization vectors, we considered
the different signs in the normalization of the longitudinal polarization vectors, see Eq.~\eqref{eq:epstrans}.
These lead to a sign-factor of $(-1)^{\lambda_1+\lambda_2}$ in the factorized scattering amplitude \eqref{eq:evba_helicity}. 
However, after on-shell projection, this sign factor is not necessarily correct and might be 
reinstalled to account for possibly significant correlations between W~helicity configurations.
Hence, we introduce the generalized helicity factor
\begin{align}
	h_{\lambda_1\lambda_2,c_4}=(-1)^{(\lambda_1+\lambda_2)c_4},
\end{align}
and analyse its effect on the EVA cross-section prediction.
For $c_4=0$, the factor simply is \mbox{$h_{\lambda_1\lambda_2,0}=1$}.

\subsubsection*{\boldmath{KS factor -- $c_5$}}

Kuss and Spiesberger (KS) pointed out in Ref.~\cite{Kuss:1995yv} 
that the longitudinal polarization vector $\varepsilon^\mu_{0,i}$
of an off-shell vector boson with momentum $k_i$,
as given in \refeq{eq:eps_off}, involves an explicit factor of
$1/\sqrt{|k_i^2|}$.
For transverse polarization vectors, there is no such $k_i^2$ dependence
in $\varepsilon^\mu_{\pm,i}$.
For final-state vector bosons, where the resonance region
$k_f^2\sim M_{V_f}^2$ is strongly dominating, there is certainly no 
extra enhancement by the $1/\sqrt{k_f^2}$ factors, but for 
initial-state vector bosons there might be some relevance
in the factors $1/\sqrt{-k_i^2}$. 
In Ref.~\cite{Kuss:1995yv}, KS propose to explicitly restore those factors
for longitudinal initial-state vector bosons by extra factors
applied to the on-shell-projected VBS subamplitudes.
However, whether these factors systematically improve the EVA
amplitude to come closer to the true off-shell
behaviour of the full amplitude is a non-trivial question because
of the presence of gauge cancellations, especially in the presence
of longitudinal gauge bosons.
To analyse this question, we optionally include these extra factors
\begin{align}
	 f^\text{KS}_{\lambda_i,c_5}=
	\begin{cases}
	\frac{M_\PW}{\sqrt{-k_i^2}}\delta_{\lambda_i0}+\delta_{\lambda_i\pm1}, & \text{if } c_5=1,\\
	1, & \text{otherwise},
	\end{cases}
	\qquad i=1,2,
\label{eq:KSfactor}
\end{align}
which are controlled by the flag $c_5$.

\subsubsection*{Summary}

In summary, we denote the amplitudes in the considered EVA versions by
\begin{align}
\mathcal{M}_{c_1c_2c_3c_4c_5}^{\text{EVA}}=
\sum_{\lambda_1,\lambda_2, \lambda_3,\lambda_4}
&\frac{f^\text{KS}_{\lambda_1,c_5}\,f^\text{KS}_{\lambda_1,c_5}\,h_{\lambda_1\lambda_2,c_4}}{K_1K_2K_3K_4}
\notag\\
&\times
\mathcal{M}_{\lambda_1,c_1}^{\text{prod}}\,\mathcal{M}_{\lambda_2,c_1}^{\text{prod}}
\times
\widehat{\mathcal{M}}^\text{VBS}_{\lambda_1\lambda_2\lambda_3\lambda_4}
\times
\mathcal{M}_{\lambda_3,c_2c_3}^{\text{decay}}\,\mathcal{M}_{\lambda_3,c_2c_3}^{\text{decay}}.
\label{eq:evba_on}
\end{align}
For instance, the KS amplitude~\cite{Kuss:1995yv} reads in our notation
\begin{align}
	\text{EVA KS}:\qquad\mathcal{M}_{ {\color{mybad}0}1111}^{\text{EVA}}.
\end{align} 
Additionally, in \refse{sec:evba_results}, we investigate the following EVA versions
\begin{align}
	\text{EVA 1\{a,b,c,d\}}&:\qquad \{\mathcal{M}_{111{\color{mybad}00}}^{\text{EVA}}, \mathcal{M}_{11{\color{mybad}000}}^{\text{EVA}}, \mathcal{M}_{1{\color{mybad}0000}}^{\text{EVA}}, 
	\mathcal{M}_{1{\color{mybad}0}1{\color{mybad}00}}^{\text{EVA}}\},\notag\\
	\text{EVA 2\{a,b,c,d\}}&:\qquad \{\mathcal{M}_{{\color{mybad}0}11{\color{mybad}00}}^{\text{EVA}}, \mathcal{M}_{{\color{mybad}0}1{\color{mybad}000}}^{\text{EVA}}, \mathcal{M}_{{\color{mybad}00000}}^{\text{EVA}}, 
	\mathcal{M}_{{\color{mybad}00}1{\color{mybad}00}}^{\text{EVA}}\},\notag\\
	\text{EVA 3\{a,b,c,d\}}&:\qquad \{\mathcal{M}_{1111{\color{mybad}0}}^{\text{EVA}}, \mathcal{M}_{11{\color{mybad}0}1{\color{mybad}0}}^{\text{EVA}}, \mathcal{M}_{1{\color{mybad}00}1{\color{mybad}0}}^{\text{EVA}}, 
	\mathcal{M}_{1{\color{mybad}0}11{\color{mybad}0}}^{\text{EVA}}\},\notag\\
	\text{EVA 4\{a,b,c,d\}}&:\qquad \{\mathcal{M}_{{\color{mybad}0}111{\color{mybad}0}}^{\text{EVA}}, \mathcal{M}_{{\color{mybad}0}1{\color{mybad}0}1{\color{mybad}0}}^{\text{EVA}}, \mathcal{M}_{{\color{mybad}000}1{\color{mybad}0}}^{\text{EVA}}, 
	\mathcal{M}_{{\color{mybad}00}11{\color{mybad}0}}^{\text{EVA}}\},
	\label{eq:evba_vesions}
\end{align}
where the approximations according to the categories $c_{1\dots5}$ are highlighted in red.

\bibliographystyle{JHEPmod}
\bibliography{refs}

\end{document}